\documentclass{article}
\usepackage{graphicx} 
\usepackage{subcaption}
\usepackage{authblk}

\usepackage[utf8]{inputenc}
\usepackage[margin = 1in]{geometry}
\usepackage{amsmath}
\usepackage{amsthm}
\usepackage{amssymb}
\usepackage{natbib}
\usepackage{xcolor}
\usepackage{algorithm}
\usepackage{algpseudocode}
\usepackage{comment}
\usepackage{cleveref}
\usepackage{appendix}
\usepackage{subcaption}
\newtheorem{example}{Example}[section]

\newtheorem{definition}{Definition}[section]

\title{Some aspects of robustness in modern Markov Chain Monte Carlo}
\author[1]{Sam Power}
\author[2]{Giorgos Vasdekis}
\affil[1]{School of Mathematics, University of Bristol}
\affil[2]{School of Mathematics, Statistics and Physics, Newcastle University}

\begin{document}

\maketitle

\begin{abstract}
    Markov Chain Monte Carlo (MCMC) is a flexible approach to approximate sampling from intractable probability distributions, with a rich theoretical foundation and comprising a wealth of exemplar algorithms. While the qualitative correctness of MCMC algorithms is often easy to ensure, their practical efficiency is contingent on the `target' distribution being reasonably well-behaved. 
    
    In this work, we concern ourself with the scenario in which this good behaviour is called into question, reviewing an emerging line of work on `robust' MCMC algorithms which can perform acceptably even in the face of certain pathologies. 
    
    We focus on two particular pathologies which, while simple, can already have dramatic effects on standard `local' algorithms. The first is \emph{roughness}, whereby the target distribution varies so rapidly that the numerical stability of the algorithm is tenuous. The second is \emph{flatness}, whereby the landscape of the target distribution is instead so barren and uninformative that one becomes lost in uninteresting parts of the state space. In each case, we formulate the pathology in concrete terms, review a range of proposed algorithmic remedies to the pathology, and outline promising directions for future research.
\end{abstract}

\section{Introduction}

Markov Chain Monte Carlo (MCMC) is a computational approach to sampling from a probability measure which is specified in terms of an unnormalised density function, a task which is ubiquitous in various facets of Statistics, Machine Learning, Signal Processing, and Computational Physics. Given a `target' measure $\pi$ living on a reasonable state space $\mathcal{X}$, the modern MCMC toolbox by now contains a range of `standard' methods for solving the sampling problem (i.e. producing approximate samples from $\pi$), with some illustrative methods including Random Walk Metropolis \cite{sherlock2010random, andrieu2024explicit}, Gibbs Sampling \cite{gelfand2000gibbs, ascolani2024entropy}, Langevin Monte Carlo \cite{roberts.tweedie:96, dalalyan2017theoretical, durmus2017nonasymptotic, durmus2019high}, Hamiltonian Monte Carlo \cite{neal2011mcmc, betancourt:18}, Hit-and-Run \cite{belisle1993hit, rudolf2013hit}, Slice Sampling \cite{neal2003slice, natarovskii2021quantitative, power2024weak}, and more. A general trend within the area is to seek general-purpose methods, which can be applied to a wide range of targets, without needing to focus on low-level details of the problem, and still expect reasonable (if not necessarily optimal) performance.

While basic MCMC often `works out of the box' for well-behaved target distributions -- indeed, this is somehow the success story of MCMC at large -- there is an increasing realisation that problems of contemporary interest are prone to deviate from this good behaviour in various ways, leading the reliability of these default algorithms to be called into question somewhat. The present paper thus seeks to review current practice in MCMC, with a focus on modern developments in `robust' algorithms which are able to perform well, even in the face of somewhat `pathological' target distributions.

The format of the paper is then as follows: in Section \ref{sec:notation} we present the basic notation we will be using, in Section \ref{sec:Standard.MCMC:00}, we recall the construction of some prominent MCMC methods, and the associated principles for designing such methods. We then give a brief interlude describing some pathologies which can cause these prominent methods to break down. In Section \ref{sec:roughtness:00}, we focus on the pathology of `roughness' of the target distribution, illustrating how it can arise, and reviewing a number of strategies for resolving it in practice. In Section \ref{sec:heavy:00}, we focus on the sibling pathology of `heavy-tailedness' of the target distribution, similarly treating how and why it arises, and how and why it can be addressed. We then conclude with some review of the area as a whole, pointing to some meaningful open problems and suggestions for future research.

\subsection{Notation}\label{sec:notation}
We will be using $\pi$ to denote a probability measure on $\mathbf{R}^d$ and when the distinction is clear, we will be abusing notation, using $\pi$ to denote the density of the measure with respect to the Lebesgue measure (assuming it exists).  With that in mind, we will be writing
\begin{equation*}
    \pi(x)=\frac{1}{Z} \exp \left( -U(x) \right),
\end{equation*}
where $U$ is defined up to an additive constant, and $Z$ is the normalising constant, which in most settings will be assumed to be unknown, or at least difficult to calculate. To indicate this, we will sometimes be writing
\begin{equation*}
    \pi(x) \propto \exp \left(  -U(x) \right)
\end{equation*}
to indicate that $\pi$ is known up to a multiplicative constant. $U$ will sometimes be called {\it the potential}. $\mu$ will typically denote a measure on $\mathbf{R}^{2d}$, having $\pi$ as its first marginal. Typically, $\mu$ will be constructed by augmenting $\mathbf{R}^d$ with an auxiliary variable $v \in \mathbf{R}^d$; in that setting, we will write $\psi$ for the law of $v$, then writing $K = - \log \psi$ for the associated potential. $P$, $Q$ will generally denote Markov kernels on $\mathbf{R}^d$, which we will again occasionally conflate with their densities with respect to the Lebesgue measure. We write $\mathsf{TV}\left( \nu, \nu^\prime \right)$ to denote the total variation distance between probability measures $\nu$ and $\nu^\prime$. $\mathcal{N} \left( m, C \right)$ denotes the Gaussian distribution with mean vector $m$ and covariance matrix $C$.

$\| \cdot \|$ will denote the Euclidean norm in $\mathbf{R}^d$, and $\| \cdot \|_1$ will denote the $1$-norm.

For a continuous-time path $(x_t)_{t \geq 0} \in \mathbf{R}^d$, we write $\dot{x}_t$ or $\dot{x}$ to denote the derivative with respect to time $t$. $\nabla$ will be used to denote the usual Euclidean gradient, with $\mathrm{D}$ sometimes being used for the same notion in the univariate case.

\section{Design of MCMC Procedures}\label{sec:Standard.MCMC:00}

\subsection{MCMC Basics}

In concrete terms, the setup and goals of MCMC are as follows, focusing on the (dominant) setting of sampling on $\mathcal{X} = \mathbf{R}^d$: one is interested in approximate sampling from the probability measure $\pi \in \mathcal{P} \left( \mathbf{R}^d \right)$, whose density with respect to Lebesgue measure is known, at least up to a possibly-unknown normalising constant. As a standard abuse of notation, we will also use $\pi$ for this density. 


Given $\pi$, the MCMC architecture should design a Markov kernel $P$ such that the Markov chain on $\mathbf{R}^d$ which is driven by $P$ will converge in distribution to this $\pi$. Typically, this is achieved by at least imposing that the kernel $P$ leaves $\pi$ \emph{invariant}, i.e.
\begin{align*}
    x \sim \pi, \quad y \sim P \left( x, \cdot \right) \implies y \sim \pi \qquad \text{marginally},
\end{align*}
and in practice, it is common to even impose that the kernel $P$ be \emph{reversible} with respect to $\pi$, in the sense that
\begin{align*}
    x \sim \pi, \quad y \sim P \left( x, \cdot \right) \implies \left( x, y \right) \overset{\mathrm{d}}{=} \left( y, x \right),
\end{align*}
or at least that some similarly `local' invariance condition is satisfied. Of course, each of these properties is satisfied by the trivial kernel $P \left( x, \mathrm{d} y \right) = \delta \left( x, \mathrm{d} y \right)$ (which is certainly not fit for purpose as an MCMC kernel!), and so practical kernels must generally also satisfy some non-degeneracy conditions, to the tune of irreducibility, aperiodicity, and a meaningful form of \emph{ergodicity}. For the latter point, a minimal requirement is that for a suitably large set of $x \in \mathbf{R}^d$, it holds that
\begin{align*}
    \text{as}\, n\to\infty,\qquad \mathsf{TV} \left( P^n \left( x, \cdot \right), \pi \right) \to 0,
\end{align*}
i.e. that for more-or-less arbitrary initialisations, the chain will indeed converge towards $\pi$ in total variation distance. Beyond this qualitative ergodicity, a more demanding condition which should be satisfied by `good' MCMC kernels is \emph{exponential ergodicity}, whereby for a similarly large set of $x \in \mathbf{R}^d$, one can even write that
\begin{equation}\label{gemoetric.ergo:00}
    \text{as}\, n\to\infty,\qquad \mathsf{TV} \left( P^n \left( x, \cdot \right), \pi \right) \leq \rho^n \cdot V \left(x\right),
\end{equation}
where the implied rate constant $\rho$ is taken uniform in $x$, but the prefactor $V \left( x \right)$ is allowed to depend on $x$. 

Given an exponentially-ergodic Markov kernel $P$ which is $\pi$-invariant, a number of desirable consequences hold, e.g. for ergodic averages along the MCMC trace, one can obtain Laws of Large Numbers, Central Limit Theorems, Concentration Inequalities for suitable expectands, and so on. In this respect, exponential ergodicity reassures the MCMC user that their method is fit for purpose, at least in principle (of course, in practice, if the constant $\rho$ is unacceptably close to $1$, then these reassurances are of little use). In this regard, we view exponential ergodicity as a `gold standard' of sorts for separating `bad' and `good' kernels, coarse as the distinction may be.

While there are many possible ways in which to construct such a process, a prevailing strategy for constructing such processes in practice is to identify some continuum process which is known to be precisely $\pi$-invariant, and then find some way to simulate it in practice. Of course, other principles for deriving MCMC algorithms do exist, but cannot be our focus here.

Upon identifying such a continuum process, a number of options arise. 

In certain situations, the continuum process is not just an idealised object, but can actually be realised algorithmically. A simple instance of this comes with Markov jump processes (MJPs), which can typically be simulated exactly by use of the Doob-Gillespie algorithm (initially introduced as a practical method in \cite{gillespie1977exact} for biochemical simulations; see also \cite{anderson2015stochastic}). For Piecewise-Deterministic Markov Processes (PDMPs), exact simulation is possible in some cases, provided that the dynamics are tractable and that certain inhomogeneous Poisson point processes are feasible to simulate (following e.g. \cite{lewis1978simulation}; see also discussion in \cite{bouchard2018bouncy, bierkens2019zig} and similar works). For Stochastic Differential Equations (SDEs) of It\^{o} type, exact simulation tends to only be possible for rather stylised models, either those with substantial analytic tractability (e.g. Gauss-Markov processes), or satisfying strong a priori bounds; see e.g. \cite{beskos2005exact, beskos2006retrospective, blanchet2020exact}. For SDEs with more general driving noises, the situation is typically even worse.

Moving beyond the somewhat restrictive class of processes for which exact simulation is feasible, the world of numerical analysis comes to our rescue, and provides us with an ample set of tools for approximate-yet-accurate simulation of a much wider class of processes. While these approximations typically incur some bias in the invariant measure of the numerical process, there are various strategies for quantifying and correcting for this bias, at least to the point that the impact of the bias falls below that of the Monte Carlo variance. Among such strategies, we highlight Multilevel Monte Carlo \cite{giles2015multilevel} as an archetypal example with wide and relatively straightforward applicability. For large MJPs, Gillespie's algorithm can become costly to implement exactly, and so approximation strategies such as tau-leaping \cite{gillespie2001approximate} become appealing (see also \cite{hird.livingstone.zanella:22} for other ways of approximating a certain relevant class of MJPs). For PDMPs, circumstances can dictate that an approximate treatment is necessitated for either the dynamics, the jumps, or both of them. When the dynamics are tractable, it is natural to consider e.g. piecewise-constant or piecewise-linear approximations to the event rates (as in \cite{pagani2024nuzz, corbella2022automatic, andral2024automated}). When the dynamics are also intractable, then the use of numerical methods for ODEs like splitting schemes becomes appealing; this was pursued in \cite{bertazzi.bierkens.dobson:22}. For SDEs, the literature on numerical approximation is yet more vast (see e.g. \cite{platen1999introduction, higham2001algorithmic, milstein2004stochastic, e2021applied, higham2021introduction}), though it is fair to say that in the context of MCMC simulation, the set of popular strategies is plainly dominated by the Euler-Maruyama method and various splitting schemes. 

Within certain communities (the Bayesian statistical community perhaps chief among them), there is a cultural preference to avoid questions of asymptotic bias entirely by constructing implementable Markov kernels whose algorithmic invariant measure coincides exactly with the intended target measure. A generic approach to this is furnished by the `propose-accept/reject' paradigm, as exemplified by \cite{metropolis1953equation, hastings1970monte, tierney1998note, andrieu2020general}. In this paradigm, the user `proposes' a move according to some exactly-implementable Markov kernel, evaluates the favourability of the move according to some carefully-designed quantitative criterion, and then either `accepts' the move to this proposed point, or `rejects' it, staying in place. This strategy (which is known as the \emph{Metropolis-Hastings filter} or simply \emph{``Metropolisation"})  provides a rather generic approach to converting approximate numerical dynamics into dynamics which are genuinely ergodic with respect to $\pi$, eliminating concerns of asymptotic bias, albeit at the cost of slowing down the original dynamics with this conservatism; we term this the \emph{Monte Carlo-Exact} paradigm for MCMC.

\subsection{From Continuum Processes to Practical Algorithms}

Having outlined these three high-level approaches to the design of MCMC kernels, in this sub-section, we review some continuum processes of interest, and detail some ways in which these processes have been translated into practical algorithms, following this rubric. 

The aim is not to cover all possible processes which could be considered, nor to review all algorithms which have been derived in this manner. Our goal is to instead focus on processes and algorithms which are of clear popular interest, have been adopted as `default' methods to some extent or another, and whose merits and limitations are relatively well-understood at a rigorous level.

\subsubsection{The Overdamped Langevin Diffusion}

Given a continuously-differentiable distribution $\pi$ on a Euclidean space $\mathbf{R}^d$, the \emph{overdamped Langevin diffusion} targeting $\pi$ is the It\^{o} SDE given by
\begin{equation}\label{overdamped.langevin:00}
    \mathrm{d} X_t = \nabla \log \pi \left( X_t \right) \,\mathrm{d}t + \sqrt{2} \,\mathrm{d} W_t
\end{equation}
where $W_t$ is a usual Wiener process on $\mathbf{R}^d$. Under rather mild conditions on $\pi$, this process is both reversible and ergodic with respect to $\pi$, and exponential ergodicity is assured under various possible conditions, the most restrictive (and informative) of which roughly asks that $\pi$ be asymptotically approximately log-concave at infinity, or at least similarly confining.

Outside of a handful of rather simple cases, exact simulation of this process is challenging. The most general treatments tend to impose some rather strong conditions on the growth of $\nabla \log \pi$, and tend to only apply in settings of quite low (effective) dimension; these are not really seen to be viable approaches in the MCMC context (though have found value in other contexts of statistical inference; see e.g. \cite{beskos2006exact, fearnhead2008particle, papaspiliopoulos2009methodological}).

As concerns approximate numerical simulation, the options become somewhat more appealing. Application of a straightforward Euler-Maruyama discretisation to the Langevin process leads to the `Unadjusted Langevin Algorithm' (ULA), also `Langevin Monte Carlo' (LMC), which takes the form
\begin{equation}\label{ULA:00}
  \textit{{\bf ULA:}} \ \   X_n = X_{n-1} + h \cdot \nabla \log \pi(X_{n-1}) + \sqrt{2 \cdot h} \cdot \xi_n , \ \ \xi_n \sim \mathcal{N} \left( 0, \mathbf{I}_d \right)
\end{equation}
(with $h > 0$ a step-size parameter). This has been studied extensively (see e.g. \cite{dalalyan2017theoretical, durmus2017nonasymptotic, durmus2019high, chewi2024analysis, durmus2024asymptotic}), and represents a reliable baseline method in various contexts where high accuracy is not a first priority.

Finally, to take the Langevin diffusion into the Monte Carlo-Exact paradigm, application of the Metropolis-Hastings device to the aforementioned Euler-Maruyama discretisation yields the so-called \emph{Metropolis-Adjusted Langevin Algorithm (MALA)} (introduced variously in \cite{rossky1978brownian, besag1994comments}), well-studied in works including \cite{roberts.tweedie:96, bou2013nonasymptotic, chewi2021optimal, wu2022minimax}. In this method, writing $Q$ for the transition kernel corresponding to the ULA update, one simulates a `proposal' move $Y_n \sim Q \left( X_{n-1}, \cdot \right)$, evaluates the \textit{Metropolis-Hastings} acceptance probability $\alpha \left( X_{n-1}, Y_n \right)$, where
\begin{align}\label{eq:Metropolis.accept.reject.probability}
    \alpha \left( x, y \right) = \min \left\{ 1, \frac{\pi \left( y \right) \cdot Q \left( y, x \right)}{\pi \left( x \right) \cdot Q \left( x, y\right)} \right\},
\end{align}
and then moves to $Y_n$ with this probability, otherwise remaining at $X_{n-1}$ (that is $X_n$ is either set to be $Y_n$ or $X_{n-1}$, with probabilities $\alpha(X_{n-1},Y_n)$ and $1-\alpha(X_{n-1},Y_n)$ respectively).

\begin{figure}[ht]
    \centering
    \begin{subfigure}[t]{0.47\textwidth}
        \includegraphics[width=\linewidth]{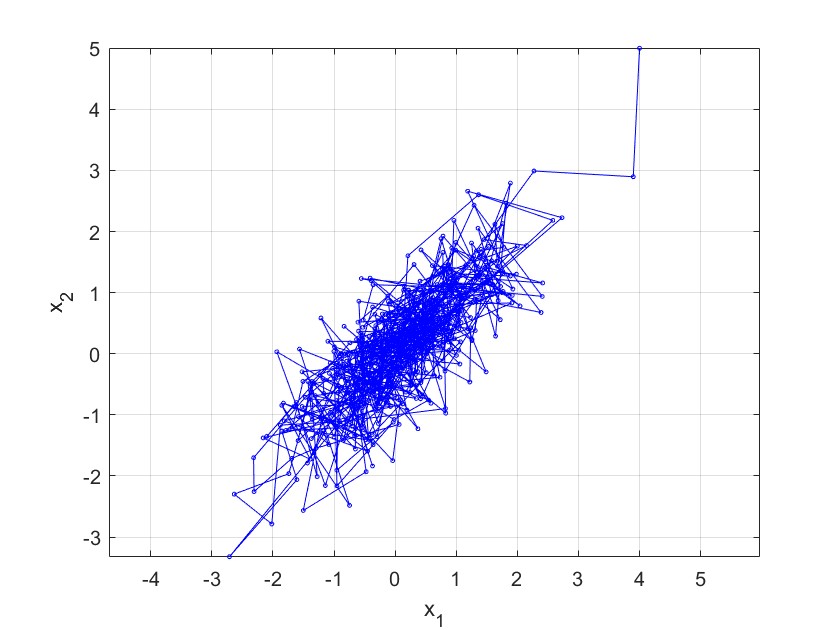}
        \caption{Scatter plot}
        \label{Fig:Gauss.MALA:part1}
    \end{subfigure}
    \hfill
    \begin{subfigure}[t]{0.47\textwidth}
        \includegraphics[width=\linewidth]{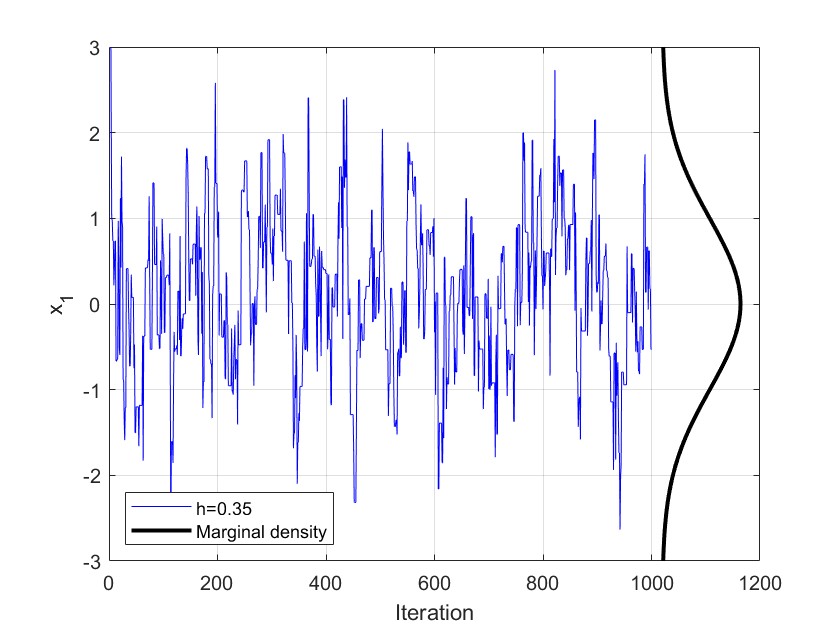}
        \caption{Traceplot of first coordinate samples}
        \label{Fig:Gauss.MALA:part2}
    \end{subfigure}
    
    \vspace{0.5cm}  

    \begin{subfigure}[t]{0.47\textwidth}
        \includegraphics[width=\linewidth]{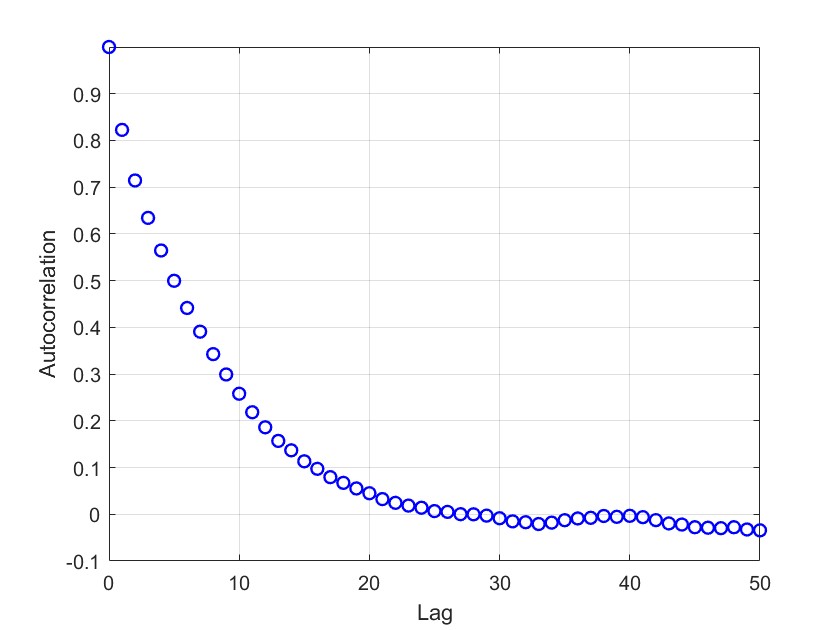}
        \caption{Autocorrelation of first coordinate samples}
        \label{Fig:Gauss.MALA:part3}
    \end{subfigure}
    \hfill
    \begin{subfigure}[t]{0.47\textwidth}
    \includegraphics[width=\linewidth]{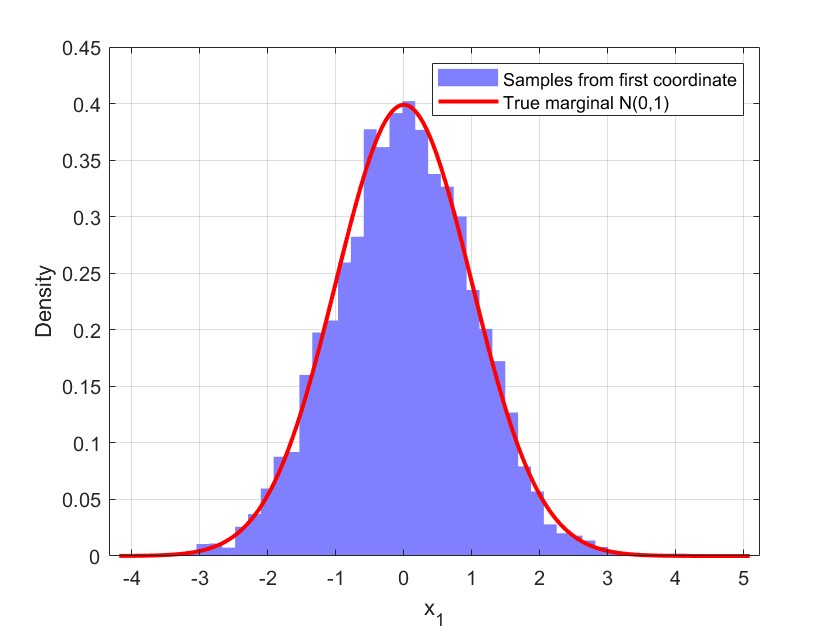}
        \caption{Histogram of first coordinate samples}
        \label{Fig:Gauss.MALA:part4}
    \end{subfigure}
    \vspace{1mm}
    \caption{MALA algorithm on a two-dimensional correlated Gaussian target (\ref{eq:2d.Gauss}). Step-size $h=0.35$. Number of iterations $N=10^4$. Starting point $x_0=(4,5)$.}
    \label{Fig:Gauss.MALA}
\end{figure}

To showcase the performance of the MALA algorithm, let us consider a two-dimensional Gaussian distribution with mean $m = [0 , 0]^\top$ and covariance matrix $C = \begin{bmatrix}1 & 0.8 \\ 0.8 & 1\end{bmatrix}$, meaning that the density is of the form
\begin{equation}\label{eq:2d.Gauss}
    \pi(x) = \frac{1}{2\pi \sqrt{\det(C)}} \cdot \exp\left( -\frac{1}{2} x^\top C^{-1} x \right), \quad x \in \mathbf{R}^2.
\end{equation}
We ran MALA with step-size $h=0.35$ on this target for $N=10^4$ iterations, starting from $x_0=(4,5)$, rather far from the mode. The step-size was chosen so that the algorithm accepts between $50-60 \%$ of the proposed jumps, as recommended by e.g. \cite{roberts.rosenthal:02}. Our results are summarised in Figure \ref{Fig:Gauss.MALA}, where we present four plots. The first is the scatter plot, which shows the path of the process in $\mathbf{R}^2$. The second plot is the traceplot of the first coordinate against time, with the marginal density plotted on the right for reference. From these two plots, we can see that the process captures the shape of the target very efficiently. The third plot is the autocorrelation plot, showing how correlated the algorithm's samples are as a function of the time increment between them. The plot shows a fairly rapid decay of this correlation to zero as time-distance between samples increases, which indicates good algorithmic performance (noting that one would ideally have i.i.d. samples from the target, with no autocorrelations whatsoever). Finally, the fourth plot is the histogram of the samples of the first coordinate, with the true marginal distribution overlaid with red colour for reference.

Finally, in terms of using numerical approximations of diffusions to motivate MCMC algorithms, we note in passing that while the seminal Random Walk Metropolis Algorithm (see e.g. \cite{sherlock2010random, andrieu2024explicit} is seldom presented as a numerical method for simulating the Langevin diffusion \emph{per se}, there exist a range of theoretical results (e.g. \cite{gelfand1991weak, gelman1997weak}) which reveal that it can be reasonably interpreted in this way.

\subsubsection{Piecewise-Deterministic Markov Processes}


In informal terms, a Piecewise-Deterministic Markov Process (PDMP) is a continuous-time Markov process evolving in a continuous space whose dynamics comprise deterministic evolution along the flow of an ordinary differential equation, punctuated by instantaneous Markovian jumps. Although initially introduced by \cite{davis1984piecewise} as a model for problems in Operations Research and Control, this class of processes has undergone a sort of renaissance in the last decade, as their application to MCMC simulation has been examined in greater depth.

For the purposes of this review, we focus our attention on three prototypical PDMPs for Monte Carlo applications: \emph{Refreshed Hamiltonian Dynamics}, the \emph{Bouncy Particle Sampler}, and the \emph{Zig-Zag Process}. In each of these, the state undergoing dynamics is a kinetic particle, i.e. a position accompanied by a velocity, i.e. $z = \left(x, v \right) \in \mathbf{R}^d \times \mathbf{R}^d$, designed so that at equilibrium, the particle $z$ is distributed according to $\mu \left( \mathrm{d} z \right) = \pi \left( \mathrm{d} x \right) \cdot \psi \left( \mathrm{d} v \right)$, where $\psi$ is some simple, known distribution over a suitable `velocity space'. Interpreting $v$ as the velocity of the system, we refer to $K(v):=-\log \psi(v)$ as the {\it Kinetic energy}, with $U$ being called the {\it potential} energy.

\textbf{Refreshed Hamiltonian Dynamics}

For Refreshed Hamiltonian Dynamics (introduced by \cite{bou2017randomized} as `\emph{Randomised Hamiltonian Monte Carlo (RHMC)}'\footnote{We prefer here to distinguish the idealised physical process with its application to simulation, cf. the use of `Langevin diffusion' and `Langevin Monte Carlo'.}), one takes $\psi \left( \mathrm{d} v \right) = \mathcal{N} \left( \mathrm{d} v ; 0, \mathbf{I}_d \right)$, meaning that $K(v)=\frac{1}{2}v^2$, and the particle moves according to Hamilton's equations with `Hamiltonian' $\mathcal{H} \left( z \right) = - \log \mu \left( z \right)$, i.e.
\begin{equation}\label{eq:Hamiltonian.Dynamics}
    \dot{x} = v, \qquad \dot{v} = \nabla \log \pi \left( x \right),
\end{equation}
punctuated by jumps in the velocity of the form
\begin{align*}
    v^\prime \sim \mathcal{N} \left( \mathrm{d} v^\prime ; \rho \cdot v , \left( 1 - \rho^2 \right) \cdot \mathbf{I}_d \right)
\end{align*}
for some correlation parameter $\rho \in \left[ -1, 1 \right]$, occurring at some time-homogeneous Poisson rate $\lambda > 0$.

Outside of very simple cases, exact simulation of the Hamiltonian dynamics is intractable, and so this is rarely treated in practice. Fortunately, numerical methods for Hamiltonian dynamics are rather well-developed, and have been especially well-served by the field of Geometric Numerical Integration \cite{hairer2003geometric}. The dominant approach to approximate simulation of Refreshed Hamiltonian Dynamics involves use of the splitting integrator of (St\o{}rmer-Verlet, Leapfrog, Strang, etc.) to solve the dynamics, with the jump times and events handled exactly by simulating exponential random variables.

Monte Carlo-Exact resolutions of Refreshed Hamiltonian Dynamics are ultimately rather similar in character; one again simulates the dynamics approximately using the Verlet integrator (with step-size $h$) for a randomly-chosen number of steps (e.g. $L \sim \mathsf{Poisson} \left( \lambda \cdot h^{-1} \right)$) and then accepting or rejecting the final state $z_L$ with a suitable probability, following the Metropolis-Hastings prescription (as in (\ref{eq:Metropolis.accept.reject.probability})); this is typically what is meant by (Metropolised) `Hamiltonian Monte Carlo'. A number of variations of Hamiltonian Monte Carlo exist, with many intriguing idiosyncrasies, but for the cohesion of presentation, we focus on this one.

\begin{figure}[ht]
    \centering
    \begin{subfigure}[t]{0.47\textwidth}
        \includegraphics[width=\linewidth]{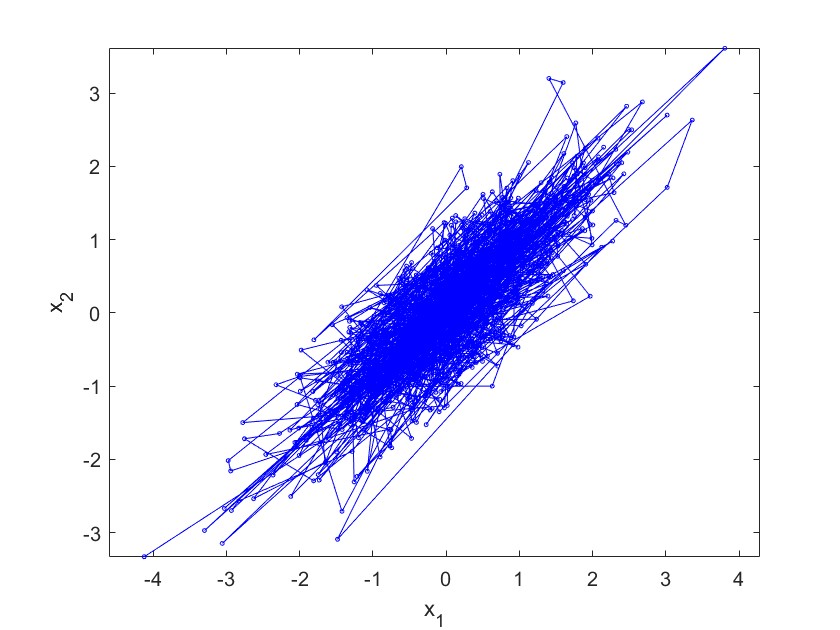}
        \caption{Scatter plot}
        \label{Fig:Gauss.RHMC:part1}
    \end{subfigure}
    \hfill
    \begin{subfigure}[t]{0.47\textwidth}
        \includegraphics[width=\linewidth]{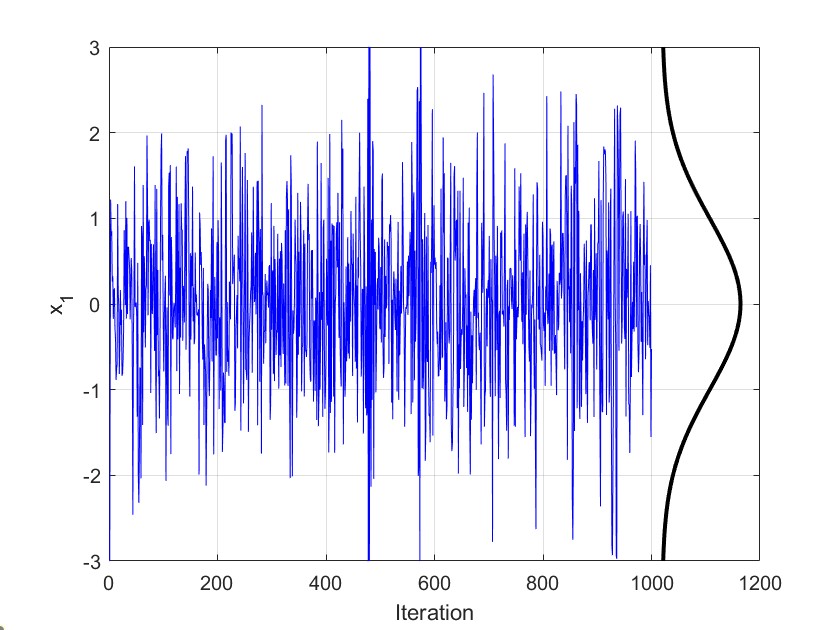}
        \caption{Traceplot of first coordinate samples}
        \label{Fig:Gauss.RHMC:part2}
    \end{subfigure}
    
    \vspace{0.5cm}  

    \begin{subfigure}[t]{0.47\textwidth}
        \includegraphics[width=\linewidth]{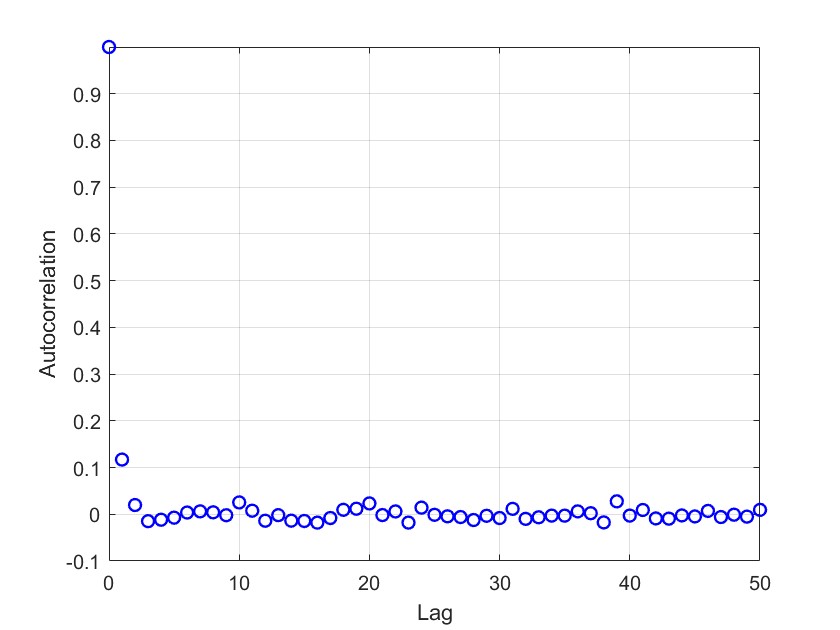}
        \caption{Autocorrelation of first coordinate samples}
        \label{Fig:Gauss.RHMC:part3}
    \end{subfigure}
    \hfill
    \begin{subfigure}[t]{0.47\textwidth}
    \includegraphics[width=\linewidth]{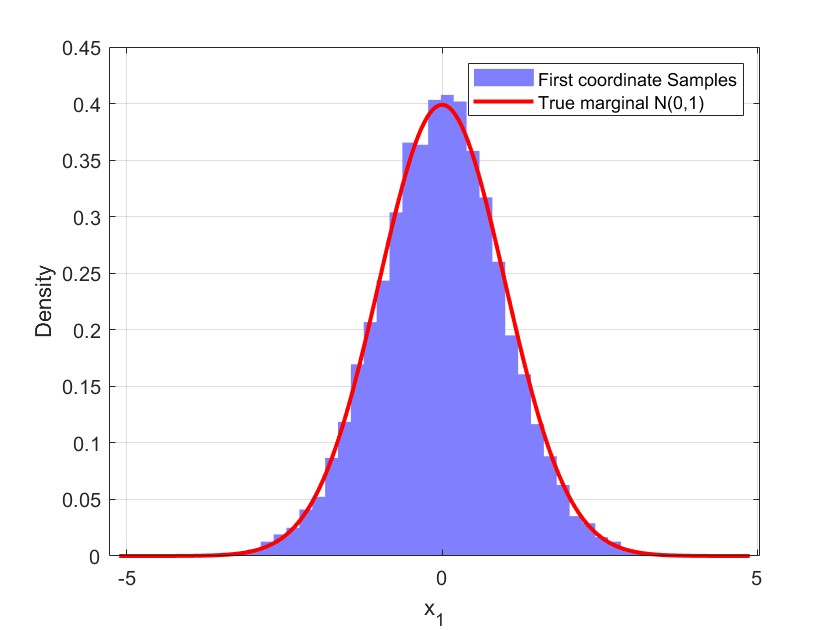}
        \caption{Histogram of first coordinate samples}
        \label{Fig:Gauss.RHMC:part4}
    \end{subfigure}
    \vspace{1mm}
    \caption{Randomized Hamiltonian Monte Carlo (RHMC) algorithm on a two-dimensional correlated Gaussian target (\ref{eq:2d.Gauss}). Correlation parameter $\rho=0$. Rate $\lambda=0.2$. Number of iterations $N=10^4$. Starting point $x_0=(4,5)$.}
    \label{Fig:Gauss.RHMC}
\end{figure}

Below we present a numerical study targeting the two-dimensional Gaussian distribution (\ref{eq:2d.Gauss}) of the previous sub-section. In the special case of the Gaussian distribution, the solution to (\ref{eq:Hamiltonian.Dynamics}) is given in closed form, and one can directly implement the Randomised Hamiltonian Monte Carlo (RHMC) algorithm without any need for numerical discretisation or Metropolis-adjustment. As in the previous sub-section, we ran the algorithm for $N=10^4$ iterations, starting from $x_0=(4,5)$. The correlation parameter and refreshment rate were chosen as $\rho = 0$ and $\lambda = 0.2$ respectively. We present our results in Figure \ref{Fig:Gauss.RHMC}, which include the scatter plot and the traceplot of the first coordinate, along with the autocorrelation plot and the histogram of the first coordinate samples. Interestingly, the correlations between samples appear to decrease significantly faster than in MALA; at the intuitive level, one could attribute this to the `momentum-driven' dynamics of RHMC, which can allow the process to explore the space faster.  

\textbf{Bouncy Particle Sampler and the Zig-Zag Process}

By contrast to Refreshed Hamiltonian Dynamics, the Bouncy Particle Sampler (BPS) instead works with ODE dynamics which are agnostic to the target distribution $\pi$, but introduces jump rates and jump types which depend intimately on the details of $\pi$. To be precise, the dynamics in question are simple free transport, i.e. 
\begin{align*}
    \dot{x} = v, \qquad \dot{v} = 0.
\end{align*}
The jump rates are no longer time-homogeneous, but they depend on the instantaneous position and velocity of the process, influenced by the rate at which the log-density of the target is changing along these dynamics. In particular, the rate at $(x,v)$ is given by
\begin{align*}
    \lambda \left( x, v\right) = \max \left\{ 0, \left\langle v, -\nabla \log \pi \left( x \right) \right\rangle \right\}.
\end{align*}
Upon the occurrence of a jump event, the position $x$ stays in place, but the velocity $v$ undergoes a specular reflection against the level sets of the log-density, i.e.
\begin{align*}
    v^\prime = \left( \mathbf{I}_d - 2 \cdot \frac{\left( \nabla \log \pi \left(x\right) \right) \left( \nabla \log \pi \left(x\right) \right)^\top}{\left( \nabla \log \pi \left(x\right) \right)^\top \left( \nabla \log \pi \left(x\right) \right)} \right) \cdot v,
\end{align*}
which forms the eponymous `bounce' and preserves the uniform measure on the sphere $\mathbf{S}^{d - 1}$ according to which the velocities are typically distributed.

The Zig-Zag Process has a similar character, using the same deterministic dynamics, but differing in the domain of the velocities and the nature of the jumps. In particular, the velocity of the Zig-Zag Process takes values in $\left\{ \pm 1 \right\}^d$, and each jump event simply `flips' one of the $d$ components. More precisely, for $1 \leq i \leq d$, at rate 
\begin{align*}
    \lambda_i \left( x, v_i \right) = \max \left\{ 0, - v_i \cdot \partial_i \log \pi \left( x \right) \right\},
\end{align*}
one observes a `jump event of type $i$', at which the position $x$ stays in place, the values $\left\{ v_j : 1 \leq j \leq d, j \neq i \right\}$ stay in place, and the $i$th component $v_i$ negates its own value, jumping to $-v_i$. This leads to the eponymous `zig-zagging' trajectories from which the process takes its name.

The two processes occupy rather similar locations vis-\`{a}-vis their practical implementability. Due to the simplicity of the deterministic dynamics, exact simulation of both the BPS and the ZZP is rather more feasible than for Refreshed Hamiltonian Dynamics. The practical difficulty arises instead in the simulation of these jump events, which requires the simulation of an inhomogeneous-in-time Poisson process. While there are a range of target distributions for which this task is achievable, in general, one requires some a priori understanding of the shape of $\log \pi$ to make this efficient. A number of numerical discretisation strategies have been proposed (see e.g. \cite{pagani2024nuzz, corbella2022automatic, andral2024automated}) which are more readily applicable to generic problems, and can be observed to incur an acceptable discretisation bias in many cases. The recent work of \cite{chevallier2024pdmp} proposes a Metropolis-type wrapper which allows to `sanitise' some of these discretisations for an audience who demand Monte Carlo-Exactness, removing the bias in the invariant measure with a suitable accept-reject mechanism.

\begin{figure}[ht]
    \centering

    \begin{subfigure}[t]{0.48\textwidth}
        \centering
        \includegraphics[width=\textwidth]{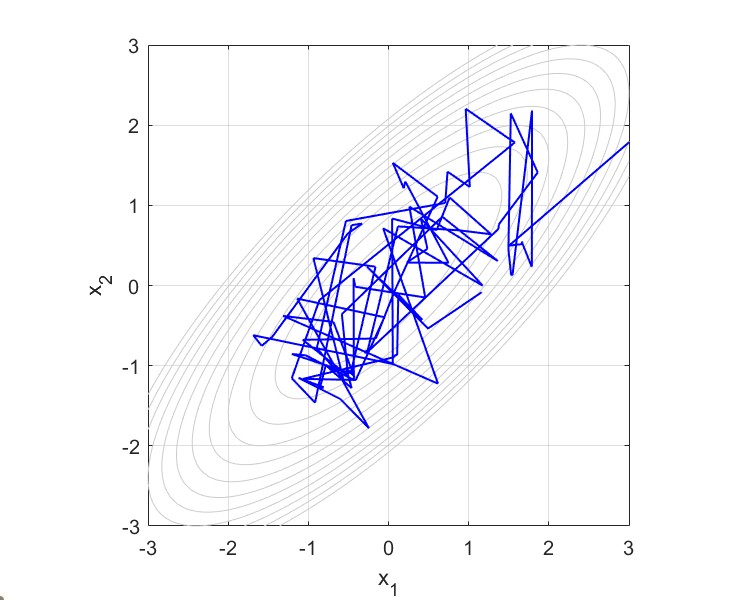}
        \caption{Scatter plot of the first 100 direction switches. Level sets on the background.}
    \end{subfigure}
    \hfill
    \begin{subfigure}[t]{0.48\textwidth}
        \centering
        \includegraphics[width=\textwidth]{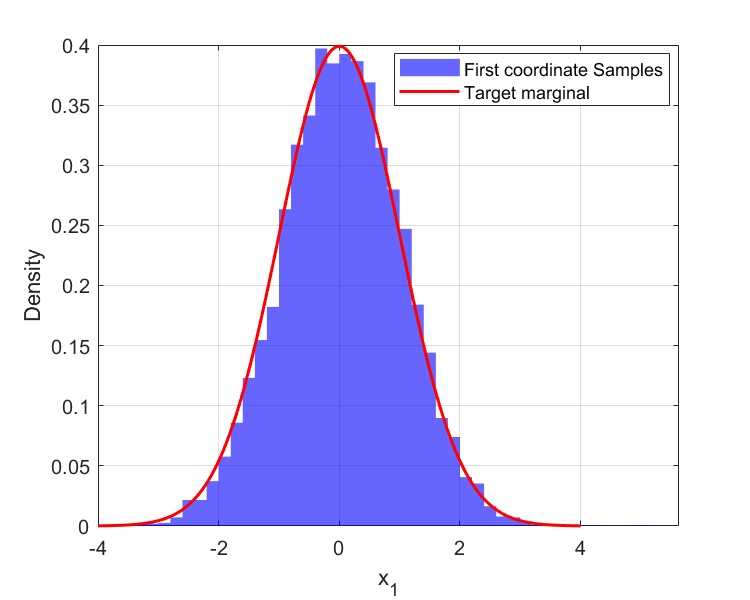}
        \caption{Histogram of the first coordinate}
    \end{subfigure}

    \caption{Bouncy Particle Sampler (BPS) on a bivariate correlated Gaussian target (\ref{eq:2d.Gauss}). Number of direction switches: $N = 10^4$. Starting point: $x_0 = (4, 5)$. Refresh rate $\lambda = 0.66$.}\label{fig:BPS.2d_Gauss}
\end{figure}

\begin{figure}[ht]
    \centering

    \begin{subfigure}[t]{0.48\textwidth}
        \centering
        \includegraphics[width=\textwidth]{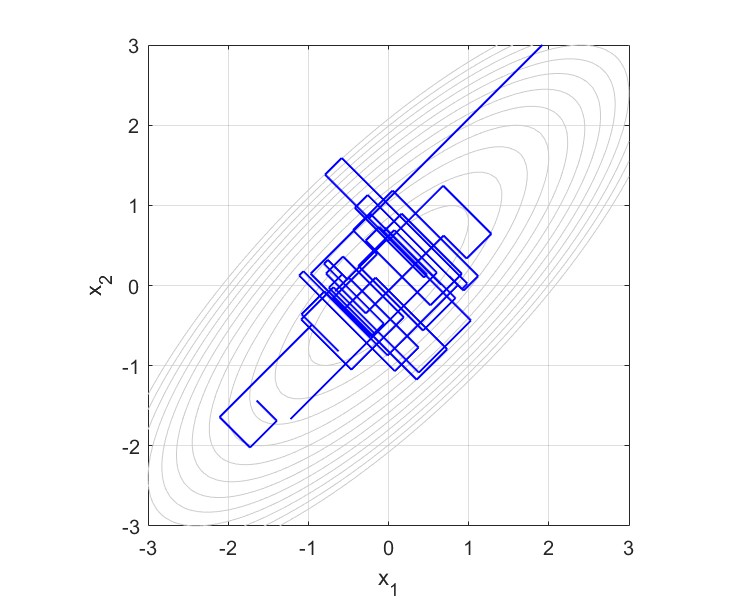}
        \caption{Scatter plot of the first 100 direction switches. Level sets on the background.}
    \end{subfigure}
    \hfill
    \begin{subfigure}[t]{0.48\textwidth}
        \centering
        \includegraphics[width=\textwidth]{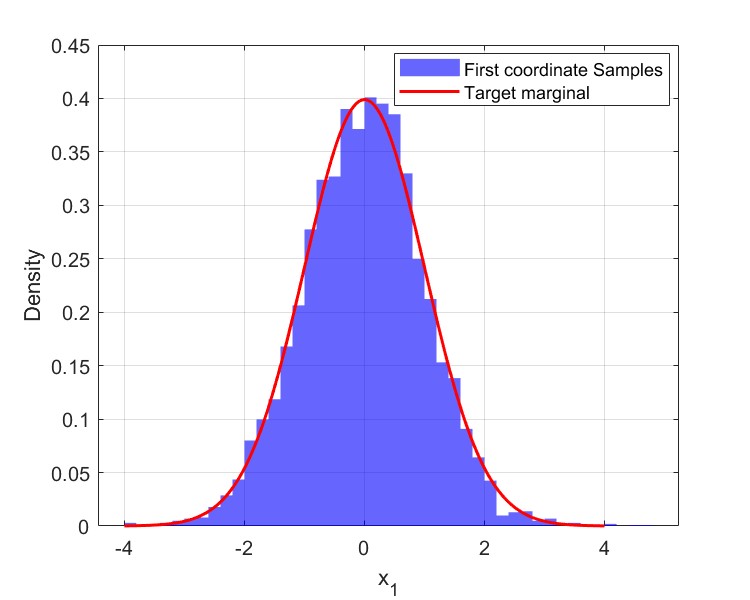}
        \caption{Histogram of the first coordinate}
    \end{subfigure}

    \caption{Zig-Zag Sampler (ZZS) on a two-dimensional correlated Gaussian
target (\ref{eq:2d.Gauss}). Number of direction switches: $N = 10^4$. Starting point: $x_0 = (4, 5)$.}\label{fig:ZZS.2d_Gauss}
\end{figure}

We consider the two-dimensional correlated Gaussian (\ref{eq:2d.Gauss}) of the previous sub-sections to showcase the performance of the Bouncy Particle Sampler (BPS) and the Zig-Zag Sampler (ZZS) via a numerical simulation. We ran both algorithms for $N=10^4$ direction switches, starting from $x_0=(4,5)$. For the BPS we used refresh rate $\lambda=0.66$ (using methodological guidance from \cite{bouchard2018bouncy}), while for ZZS we used $\lambda=0$, which is conjectured to be optimal in terms of minimising the asymptotic variance (e.g. \cite{bierkens.duncan:17}). In Figure \ref{fig:BPS.2d_Gauss} we present the scatter plot of the BPS for the first 100 direction switches, and the histogram of the first coordinate samples with the marginal target density overlaid. For the scatter plot, we also add the level sets of the target on the background to indicate how the direction switches work as a result of bouncing on the level sets. We present the same plots for the Zig-Zag process in Figure \ref{fig:ZZS.2d_Gauss}.

\subsubsection{Omissions}

We pause for a minute here to discuss some other prominent MCMC algorithms which will not feature as explicitly in our subsequent discussion. 

The first omission is the Random Walk Metropolis (RWM) algorithm \cite{metropolis1953equation, sherlock2010random, andrieu2024explicit}. Due to the aforementioned interpretation of the RWM as a zeroth-order discretisation of the Langevin diffusion, one might anticipate that aside from generally moving more slowly than gradient-based discretisations, it will not really encounter any additional difficulties, and all available evidence suggests that this is indeed the case. For similar reasons, while Monte Carlo methods based around the Underdamped (also `Kinetic', `Second-Order', etc.) Langevin Diffusion \cite{cheng2018underdamped} are of practical interest, their qualitative features are arguably subsumed by a combination of the Overdamped Langevin Diffusion and the Refreshed Hamiltonian Dynamics, and so neither do they present unique challenges in terms of robustness. The same comments apply equally to other methods based upon similar ODEs and SDEs.

The other class of notable omissions will be MCMC methods whose dynamics are `non-local' in some sense. The chief example of this class is the Gibbs sampler \cite{gelfand2000gibbs, zanella2020informed}, and its many descendents based on principles of iterative conditional simulation, including Hit-and-Run \cite{belisle1993hit, rudolf2013hit}, Slice Sampling \cite{neal2003slice, rudolf2013hit}, and various other `auxiliary variable' methods. We choose to spare these methods from discussion on the grounds that the qualitative picture is rather different here, to the effect that non-local samplers face different challenges to their robustness, thus requiring rather different solutions. As such, rather than introducing further notational and conceptual overhead to this chapter, we prefer to confine our discussion to methods based on local dynamics derived from the Langevin diffusion and the Refreshed Hamiltonian Dynamics.


\subsection{Interlude}

These methods are general-purpose, widely-used, and theoretically well-understood to work well under reasonable conditions. Conventional results tend to operate under the rather strong assumptions that the target distribution $\pi$ be strongly log-concave, and that $\log \pi$ have a Lipschitz-continuous gradient. More refined results have been obtained under various weakenings of these assumptions, usually amounting to some sort of light-tailedness property for $\pi$ combined with some relaxed notion of quantitative continuity for $\nabla \log \pi$. While it is difficult to obtain a tight characterisation of conditions on $\pi$ which ensure quantitatively efficient sampling, experience dictates that conditions of this form are qualitatively appropriate.

Taking this as given, it bears mentioning that these assumptions can fail in contemporary statistical problems, even those which are relatively conventional. Sometimes, this is rather benign, and the practical performance of the methods remains acceptable, even in the absence of theoretical guarantees. Sometimes, however, the failure of these assumptions leads to a genuine breakdown of the methods, which fail to produce an acceptable output in a reasonable time-frame. This can come in various forms: the cost of each iteration of the method might blow up, the numerical stability of the method might fail, the range of stable step-sizes for a numerical discretisation might collapse, the rate of convergence as a function of the number of iterations might slow to a crawl, and so on. 

This has prompted an emerging line of MCMC research that seeks to develop sampling algorithms that retain the general-purpose character of well-established methods, but have an additional character of `robustness'. In particular, a key aim is for the algorithm to perform effectively when the classical assumptions on $\pi$ are satisfied, and to degrade gracefully when these assumptions fail. This should be understood less as a task of `acceleration', but more of devising simple, lightweight strategies for avoiding potentially-catastrophic pitfalls.

\begin{figure}[ht]
    \centering
    \begin{subfigure}[t]{0.47\textwidth}
        \includegraphics[width=\linewidth]{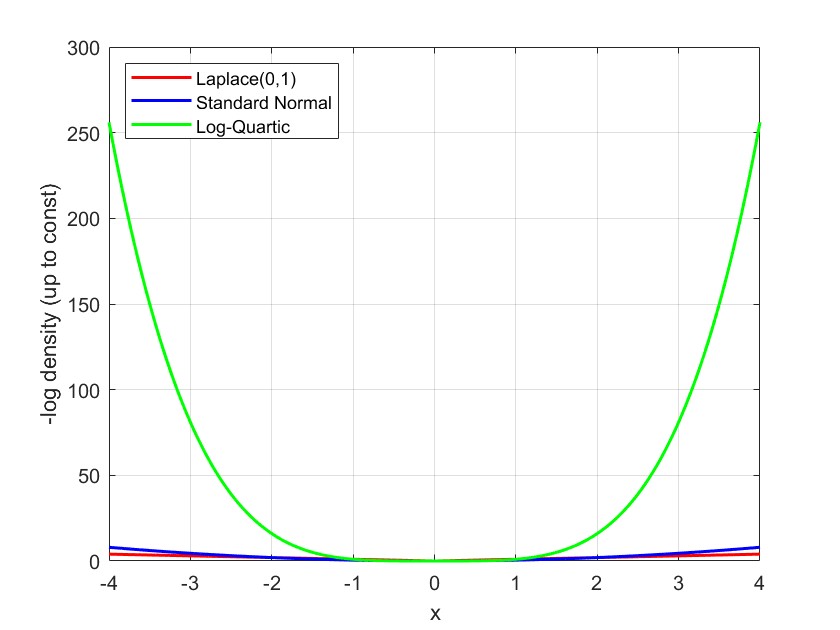}
        \caption{Light tailed potential.}
        \label{fig:Plot.Light}
    \end{subfigure}
    \hfill
    \begin{subfigure}[t]{0.47\textwidth}
    \includegraphics[width=\linewidth]{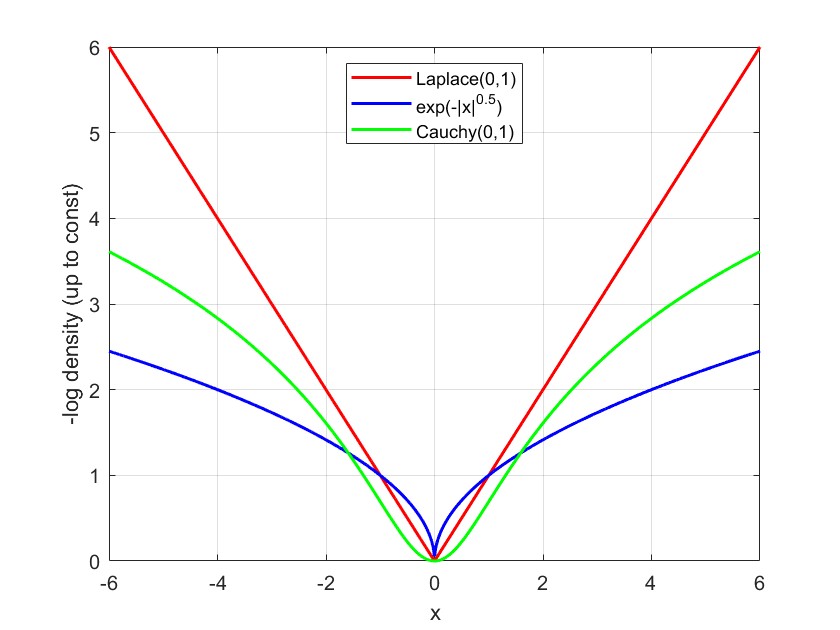}
        \caption{Heavy tailed potential}
        \label{fig:Plot.Heavy}
    \end{subfigure}
    \vspace{1mm}
    \caption{Plots of the negative log-densities of various densities. The left plot shows the growth of Laplace potential and of other densities with lighter tails. The right plot shows the growth for heavier tails.}
    \label{fig:light.heavy}
\end{figure}

In the next two sections, we will focus on two different types of pathologies that often arise in various practical statistical settings. The first one, `roughness', relates to settings in which the target distribution has reduced regularity (e.g. when $\nabla \log \pi$ is non-Lipschitz, non-differentiable, non-continuous, or similar). The second one, `heavy-tailedness' concerns settings in which the target fails to concentrate its probability mass well in the bulk of the space, with substantial mass at a large distance from e.g. the mode. One can characterise the former as referring to bad `local' behaviour of $\pi$, whereas the latter refers to bad `global' behaviour of $\pi$. To indicate differences between these two types of distributions, in Figure \ref{fig:light.heavy} we show the growth of the negative log-density (also known as potential) for various light (typically tails lighter than Gaussian) and heavy (tails heavier than Laplace) densities.

\section{Roughness}\label{sec:roughtness:00}

\subsection{Formulation of Pathology, Basic Examples}






Our starting point in this section is that for target distributions $\pi$ with good smoothness properties, if one starts with some $\pi$-invariant continuum process, then numerical discretisation essentially `works'. Broadly speaking, this entails that the numerical process inherits the pertinent stability properties of the continuum process, and that by taking sufficiently fine discretisations, the discretisation error can be reduced arbitrarily. In the context of Monte Carlo-Exact procedures, the consequences are similar, in that for sufficiently small step-sizes, acceptance probabilities can be increased to an appropriate level, and the resulting Markov chain is not too `sticky', i.e. the chain `actually moves' on most steps. This is somehow a minimal requirement for a discrete-time Markov chain to be performant; various elementary arguments can quantify the intuition that a chain which moves only rarely can only converge slowly.

Concretely, `good smoothness properties' tend to enforce that the target distribution varies locally in predictable ways. A usual assumption to this effect is `$L$-Smoothness' of the potential $U = -\log \pi$, or equivalently, that the `force' $\nabla U$ is $L$-Lipschitz, i.e.
\begin{align*}
    x, y \in \mathbf{R}^d \implies \left\| \nabla U \left( x \right) - \nabla U \left( y \right) \right\| \leq L \cdot \left\| x - y \right\|
\end{align*}
for some $L \in \left( 0, \infty \right)$. This is sometimes weakened to a more general modulus of continuity assumption, i.e. that for some suitable $\psi : \mathbf{R}_+ \to \mathbf{R}_+$, it holds that
\begin{align*}
    x, y \in \mathbf{R}^d \implies \left\| \nabla U \left( x \right) - \nabla U \left( y \right) \right\| \leq \psi \left( \left\| x - y \right\| \right).
\end{align*}
Under an assumption of this form, various quantitative consequences can be derived. As an example, under these assumptions, \cite{andrieu2023weak} shows that in the Random Walk Metropolis algorithm, by tuning the step-size based on $\psi$ and the ambient dimension $d$, one can ensure that uniformly over the state space, the acceptance rates out of any state can be uniformly bounded from below. Scaling analysis of a more specific model problem in \cite{vogrinc.kendall:19} suggests that these estimates do not need to be overly conservative. Along similar lines, \cite{chewi2024analysis} show that for Langevin Monte Carlo, for polynomial $\psi$ (corresponding to smoothness of H\"{o}lder type), one can design step-sizes so that the asymptotic bias of the numerical invariant measure is controlled well. In each case, one sees that `rougher' target distributions are subject to more stringent assumptions on viable step-sizes, often in a dimension-dependent way.

Although this `modulus of continuity' approach to smoothness is rather flexible, it does preclude certain interesting classes of target distribution. To this point, observe that although the modulus $\psi$ is a priori quite general, there are some implied constraints on both its form and growth. In particular, note that any nontrivial control of the form 
\begin{align*}
    \left\| x - y \right\| \leq r \implies \left\| \nabla U \left( x \right) - \nabla U \left( y \right) \right\| \leq \Lambda
\end{align*}
can be inductively bootstrapped into the global estimate
\begin{align*}
    \left\| \nabla U \left( x \right) - \nabla U \left( y \right) \right\| \leq \Lambda \cdot \left\lceil \frac{\| x - y \|}{r} \right\rceil \leq \frac{\Lambda}{r} \cdot \left\{ r + \left\| x - y\right\| \right\}.
\end{align*}
As such, any $\nabla U$ satisfying such an estimate is necessarily of at most \textit{linear growth} at infinity, and hence corresponds to potentials of at most quadratic growth, i.e. with tails no lighter than Gaussian. One is then led to wonder what this implies about the performance of such algorithms when applied to such light-tailed targets: shall we observe reasonable performance which happens to be unsupported by the existing theory, or shall we observe genuine problems?

To this end, we introduce an illustrative toy example which will help demonstrate some of these features.

\begin{example}[`Polynomially-Steep' Potential]\label{ex:steep.gradient}
    Noting that any potential $U$ for which $\nabla U$ admits a non-trivial modulus of continuity is of at most quadratic growth at infinity, it is natural to seek counter-examples by constructing potentials of super-quadratic (but still polynomial) growth at infinity, e.g. $U \left( x \right) = \left\| x \right\|^q$ for some $q > 2$. For concreteness (and to avoid issues about the existence and regularity of higher-order derivatives), we will generally focus on the quartic potential $U \left( x \right) = \| x \|^4$, i.e. the choice $q = 4$, and the induced target distribution $\pi \left( x \right) \propto \exp \left( - \left\| x\right\|^4 \right)$. Note that this distribution remains log-concave and very well-confined, with tails much lighter than those of the Gaussian distribution.
\end{example}

\begin{figure}[ht]
    \centering
    \includegraphics[width=0.6\linewidth]{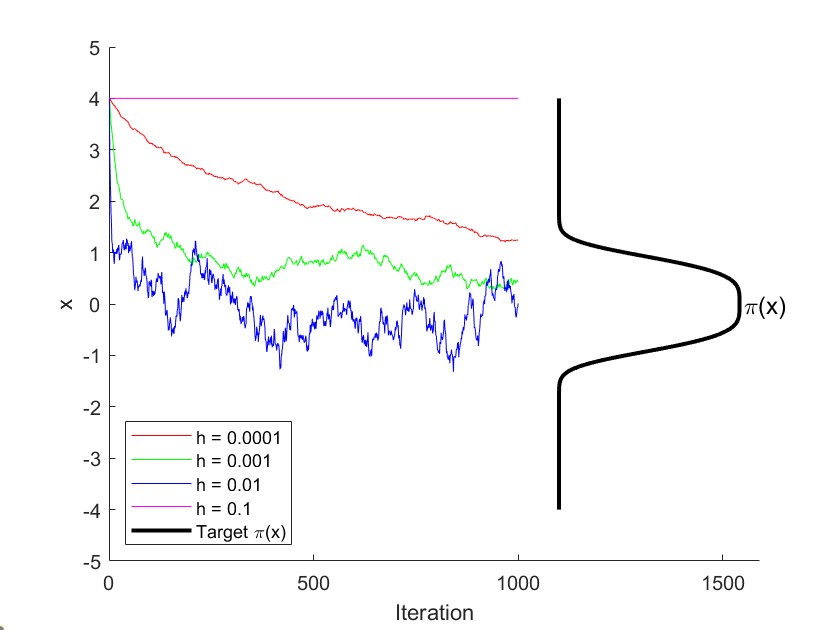}
    \caption{MALA trace plots with various step-sizes. The target density ($\pi(x) \propto \exp \left( -x^4 \right)$) is overlaid on the right hand side of the graph.}
    \label{fig:MALA.stepsizes}
\end{figure}

To this end, we perform some illustrative simulations on the target distribution associated to the quartic potential in $d = 1$ (see Figure \ref{fig:MALA.stepsizes}).
We observe that when the step-size is not appropriately small ($h = 0.1$, magenta line), the Metropolis-Adjusted Langevin Algorithm fails to explore the state space, rapidly becoming `stuck', rejecting every proposed move in our simulation run. Use of smaller step-sizes can delay this pathology to some extent, but any eventual excursion of the process to a larger value of $x$ will lead to similar issues. At the same time, use of very small step-sizes ($h=10^{-4}$ red line, $h=10^{-3}$ green line) will lead to very slow space exploration, as indicated in Figure \ref{fig:MALA.stepsizes}. This serves as strong empirical evidence that `steep' gradients can be problematic for standard gradient-based MCMC algorithms, which is validated theoretically in other works; see e.g. \cite{roberts.tweedie:96, bou2013nonasymptotic}.

At the other end of the spectrum, one might ask what happens when the force does admit a bounded modulus of continuity, but one which does not vanish continuously at $0$. 

\begin{example}[`Locally-Sharp' Potential]\label{example:Laplace}
    When the obstruction to smoothness is `local' roughness rather than growth at infinity, an archetypal example is the `Laplace-type' potential associated to the $\ell^1$ norm, i.e. $U \left( x \right) = \left\| x \right\|_1$ for $x \in \mathbf{R}^d$. In this setting, one computes that away from the coordinate axes, $\nabla U \left( x \right) = \text{sign} \left( x \right)$. By considering points in a neighbourhood of the origin, one sees that the best possible $\psi$ is $\psi \left( r \right) = 2 d$ for $r > 0$. While this is favourable in terms of uniform boundedness, it is problematic in that one has no immediate guarantees to generate acceptable moves, even at a small step-size. Moreover, one sees that high dimensionality of $x$ apparently makes matters even worse.
\end{example}

\begin{figure}[ht]
    \centering
    \begin{subfigure}[t]{0.47\textwidth}
        \includegraphics[width=\linewidth]{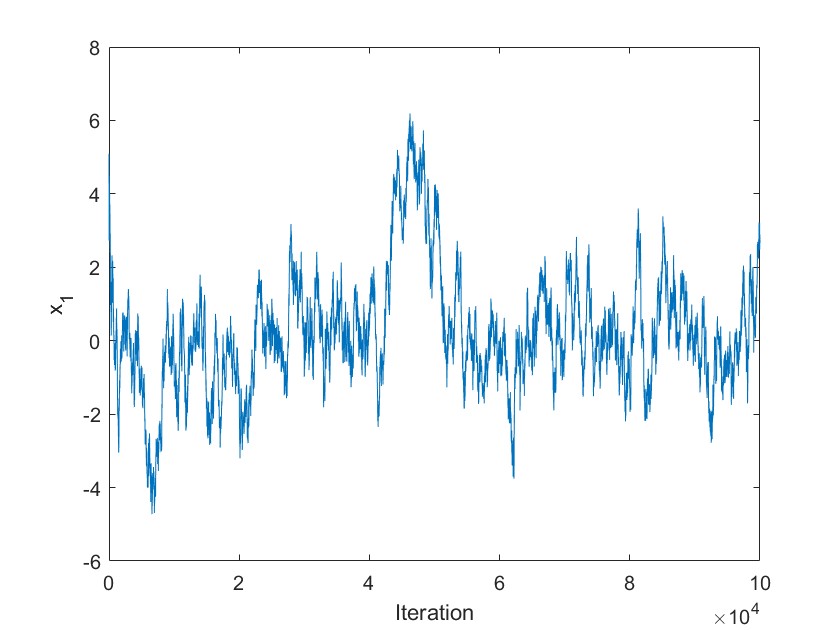}
        \caption{Traceplot of first coordinate. Step-size $h=0.01$.}
        \label{fig:MALA.Laplace:part1}
    \end{subfigure}
    \hfill
    \begin{subfigure}[t]{0.47\textwidth}
        \includegraphics[width=\linewidth]{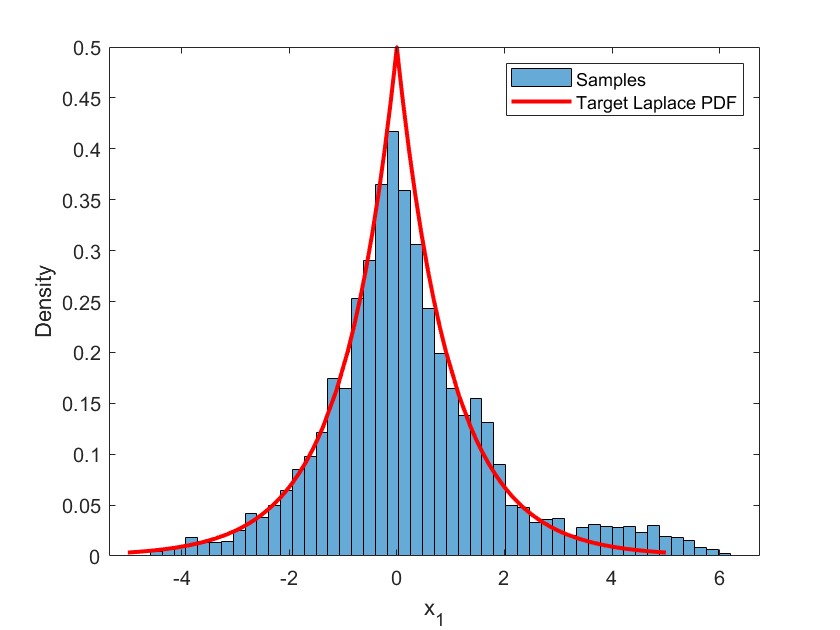}
        \caption{Histogram of first coordinate samples. Step-size $h=0.01$.}
        \label{fig:MALA.Laplace:part2}
    \end{subfigure}
    
    \vspace{0.5cm}  

    \begin{subfigure}[t]{0.47\textwidth}
        \includegraphics[width=\linewidth]{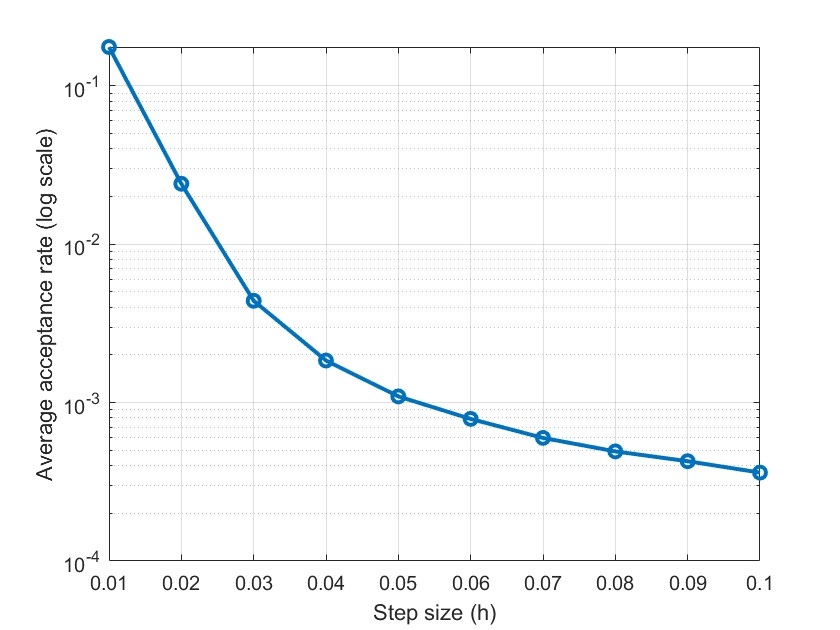}
        \caption{Average acceptance probability over step-sizes.}
        \label{fig:MALA.Laplace:part3}
    \end{subfigure}
    \hfill
    \begin{subfigure}[t]{0.47\textwidth}
    \includegraphics[width=\linewidth]{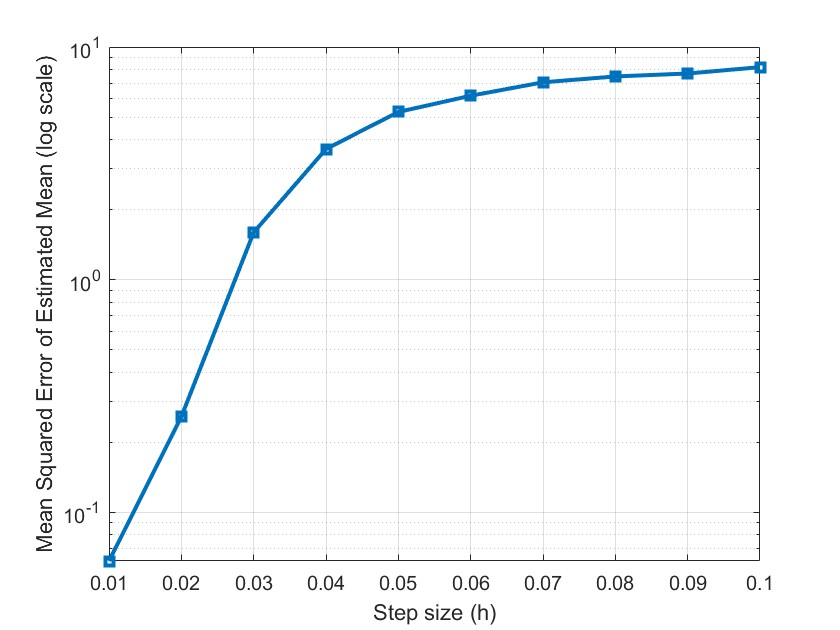}
        \caption{Mean square error of target expectation over step-sizes}
        \label{fig:MALA.Laplace:part4}
    \end{subfigure}
    \vspace{1mm}
    \caption{MALA on the $10^4$-dimensional Laplace target (\ref{Laplace.target:1}). First row: traceplot and histogram with optimally tuned step-size  $h=0.01$. Second row: Average acceptance probability and Mean squared error (estimating the target expectation) over step-size.}
    \label{fig:MALA.Laplace}
\end{figure}

To demonstrate some of these features, we consider the application of MALA to the associated Laplace target in dimension $d = 10^4$, i.e.
\begin{equation}\label{Laplace.target:1}
\pi(x)\propto \prod_{i=1}^{10^4} \exp \left( - \left| x_i \right| \right).
\end{equation}
Figure \ref{fig:MALA.Laplace} documents the results, showing the traceplot and histogram of the first coordinate of the algorithm run for $10^5$ iterations. We considered two different step-sizes. The first row presents the results with step-size $h=0.01$, chosen so that the algorithm accepts between $50-60 \%$ of the proposed jumps, as suggested by the literature (see e.g. \cite{pillai.stuart.thiery:12}). From the trace plot, the algorithm seems to be performing sufficiently well. However, when one changes the step-size, the algorithmic behaviour quickly deteriorates. To showcase this, we ran the algorithm for different step-sizes $h \in \{ 0.01 , 0.02 , \dots , 0.1 \}$. For each step-size we ran $100$ i.i.d. copies of the algorithm for $N=10^5$ iterations. We report the average (over iterations and algorithmic configurations) Acceptance Probability and Mean Squared Error when estimating the target expectation (in this case the true quantity is zero) as functions of step-sizes. It is evident that different step-sizes lead to drastically different behaviour, with any step-size larger than $h=0.01$ leading to poor algorithmic performance.

Finally, there are various other intermediate model problems which exhibit qualitatively similar pathologies, with densities which blow up at some natural boundary, as one observes in Beta-type distributions. In a similar category, there are targets which are subject to hard boundaries, e.g. Gaussian distributions subject to non-negativity constraints. These boundaries typically then create computational issues for MCMC algorithms such as MALA. Most algorithms fail to behave appropriately when close to boundary, either by not visiting it enough, or by proposing moves outside the bounded state space, which are ultimately rejected. To illustrate the problem, we introduce a stylised model problem.
\begin{example}[Divergent Potential with Natural Boundary]
    Fix the domain $\mathcal{B} = \left\{ x \in \mathbf{R}^d : \left\| x \right\| <1 \right\}$, and consider the potential $U : \mathcal{B} \to \mathbf{R}$ which is given by $U \left( x \right) = \log \left( \frac{1}{1 - \| x \|^2} \right)$. One sees immediately that $U$ blows up near to the boundary of $\mathcal{B}$, as do all derivatives of $U$. When considering the associated probability measure `at positive temperature', this blow-up corresponds to the vanishing of the target density; at `negative temperature', it instead reflects a concentration of mass near this boundary (cf. Beta distributions with parameters in $\left( 0, 1 \right)$). In either case, the blow-up of $U$ and its derivatives can be expected to cause difficulties when discretising $\pi$-invariant dynamics.
\end{example}

\begin{figure}[ht]
    \centering
    \includegraphics[width=0.48\textwidth]{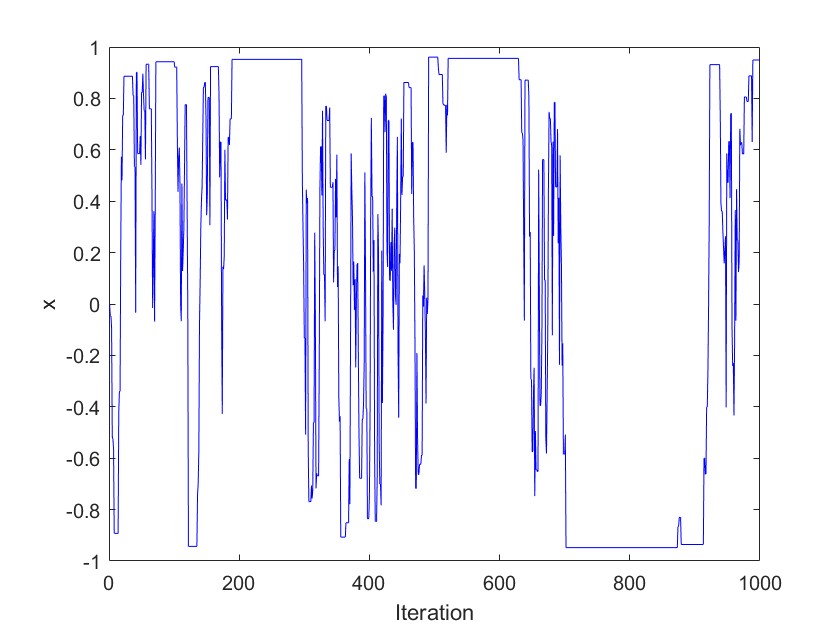}
    \hfill
    \includegraphics[width=0.48\textwidth]{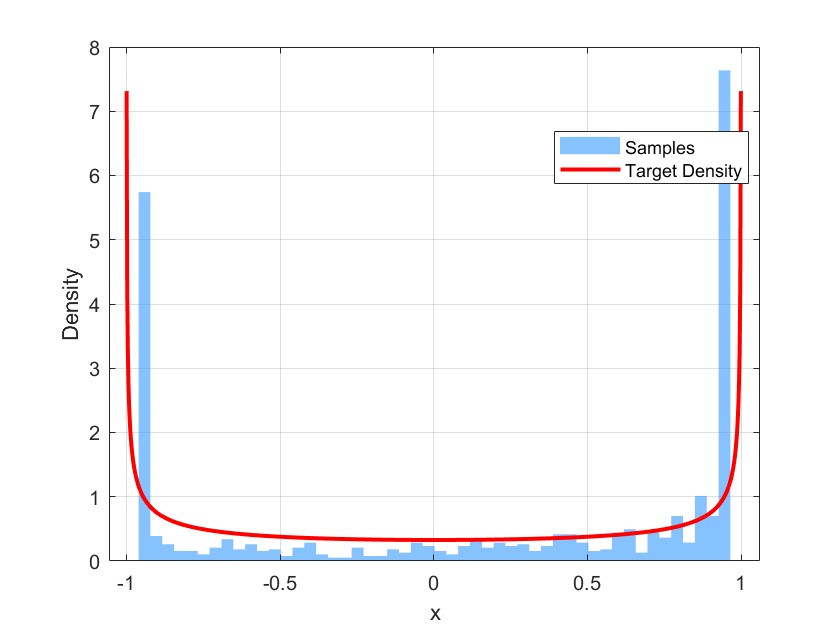}
    \caption{MALA traceplot and histogram on the one-dimensional target with exploding boundary at $-1$ and $1$, i.e. $\pi(x) \propto \left( 1-x^2 \right)^{-\frac{1}{2}}$. Step-size $h=0.1$.}\label{fig:MALA.Exploding.Boundary}
\end{figure}

We present now a numerical example for this divergent potential in the univariate setting, at `temperature' $\tau = -2$, yielding the specific density 
\begin{equation}\label{target.exploding.boundary:1}
    \pi(x) \propto \left( 1- x^2 \right)^{-\frac{1}{2}}, \qquad x \in \left( -1, 1 \right),
\end{equation}
which blows up as one approaches $x = \pm 1$. The algorithm runs for $N = 1000$ iterations, with step-size $h = 0.1$ (again chosen so that around $50-60 \%$ of the proposed jumps are accepted), starting  from $x_0 = 0.5$. We present the traceplot and histogram of the MALA algorithm in Figure \ref{fig:MALA.Exploding.Boundary}. It is evident that the process has not captured the target's behaviour sufficiently well, and there are areas close to the boundary that are not explored at all. The traceplot also indicates that the process can get quite stuck close to the boundary. Closer examination of the algorithm output reveals that this results from frequent proposals outside of the domain $\left( -1, 1 \right)$, which result in immediate rejections. 

A closely-related scenario, not uncommon in applications, is when the potential itself is `nice' \emph{per se}, but the domain is constrained for other reasons. 

\begin{example}[Nice Potential, Artificial Boundary]\label{example:Truncated.Gauss}
    A simple example of this setting would be a Gaussian distribution, subject to a collection of affine constraints, i.e. the distribution of $X \sim \mathcal{N} \left( 0, \mathsf{C}\right)$, subject to the constraints $a_j^\top X\leq b_j$ for $j$ in some finite index set $\mathcal{J}$, with $a_j$ and $b_j$ a suitable collection of vectors and scalars respectively. Such distributions arise naturally in the study of shape-constrained Gaussian processes, as well as for algorithmic reasons in data augmentation strategies for certain generalised linear models (GLMs). A particularly tractable instance (which nevertheless exhibits a number of relevant pathologies) is to consider a spherical Gaussian random variable $X \sim \mathcal{N} \left( 0, \sigma^2 \cdot \mathbf{I}_d \right)$, constrained to the $\ell^\infty$ box $\mathsf{B} = \left\{ x \in \mathbf{R}^d : -1 \leq x_i \leq 1 \, \text{ for all } i \right\}$. We refer to this as the `box-constrained Gaussian' target.
\end{example}

To illustrate, we run some simulations on a box-constrained Gaussian target in $d = 10^4$ dimensions with $\sigma^2 = 1$, i.e.

\begin{equation}\label{Truncated.Gaussian:1}
    \pi(x) \propto \exp \left( -
    \frac{1}{2} \left( x_1^2 + x_2^2 + \dots x_{d}^2 \right) \right), \qquad x \in [-1,1]^{10^4}.
\end{equation}

\begin{figure}[ht]
    \centering
    \begin{subfigure}[t]{0.47\textwidth}
        \includegraphics[width=\linewidth]{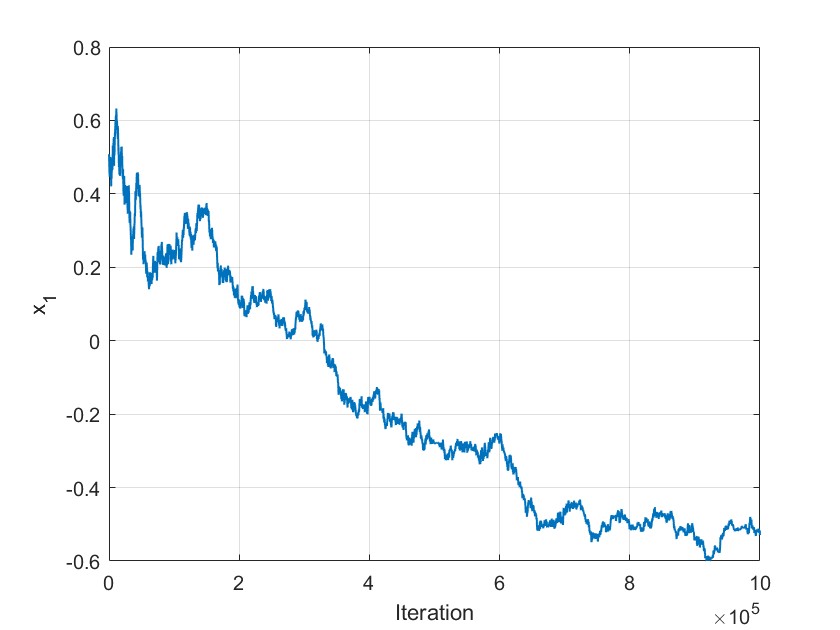}
        \caption{Traceplot of first coordinate. Step-size $h= 5 \cdot 10^{-7}$.}
        \label{fig:MALA.Truncated.Gauss:part1}
    \end{subfigure}
    \hfill
    \begin{subfigure}[t]{0.47\textwidth}
        \includegraphics[width=\linewidth]{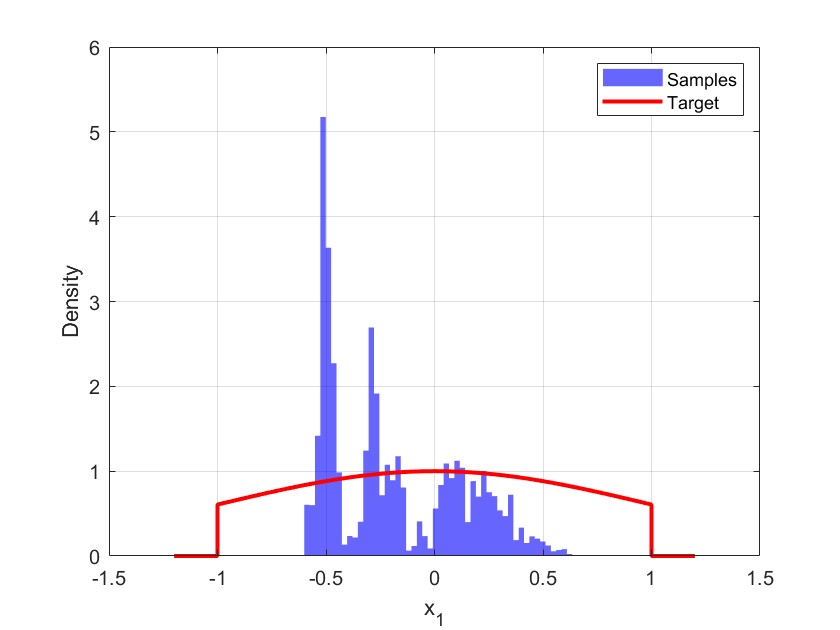}
        \caption{Histogram of first coordinate samples. Step-size $h=5 \cdot 10^{-7}$.}\label{fig:MALA.Truncated.Gauss:part2}
    \end{subfigure}
    
    \vspace{0.5cm}  

    \begin{subfigure}[t]{0.47\textwidth}
        \includegraphics[width=\linewidth]{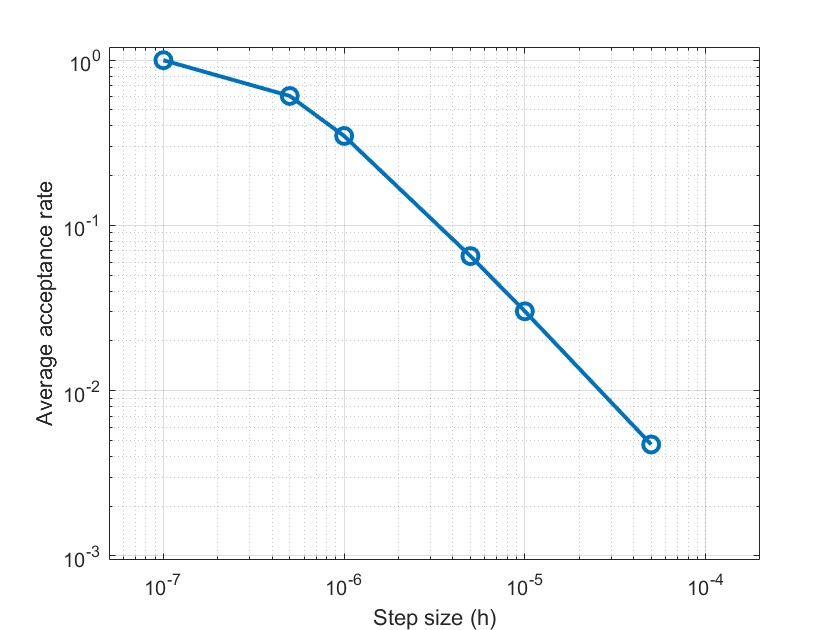}
        \caption{Average acceptance probability over step-sizes.}
        \label{fig:MALA.Truncated.Gauss:part3}
    \end{subfigure}
    \hfill
    \begin{subfigure}[t]{0.47\textwidth}
    \includegraphics[width=\linewidth]{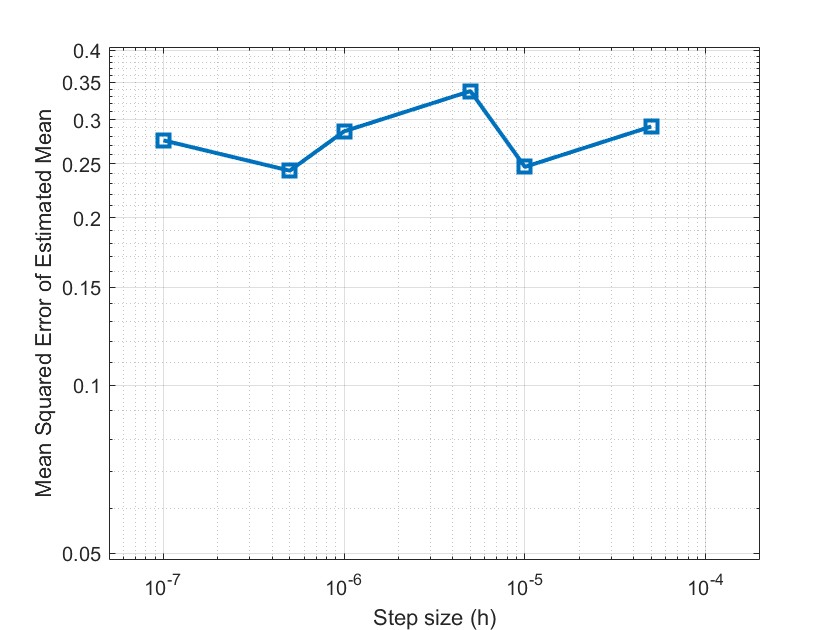}
        \caption{Mean square error of target expectation over step-sizes}
        \label{fig:MALA.Truncated.Gauss:part4}
    \end{subfigure}
    \vspace{1mm}
    \caption{MALA on the $10^4$-dimensional box-constrained Gaussian target (\ref{Truncated.Gaussian:1}), ran for $N=10^5$ iterations. First row: traceplot and histogram with optimally tuned step-size  $h=5\cdot 10^{-7}$. Second row: Average acceptance probability and mean squared error (estimating the target expectation) over step-size.}
\label{fig:MALA.Truncated.Gauss}
\end{figure}

Note this is a significant constraint; due to the high dimensionality of the target, a sample from the \emph{unconstrained} Gaussian distribution will very rarely fall inside this box. 
We ran the MALA algorithm on this target for $N = 10^5$ iterations, starting from the point $(0.5,0.5,\dots , 0.5)$ and we present our result in Figure \ref{fig:MALA.Truncated.Gauss}. In the first row we present the traceplot of the first coordinate and its histogram with a step-size chosen to be $h=5 \cdot 10^{-7}$ so that around $50-60 \%$ of the proposed jumps are accepted. It is evident from the histogram that the algorithm needs more time to converge and has not fully captured the shape of the target, while from the traceplot, we see that the algorithm moves around very slowly. Furthermore, in the second row, we present the results of an analysis similar to Figure \ref{fig:MALA.Laplace}. We ran the process for different step-sizes $h \in \{ 10^{-7}, 5 \cdot 10^{-7}, 10^{-6}, 5 \cdot 10^{-6}, 10^{-5}, 5 \cdot 10^{-5} \}$, we ran 100 copies of each process for $=10^5$ iterations, and we present the average acceptance probability and the Mean Squared Error (when estimating the target expectation, which in this case is zero) as a function of step-sizes. It is evident from the plots that the Mean Squared Errors are significantly large, given that the range of values each coordinate can take is limited to $[-1,1]$. The acceptance probability heavily depends on the chosen step-size, and the algorithm's performance rapidly deteriorates at larger step-sizes.

Given the problems that state-of-the-art algorithms may experience in the presence of roughness, we now proceed to describe some resolutions to these pathologies.

\subsection{Proposed Solutions}

Returning to the original approach of designing MCMC algorithms by taking inspiration from a suitable continuum process, we describe two popular strategies for `robustifying' MCMC in the face of roughness. A first strategy is to focus on discretising the process more carefully, perhaps taking inspiration from numerical-analytic approaches to stiff dynamics. A second strategy is to concoct entirely new continuum processes which admit more stable discretisations `by design', or for which discretisation can somehow be avoided entirely. In this section, we will review some individual proposals which tackle roughness through some implementation of these principles, focusing more on their conceptual genesis than upon their application to specific examples of practical relevance and scale.

\subsubsection{Truncated Langevin Monte Carlo}\label{sec:Truncated:00}

In the setting of `steep' gradients, whereby $\nabla U$ is perhaps locally-Lipschitz without being globally-Lipschitz, a natural strategy is to perform some sort of truncation. Indeed, in many settings, steepness of gradients is associated with light tails for the invariant measure (it is relatively straightforward to concoct such examples in e.g. the case of convex $U$), and so given $\delta \in \left( 0, 1 \right)$, one can often find some rather moderately-sized $R = R_\delta \in \left( 0, \infty \right)$ so that
\begin{align*}
    x \sim \pi \implies \text{with probability } \geq 1 - \delta, \quad \left\| \nabla U \left( x \right) \right\| \leq R.
\end{align*}
As such, in e.g. a Langevin-type algorithm, one could envision replacing all instances of the `raw' drift $\nabla U \left( x \right)$ with the $R$-truncated drift $\tau_R \circ \nabla U$, where
\begin{align*}
    \tau_R \left( g \right) :=
    \begin{cases}
        g &\text{if } \left\| g \right\| \leq R \\
        R \cdot \frac{g}{\left\| g \right\|} &\text{if } \left\|  g \right\| > R
    \end{cases}
\end{align*}
and obtain a process with desirable long-time behaviour (in terms of converging to a sensible invariant measure, even without Metropolis-adjustment) while enjoying good numerical stability properties. In particular, for most `reasonable' targets, the truncated drift will be uniformly Lipschitz, and so `standard' analyses will apply quite directly.

In the context of Monte Carlo-Exact methods, this strategy appears to have first been introduced by \cite{roberts.tweedie:96} as the `Metropolis Adjusted Langevin Truncated Algorithm' (MALTA). \cite{atchade2006adaptive} later studied this approach in the context of Adaptive MCMC, observing that the intrinsic stability of the truncated algorithm allows for well-behaved adaptation of algorithmic hyperparameters. Additionally, \cite{bou2010pathwise} study the numerical-analytic properties of truncated and Metropolis-adjusted procedures when viewed as numerical integrators for SDEs, again observing favourable properties.

We note in passing that such strategies might be characterised more precisely as `drift-truncated', to contrast with other methods which explicitly truncate the state space; see e.g. \cite{milstein2005numerical}. Such methods have apparently seen less widespread use in the MCMC world.

\begin{figure}[ht]
    \centering
    \includegraphics[width=0.5\textwidth]{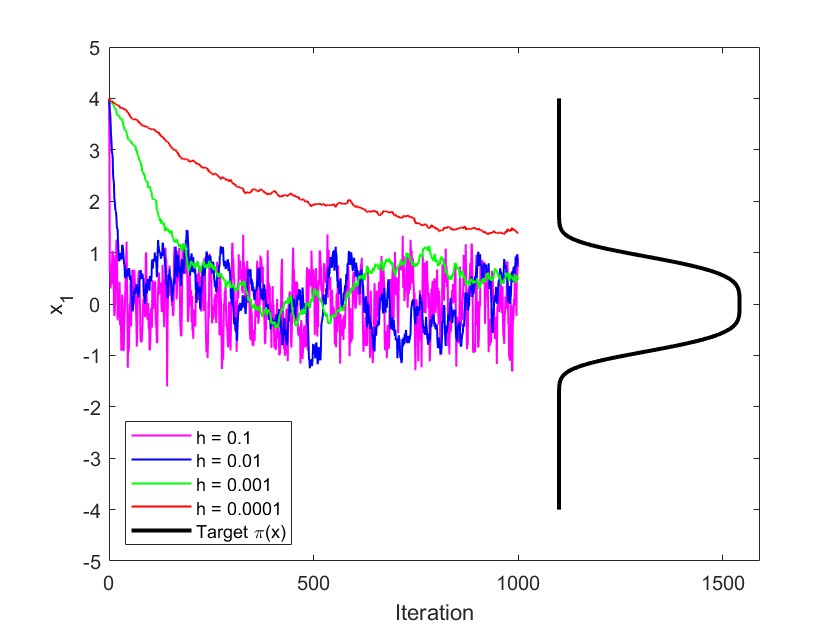}
    \caption{Truncated MALA traceplots with various step-sizes. The target density ($\pi(x) \propto \exp \left( -x^4 \right)$) is overlaid on the right hand side of the graph.}\label{fig:Truncated.MALA.SuperLight}
\end{figure}

In Figure \ref{fig:Truncated.MALA.SuperLight}, we present a numerical simulation of the Truncated MALA algorithm applied on the one-dimensional log-quartic target from Example \ref{ex:steep.gradient} for various step-sizes. Comparing the trace plots with those of MALA in Figure \ref{fig:MALA.stepsizes}, we see that the algorithmic performance is critically improved for large step-sizes. Of particular note is the step-size $h=0.1$ where the Truncated algorithm is exploring the space in an efficient manner, whereas the same step-size for MALA was inducing an algorithm that did not move at all. For smaller step-sizes the performance seems similar to MALA; this is to be expected, as at small step-size, both schemes converge to the Langevin diffusion. Similar improvement in algorithmic performance was observed in simulations for higher-dimensional targets.

\subsubsection{Tamed Langevin Monte Carlo}

The `tamed' approach to MCMC has its genesis in the literature on the numerical analysis of stochastic differential equations. A result of \cite{hutzenthaler2011strong} established that for a variety of SDEs with non-Lipschitz drift, the standard Euler-Maruyama scheme is almost-surely explosive, and hence fails to be fit for purpose as a discretisation method. Subsequent work by the same authors \cite{hutzenthaler2012strong} introduced the `Tamed Euler' scheme, which `tames' the original drift to be better-behaved, performing a sort of `soft truncation' by analogy with the previous section. We follow here a simplified presentation of a perspective on taming given in \cite{szpruch2013v}, focusing for simplicity on the case of SDEs with additive noise. 

In general, given the SDE 
\begin{align*}
    \mathrm{d} X_t = b \left( X_t \right) \, \mathrm{d} t + \sigma \, \mathrm{d} W_t,
\end{align*}
`taming' is accomplished by introducing some step-size-dependent function $G^h \left( x \right)$ which vanishes as $h \to 0^+$ (e.g. $G^h(x)=h^\alpha \cdot \left\| b \left( x \right) \right\|$ for some $\alpha > 0$), defining
\begin{align*}
    b^h \left( x \right) = \left( 1 + G^h \left( x \right) \right)^{-1} \cdot b \left( x \right),
\end{align*}
and solving the `tamed' SDE
\begin{equation}\label{taming.equation:00}
    \mathrm{d} X_t = b^h \left( X_t \right) \, \mathrm{d} t + \sigma \, \mathrm{d} W_t,
\end{equation}
with a conventional Euler-Maruyama scheme of step-size $h$. For a well-chosen $G^h$, the drift $b^h$ can have much milder growth at infinity than the original $b$, and so the stability properties of this composite method can be obtained more directly. Moreover, since $G^h \left( x \right)$ vanishes as $h \to 0^+$, for small $h$, the two SDEs should follow one another reasonably closely. Indeed, \cite{hutzenthaler2012strong} establish strong convergence guarantees for the `Tamed Euler' scheme in some generality, allowing for results to be obtained in settings for which the basic Euler scheme is known to fail, and for which implicit schemes can be rather costly to implement.

In the context of Monte Carlo simulation, the approach of taming has largely been used to stabilise numerical approximations to the Overdamped Langevin Diffusion (\ref{overdamped.langevin:00}). \cite{brosse2019tamed} introduced the `Tamed ULA' approach (including a variant with `coordinate-wise' taming, natural for high-dimensional applications with meaningful coordinate structure), proving some theoretical guarantees, and performing some exploration of a Metropolis-adjusted variant. Various follow-up works (\cite{lytras2024tamed, lytras2025ktula, lytras2025taming}) have continued to complete the theoretical picture for these approaches, focusing predominantly on the unadjusted setting.
\begin{figure}[ht]
    \centering

    \begin{subfigure}[t]{0.48\textwidth}
        \centering
        \includegraphics[width=\textwidth]{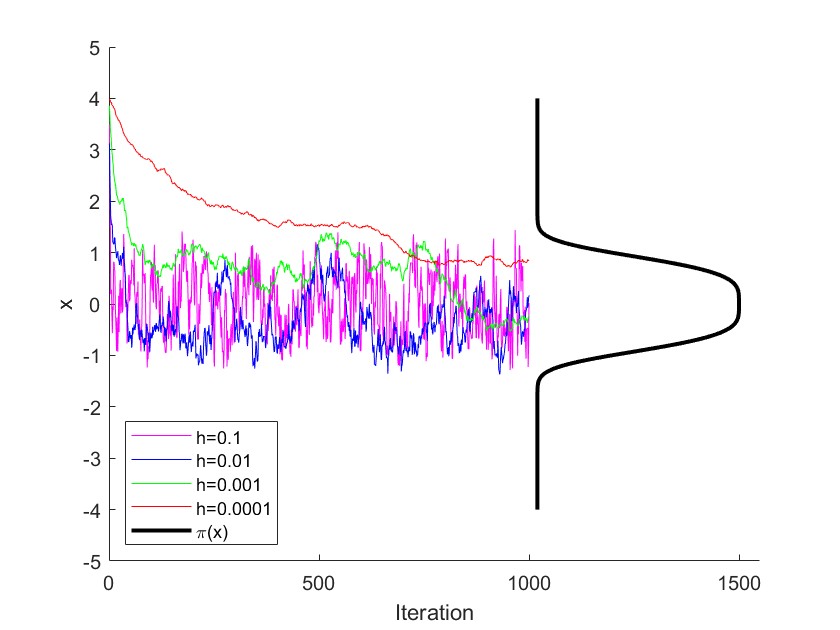}
        \caption{$\alpha = 1$}
    \end{subfigure}
    \hfill
    \begin{subfigure}[t]{0.48\textwidth}
        \centering
        \includegraphics[width=\textwidth]{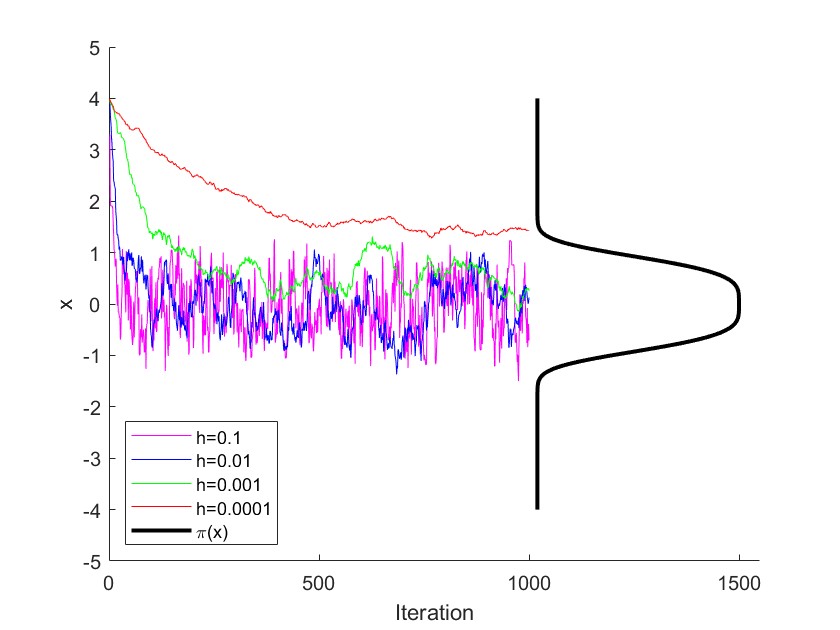}
        \caption{$\alpha = 0.5$}
    \end{subfigure}

    \caption{Tamed Langevin Monte Carlo with various step-sizes. The target density $\pi(x) \propto \exp \left( -x^4 \right)$ is overlaid on the right-hand side of each graph.}
    \label{fig:Tamed.MALA.Light}
\end{figure}

In Figure \ref{fig:Tamed.MALA.Light} we present a numerical simulation of the Tamed Metropolis Adjusted Langevin algorithm. We consider again the one-dimensional log-quartic target $\pi(x) \propto \exp \left( -x^4 \right)$, and use the Tamed scheme that aims to discretise the Langevin equation
\begin{align*}
    \mathrm{d} X_t &= \nabla \log \pi(X_t) \, \mathrm{d}t + \sqrt{2} \, \mathrm{d} W_t \\
    &= - X_t^3 \, \mathrm{d}t + \sqrt{2} \, \mathrm{d} W_t
\end{align*}
using an update of the form (\ref{taming.equation:00}), with 
\begin{align}
G^h \left( x \right) &= h^\alpha \cdot \left\| \nabla \log \pi \left( x \right) \right\| \\
&= h^\alpha \cdot \left\| x \right\|^3
\end{align}
The left plot considers the case where $\alpha = 1$, while the right one has $\alpha = 0.5$. Both plots consider different step-sizes, just as in the previous section. Comparing the trace plots with those of MALA
in Figure \ref{fig:MALA.stepsizes}, we once again see that the algorithmic performance is critically improved for large step-sizes. The algorithmic performance seems similar to the Truncated algorithm, discussed in Section \ref{sec:Truncated:00}.

\begin{figure}[ht]
    \centering

    \begin{subfigure}[t]{0.48\textwidth}
        \centering
        \includegraphics[width=\textwidth]{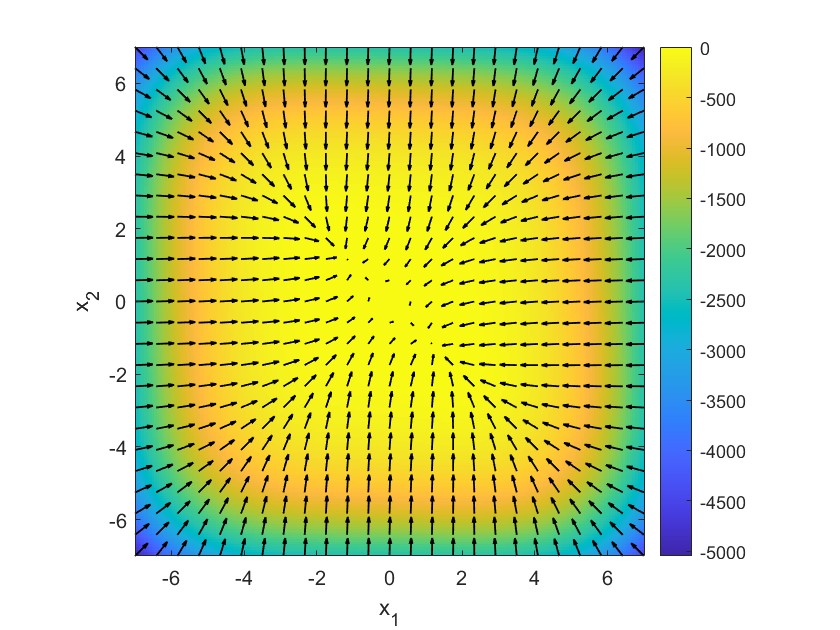}
        \caption{MALA}
    \end{subfigure}
    \hfill
    \begin{subfigure}[t]{0.48\textwidth}
        \centering
        \includegraphics[width=\textwidth]{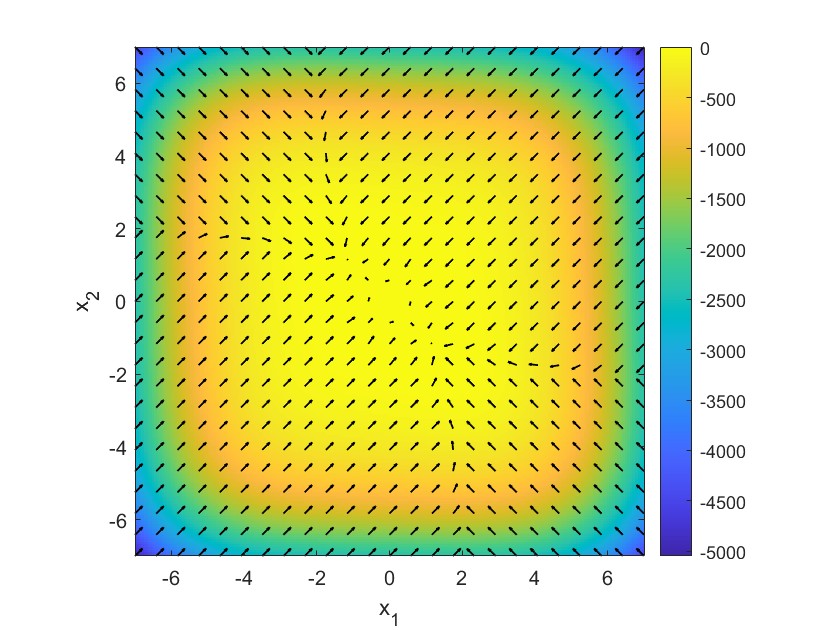}
        \caption{Truncated MALA}
    \end{subfigure}
    \hfill \begin{subfigure}[t]{0.48\textwidth}
        \centering
        \includegraphics[width=\textwidth]{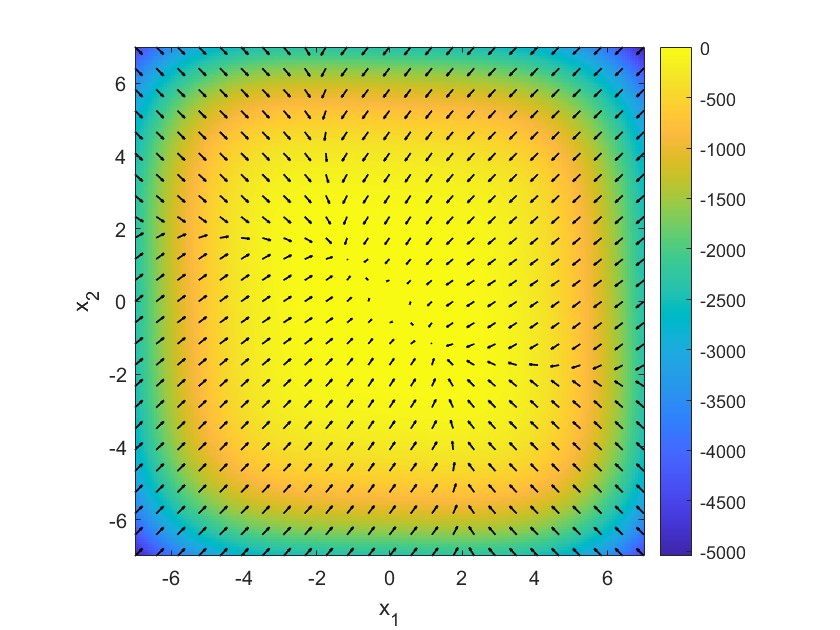}
        \caption{Tamed MALA}
    \end{subfigure}

    \caption{Vector fields of the expected proposed jump from the current position for MALA, Truncated MALA ($R=10$) and Tamed MALA. Target $\pi$ as in (\ref{target:two.d.vector.field}). Step-size $h=0.1$. Colours indicate the level sets of log-density.}
    \label{fig:Vector.Fields}
\end{figure}

Finally, to further showcase the differences between algorithms, in Figure \ref{fig:Vector.Fields} we consider the two-dimensional target of the form 
\begin{equation}\label{target:two.d.vector.field}
    \pi(x_1,x_2) \propto \exp \left( -x_1^4 - x_2^4 -5 \cdot x_1 \cdot x_2 \right)
\end{equation}
and for each of the algorithms MALA, Truncated MALA and Tamed MALA, we plot the vector field that indicates the expected proposed jump of the algorithm from each current position. The colour indicates the log density of the target at the current point and the magnitude of the arrow indicates how large the expected jump will be. By inspecting the figure, we see that away from the mode, the drift which is used in MALA is rather large in magnitude, which runs the risk of inducing numerical instability. On the other hand, for the other two algorithms, the drift tends to stay uniform in magnitude, while it still guides the process towards the mode, leading to more stable numerical behaviour.

%


\subsubsection{Proximal Langevin Monte Carlo}

A benefit of the taming approach as a numerical method is that it induces stability while retaining an explicit implementation. From a numerical-analytic perspective, another conventional approach to stability is to instead take an \textit{implicit} approach. Such approaches often observe similar stability benefits and improved accuracy, albeit at the cost of a more involved implementation.

\begin{figure}[ht]
    \centering
    \includegraphics[width=0.5\textwidth]{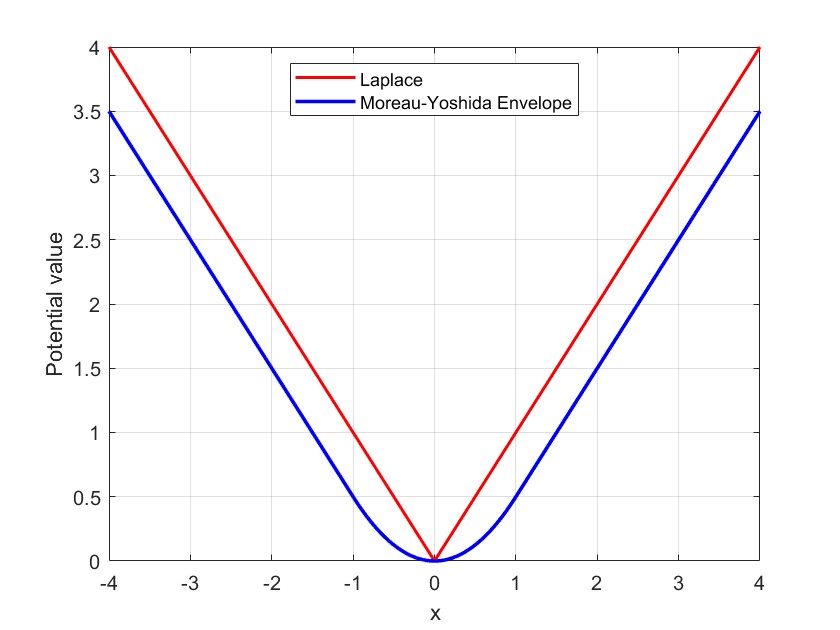}
    \caption{Laplace potential and its associated Moreau-Yosida envelope with $\lambda =1$}\label{fig:Moreau_Yosida}
\end{figure}

As applied to the Langevin diffusion, this approach has been implemented in a couple of different ways. The original work of \cite{pereyra2016proximal} starts from the perspective of sampling from a target whose potential $U$ is convex but non-smooth. Potentials of this form are historically common in optimisation-based approaches to image processing, and gained some popularity in sampling-based Bayesian approaches to the same task. With this non-smoothness in mind, the author proposes to mollify the target potential $U$ according to the `Moreau-Yosida convolution', i.e. constructing the `Moreau-Yosida envelope' as
\begin{align*}
    U^\lambda \left( x \right) := \inf \left\{ U \left( y \right) + \frac{\left\| x - y\right\|^2}{2 \cdot \lambda} : y \in \mathbf{R}^d \right\}
\end{align*}
to obtain an approximate target with improved regularity properties. In particular (see Theorem 4.1.4  and Proposition 4.1.5 of \cite{hiriart.lemarechal:93}), assuming that $U$ is convex with a closed graph, then
\begin{equation*}
    U^{\lambda}(x) \xrightarrow{\lambda \rightarrow 0} U(x)
\end{equation*}
point-wise, while
the mollified potential $U^\lambda$ is automatically differentiable, and its gradient is automatically Lipschitz-continuous, with Lipschitz constant upper-bounded by $\lambda^{-1}$. Moreover, provided that the original $U$ is convex, so too will $U^\lambda$ be. Combining these properties, one sees that usual numerical discretisations of the gradient flow with respect to $U^\lambda$ will behave stably, even in the presence of stochastic noise and other perturbations. Bearing all of this in mind, $U^\lambda$ presents itself favourably as a candidate for more naive numerical discretisation, whereby explosivity and instability of the process are essentially ruled out by design. For a graphical representation of the Moreau-Yosida envelope, see Figure \ref{fig:Moreau_Yosida}, where it is applied with $\lambda=1$ on the (non-differentiable at zero) Laplace potential.

Following in this direction, using Euler discretisation on the Overdamped Langevin Diffusion with target measure proportional to $\exp \left\{ - U^\lambda \left( x \right) \right\}$, one obtains the `Proximal ULA' (P-ULA) proposal as
\begin{align*}
    X_n = X_{n-1} - h \cdot \nabla U^\lambda (X_{n-1}) + \sqrt{2 \cdot h} \cdot \xi_n , \qquad \xi_n \sim \mathcal{N} \left( 0, \mathbf{I}_d \right).
\end{align*}
This framing highlights an interesting point of comparison with some of the previous solutions: while the Truncated and Tamed Langevin strategies can be viewed as employing a modified numerical scheme to sample from the original target distribution (up to discretisation error), the Proximal Langevin approach instead fixes the numerical scheme, and systematically changes the target distribution. We will see in later discussion that this conceptual distinction allows for some specific functionalities which are not available to all approaches.

A priori, one could expect that computing $U^\lambda$ and $\nabla U^\lambda$ at each step might add substantial complexity to the algorithm. However, in various settings, the additional burden is relatively manageable. In particular, for small enough $\lambda$, the optimisation formulation which defines $U^\lambda$ is a strongly-convex minimisation problem, which can be rapidly solved using the methods of modern convex optimisation (see e.g. \cite{boyd2004convex, bubeck2015convex}). Moreover, the gradient $\nabla U^\lambda$ can be obtained by examining the value of $y$ which solves this minimisation problem, that is
\begin{align*}
    \nabla U^\lambda \left( x \right) &= \lambda^{-1} \cdot \left( x - \mathrm{prox}_{\lambda \cdot U} \left( x \right) \right) \\
    \mathrm{prox}_{\lambda \cdot U} \left( x \right) &:= \arg \min \left\{ U \left( y \right) + \frac{\left\| x - y\right\|^2}{2 \cdot \lambda} : y \in \mathbf{R}^d \right\},
\end{align*}
where $\mathrm{prox}_{\lambda \cdot U}$ is the so-called `proximal operator' associated to $U$. As such, both $U^\lambda$ and its gradient can be obtained `in a single pass', so to speak. Observe also that upon taking $\lambda = h$, the P-ULA proposal simplifies to
\begin{align*}
    X_n = \mathrm{prox}_{h \cdot U} \left( X_{n - 1} \right) + \sqrt{2 \cdot h} \cdot \xi_n , \qquad \xi_n \sim \mathcal{N} \left( 0, \mathbf{I}_d \right).
\end{align*}

As an aside for readers unfamiliar with convex analysis, it may be useful for one's intuition to verify that when $U$ corresponds to the characteristic function of a closed, convex set $\mathcal{K}$ (i.e. $0$ for $x \in  \mathcal{K}$ and infinite otherwise), the Moreau envelope takes the form $U^\lambda \left( x \right) = \frac{1}{2\cdot\lambda}\cdot\mathsf{dist}^2 \left(x, \mathcal{K} \right)$, and the proximal operator is the projection onto $\mathcal{K}$. In this sense, the Moreau-Yosida smoothing can be viewed as a way of generalising these notions of distance and projection from sets to arbitrary convex functions. 

Having established these preliminaries, \cite{pereyra2016proximal} identifies a range of practical problems in high-dimensional Bayesian image analysis for which basic Langevin algorithms perform poorly, the new proposals can be computed efficiently, and the resulting proximal Langevin algorithms --  whether unadjusted (P-ULA) or Metropolis-adjusted (P-MALA) -- perform favourably.

\begin{figure}[ht]
    \centering
    \begin{subfigure}[t]{0.47\textwidth}
        \includegraphics[width=\linewidth]{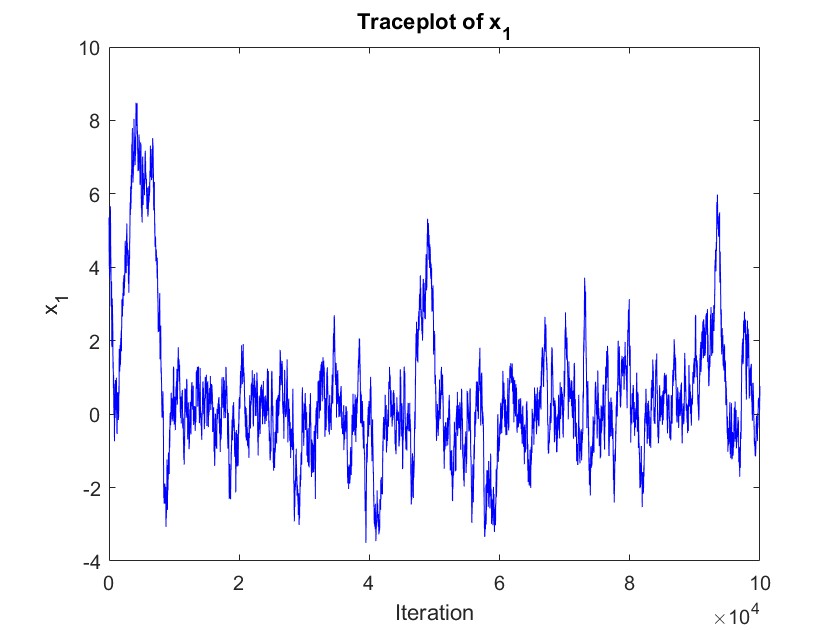}
        \caption{Traceplot of first coordinate. Step-size $h=0.01$.}
        \label{fig:PMALA.Laplace:part1}
    \end{subfigure}
    \hfill
    \begin{subfigure}[t]{0.47\textwidth}
        \includegraphics[width=\linewidth]{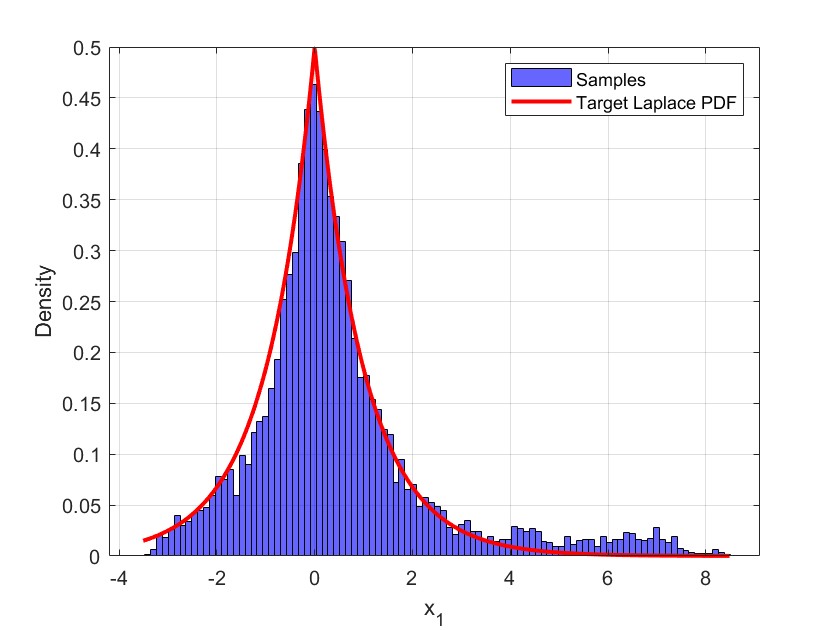}
        \caption{Histogram of first coordinate samples. Step-size $h=0.01$.}
        \label{fig:PMALA.Laplace:part2}
    \end{subfigure}
    
    \vspace{0.5cm}  

    \begin{subfigure}[t]{0.47\textwidth}
        \includegraphics[width=\linewidth]{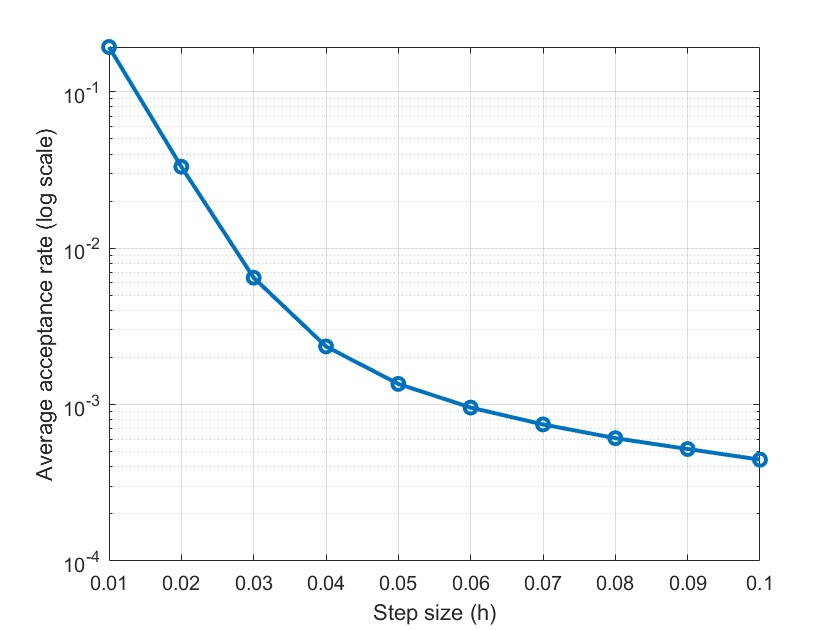}
        \caption{Average acceptance probability over step-sizes.}
        \label{fig:PMALA.Laplace:part3}
    \end{subfigure}
    \hfill
    \begin{subfigure}[t]{0.47\textwidth}
    \includegraphics[width=\linewidth]{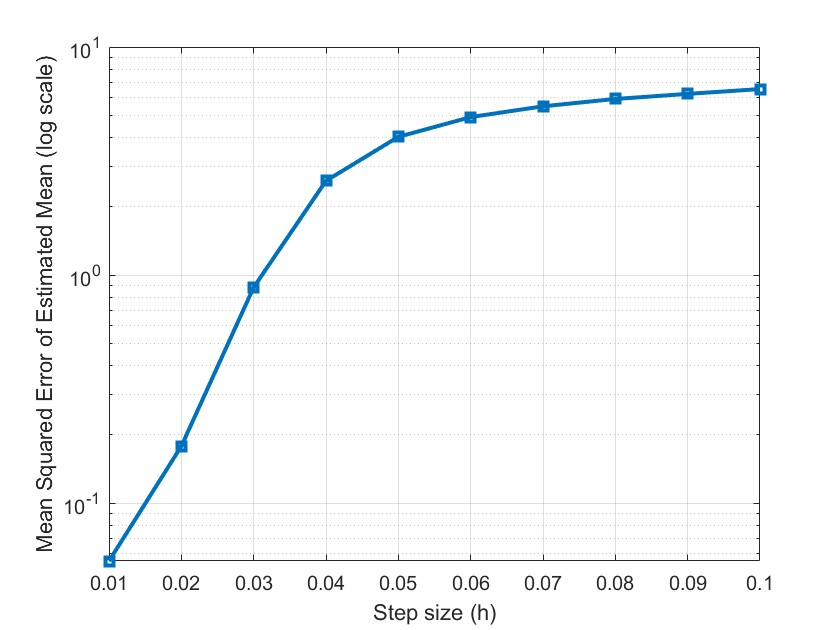}
        \caption{Mean square error of target expectation over step-sizes}\label{fig:PMALA.Laplace:part4}
    \end{subfigure}
    \vspace{1mm}
    \caption{Proximal MALA on the $10^4$-dimensional Laplace target (\ref{Laplace.target:1}). First row: traceplot and histogram with optimally tuned step-size  $h=0.01$. Second row: Average acceptance probability and Mean squared error (estimating the target expectation) over step-size.}
    \label{fig:PMALA.Laplace}
\end{figure}

To showcase the performance of P-MALA, we return to the example of a $10^4$-dimensional product-of-Laplace target distribution (\ref{Laplace.target:1}), following Example \ref{example:Laplace}. We report our results in Figure \ref{fig:PMALA.Laplace}. On the first row we report the traceplot and histogram of the first coordinate of a P-MALA algorithm, ran for $N = 10^5$, using step-size $h=0.01$. The step-size was chosen so that the average acceptance of proposed step is close to $36 \%$ (see e.g. \cite{crucinio2025optimal}). The chain was initialised at $\left( 5, 5, \dots, 5 \right)$, a significant distance from the mode. On the second row, we present a similar analysis to Figure \ref{fig:MALA.Laplace}, testing the behaviour of the algorithm over different step-sizes and reporting the average (over 100 independent runs) acceptance probability and Mean Squared Error (MSE) when estimating the target expectation (in this case zero). While the variations in acceptance probability and MSE seem similar to Figure \ref{fig:MALA.Laplace}, it can be observed that for all step-sizes, the average acceptance probability for P-MALA is larger than it is for MALA, while the MSE is always lower (albeit still quite large) for most step-sizes, indicating better algorithmic performance. Comparing with Figure \ref{fig:MALA.Laplace}, it seems that the PMALA has captured the shape of the distribution in a more efficient manner.

As a brief aside, we note that while the Moreau-Yosida convolution always delivers a potential whose gradient is Lipschitz, for small regularisation parameters $\lambda$, this Lipschitz constant can also be small, leading to residual stiffness in the dynamics of the Langevin diffusion. In this setting, it has proven fruitful \citep{pereyra2020accelerating} to make use of `explicit, stabilised' numerical integrators which are able to resolve this stiffness while retaining a large nominal step-size, at an increased but worthwhile per-iteration computational cost.

In terms of other algorithms that are based on the idea of Moreau-Yosida convolution, \cite{wibisono2019proximal} studies a related algorithm which transposes the order of computing $\mathrm{prox}_{\lambda \cdot U}$ and injecting additive Gaussian noise, and \cite{benko2025langevin} study a variant which accommodates the possibility of computing the proximal operator inexactly, each obtaining convergence results along the way.

Subsequent work of \cite{hodgkinson2021implicit} revisits these ideas from a more numerical-analytic perspective, instead framing the P-ULA proposal as a (non-standard) implicit Euler-Maruyama discretisation of the Langevin diffusion. Building on this perspective, the authors propose to generate moves by the `theta method', whereby one fixes a $\vartheta \in \left[ 0, 1 \right]$ and transitions from $X_{n-1}$ to $X_n$ by solving the nonlinear equation
\begin{align*}
    X_n = X_{n-1} - h \cdot \left\{ \vartheta \cdot \nabla U (X_{n-1}) + \bar{\vartheta} \cdot \nabla U (X_n) \right\} + \sqrt{2 \cdot h} \cdot \xi_n , \qquad \xi_n \sim \mathcal{N} \left( 0, \mathbf{I}_d \right),
\end{align*}
where $\vartheta + \bar{\vartheta} = 1$. The resulting algorithm is then termed the `Implicit Langevin Algorithm' (ILA). For $\vartheta = 0$, this is the conventional explicit ULA proposal, which can be expected to suffer from stability issues as detailed earlier. For $\vartheta > 0$, this is now an \textit{implicit} proposal, which requires the solution of an auxiliary optimisation problem, but is expected to perform favourably with a view to numerical and dynamical stability. While this formulation is superficially different from that of P-ULA, some rearrangement shows that $X_n$ can be obtained from $X_{n - 1}$ as the solution to
\begin{align*}
    X_n &= \arg\min \left\{ U \left( y \right) + \frac{\left\| y - \bar{X}_{n, \bar{\vartheta}} \right\|^2}{2 \cdot \vartheta \cdot h} : y \in \mathbf{R}^d\right\} \\
    \text{where } \bar{X}_{n, \bar{\vartheta}}  &= X_{n-1} - h \cdot \bar{\vartheta} \cdot  \nabla U \left( X_{n-1} \right) + \sqrt{2 \cdot h} \cdot \xi_n,
\end{align*}
i.e. by minimising a similar quadratic penalisation of the potential $U$. As such, the implementation complexity of ILA should be comparable to that of P-ULA. Note that ILA is superficially closer to the method proposed in \cite{wibisono2019proximal} than to P-ULA, in that the proximal operator is applied \emph{after} adding noise to the current iterate.

One might expect this methodology to perform reasonably in similar situations to P-ULA, although the work \cite{hodgkinson2021implicit} does not really examine the non-smooth case empirically, focusing instead on taking larger step-sizes in the setting of strongly-log-concave targets with Lipschitz-continuous forces, albeit with a large condition number.

An interesting hybrid approach was presented recently in \cite{shukla2025mcmc}. This approach proposes to run MCMC which genuinely targets the Moreau-smoothing $\pi^\lambda$ of the target distribution $\pi$, and then use these samples as the basis for an importance sampling correction. In practice, this amounts to using the \textit{proposal} kernel corresponding to P-ULA (or similar), but performing the Metropolis-adjustment with respect to $\pi^\lambda$ rather than $\pi$. One thus benefits from the good stability properties of the MCMC kernels, allowing for a relatively friendly implementation. Moreover, while the invariant measure of the implemented MCMC kernel is not the target distribution $\pi$, for a well-chosen $\lambda$, one expects to sample from a rather good approximation to $\pi$, and so the importance sampling adjustment should be rather gentle, leading to only a mild inflation of variance. Observe that implementing the same conceptual program with the Truncated or Tamed Langevin approach would not be entirely straightforward, as the invariant measure of the unadjusted methods is both different from $\pi$ and difficult to characterise explicitly.




\subsubsection{Barker-Langevin Monte Carlo}

The preceding methods largely have their genesis in viewing existing MCMC algorithms as discretisations of diffusion processes. The so-called `Barker proposal', as introduced in \cite{livingstone2022barker}, originates from a different perspective of `local balancing', initially put forward as a general principle in \cite{zanella2020informed}. In intuitive terms, the local balancing framework takes as given some target-agnostic `local exploration kernel' $Q \left( x, \mathrm{d} y \right)$, and seeks to correct it gently so as to obtain an improved proposal kernel. In particular, the authors considered `tilted' kernels of the form
\begin{align*}
    \bar{Q} \left( x, \mathrm{d} y \right) \propto Q \left( x, \mathrm{d} y \right) \cdot \beta \left( \frac{\pi \left( y \right)}{\pi \left( x \right)} \right)
\end{align*}
for some `balancing function' $\beta : \mathbf{R}_+ \to \mathbf{R}_+$. By taking this function to satisfy the symmetry condition
\begin{align*}
    \beta \left( r \right) = r \cdot \beta \left( r^{-1} \right),
\end{align*}
and assuming that $Q$ is symmetric (in that the law $Q(x,\cdot)$ has a density which is invariant under swapping its arguments), one obtains a kernel which is approximately $\pi$-reversible, and is hence expected to interface well with Metropolis-adjustment. In general, simulation from the kernel $\bar{Q}$ is not directly tractable, and so some additional approximations are required in order to obtain a practical kernel.

\begin{example}[Connections with ULA]\label{Barker.ULA.connection}
Focus on the case in which $Q_\sigma$ is a simple Gaussian random walk with proposal standard deviation equal to $\sigma$, and consider $\beta(r)=r^\frac{1}{2}$. One can then compute that the resulting $\bar{Q}$ takes the form
\begin{align*}
    \bar{Q}_{\sigma} \left( x, \mathrm{d} y \right) &\propto Q_{\sigma} \left( x, \mathrm{d} y \right) \cdot \sqrt{\frac{\pi \left( y \right)}{\pi \left( x \right)}} \\
    &=  Q_{\sigma} \left( x, \mathrm{d} y \right) \cdot \exp\left\{ \frac{1}{2} \left(\log \pi(y) - \log \pi(x) \right) \right\} \\
    &\approx  Q_{\sigma} \left( x, \mathrm{d} y \right) \cdot \exp \left( \frac{1}{2} \left\langle \nabla \log \pi \left( x \right), y - x \right\rangle \right).
\end{align*}
It can be verified that this final law coincides with the ULA kernel. As per our earlier discussion, while this method can perform well for regular targets, one anticipates various instabilities in the face of more serious steepness.
\end{example}

An insight of \cite{livingstone2022barker} is that one can consider the same construction, but replacing the square root with the so-called `Barker' balancing function $\beta \left( r \right) = \frac{r}{1 + r}$ (named due to a connection with an early work of \cite{barker1965monte}). Similarly to Example \ref{Barker.ULA.connection}, a first order Taylor approximation  of the form $\log \pi(y) - \log \pi(x) \approx \langle \nabla \log \pi(x), y-x \rangle $ yields
\begin{align*}
    \bar{Q}_{\sigma} \left( x, \mathrm{d} y \right) &\propto Q_{\sigma} \left( x, \mathrm{d} y \right) \cdot \frac{\pi \left( y \right)}{\pi \left( x \right) + \pi \left( y \right)} \\
    &\approx  Q_{\sigma} \left( x, \mathrm{d} y \right) \cdot s \left(  \left\langle \nabla \log \pi \left( x \right), y - x \right\rangle \right),
\end{align*}
where $s : t \mapsto \left( 1 + \exp \left( - t \right) \right)^{-1}$ is a standard sigmoid function, which also coincides with the distribution function of the standard Logistic distribution. The resulting kernel
\begin{equation}
    \hat{Q} (x, \mathrm{d}y) = 2 \cdot \frac{1}{\sqrt{2\pi \sigma^2}} \exp \left( -\frac{1}{2\sigma^2} \left| y - x \right|^2 \right) \cdot s \left(  \left\langle \nabla \log \pi \left( x \right), y - x \right\rangle \right)dy
\end{equation}
is then viable as a proposal kernel which can be corrected via a Metropolis acceptance step. Interestingly, sampling from the proposal $\hat{Q}_{\sigma}$ has a simple algorithmic interpretation: first, `pre-propose' a uninformed move by generating some standard Gaussian noise $\xi$, and by looking at the alignment of $\xi$ with $\nabla \log \pi \left( x \right)$, then decide whether it will be more fruitful to propose a `real' move to $y_+ = x + \sigma \cdot \xi$, or to its `reflection', $y_- = x - \sigma \cdot \xi$. In particular, given $\xi$, one samples $y_+$ with probability $p(x,\xi) = s \left(  \left\langle \nabla \log \pi \left( x \right), \sigma \cdot \xi \right\rangle \right)$, and $y_-$ with the complementary probability $1 - p \left( x, \xi \right)$.

A key property of this proposal is that while it is gradient-informed (and hence makes active use of the local geometry of the target distribution), it also enjoys a distinctive `bounded influence' property: even in the face of massive gradients, one does not make massive moves, as the original size of the jump $\sigma \cdot \xi$ comes from a Gaussian distribution. Instead, one simply becomes overwhelmingly likely to move in a specific direction. In this regard, the proposal circumvents one of the challenging effects of steep gradients.

In \cite{livingstone2022barker}, however, it was shown that the spectral gap of Metropolised $\hat{Q}_{\sigma}$ is no better than twice the spectral gap of the Metropolised Gaussian Random Walk $Q_{\sigma}$, which indicates that the computational gains obtained by choosing such a method are rather limited. An important practical variation of the algorithm was thus suggested. Assuming that the target lives in $\mathbf{R}^d$, instead of choosing whether to add or subtract the noise vector $\xi$ `all at once', one can instead look at each of the $d$ coordinates of the noise \emph{separately}, and consider all $2^d$ possible combinations in which this noise can be added to the current state $x$. More specifically, one first samples a Gaussian random variable $\xi \sim \mathcal{N} \left( 0, \mathbf{I}_d \right)$. For any coordinate $i \in \{ 1,...,d \}$, one then samples a random variable $b_i$ such that
\begin{align*}
    b_i = 
        \begin{cases}
        1 & \text{with probability } p_i(x,\xi), \\
        -1 & \text{with probability } 1 - p_i(x,\xi),
        \end{cases}
\end{align*}
where 
\begin{equation*}
    p_i \left( x, \xi \right) = s \left(  \frac{\partial}{\partial x_i} \log \pi \left( x \right) \cdot \sigma \cdot \xi_i \right)
\end{equation*}
and for $\mathrm{b}=(b_1,\dots,b_d) \in \left\{ \pm 1\right\}^d$, one proposes to move to $y_\mathrm{b} = x + \sigma \cdot \mathrm{b} \odot \xi$, where $\odot$ denotes the element-wise product. In other words, the algorithm decides whether to accept or reflect in the direction of each coordinate \emph{separately}, conditionally on $x$ and $\xi$. This modified procedure has become known as the ``(coordinate-wise) Barker proposal''. It preserves many of the desirable stability properties of the original $\hat{Q}$, but allows for extra flexibility and adaptivity to the information provided by the gradient, leading to improved sampling efficiency, particularly in high dimension. While the proposal is not exact in terms of exactly preserving $\pi$, it can similarly be corrected by Metropolis-adjustment without losing most of its computational efficiency; see \cite{vogrinc2023optimal} for a scaling analysis in support of this observation.

Working backwards from this construction, recent work of \cite{livingstone.nusken.vasdekis.zhang:24} proposes to use the same `propose-reflect' mechanism as a tool for numerical discretisation of general stochastic differential equations, naturally focusing on applications with similarly `steep' drifts. Interestingly, one can interpret the Barker proposal as a special case of this SDE discretisation, when applied to the Overdamped Langevin Diffusion (\ref{overdamped.langevin:00}), which further motivates the use of the Barker proposal. It also presents a number of opportunities for future developments in which Metropolis-adjustment of the dynamics is not a key priority, as is often the case for e.g. stochastic gradient methods; see e.g. \cite{mauri2024robust}.

\begin{figure}[ht]
    \centering
\includegraphics[width=0.5\textwidth]{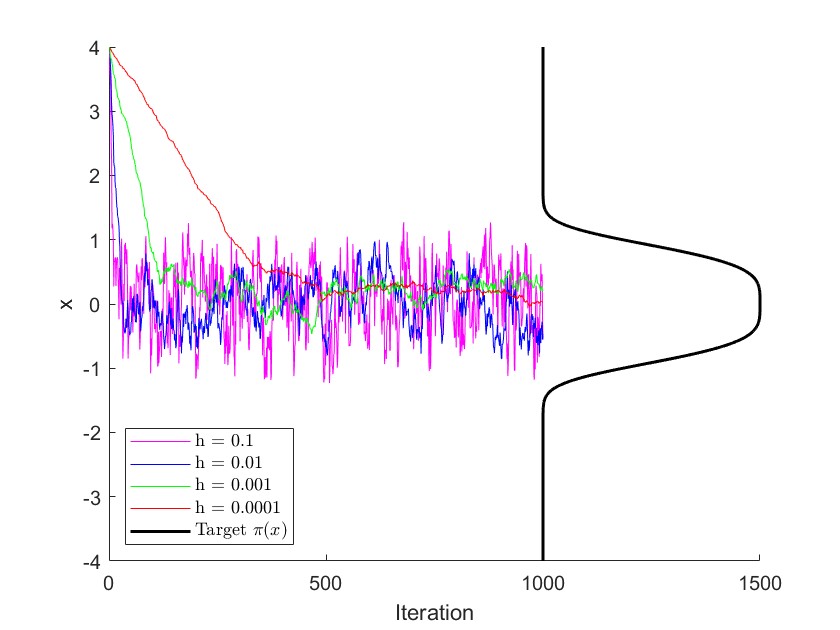}
    \caption{Barker trace plots with various step-sizes. The target density ($\pi(x) \propto \exp \left( -x^4 \right)$) is overlaid on the right hand side of the graph.}\label{fig:Barker.SuperLight}
\end{figure}

In order to showcase the performance of the Barker proposal, we consider two numerical examples. The first target is our running univariate log-quartic model problem. In Figure \ref{fig:Barker.SuperLight}, we present the traceplots of the algorithm run for $N = 1000$ iterations, starting from $x_0 = 4$ and for four different step-sizes ($h = 0.1, 0.01, 0.001 , 0.0001$). Contrary to MALA, Barker shows a remarkable robustness in the behaviour across the different step-sizes. Interestingly, for this particular example, it seems that for small step-sizes, the method is able to find the mode of the distribution faster than for the previously mentioned algorithms. We conjecture that this is due to the reflection mechanism, which allows for better-informed proposal jumps, and hence to fewer rejected moves.

\begin{figure}[ht]
    \centering
    \includegraphics[width=0.48\textwidth]{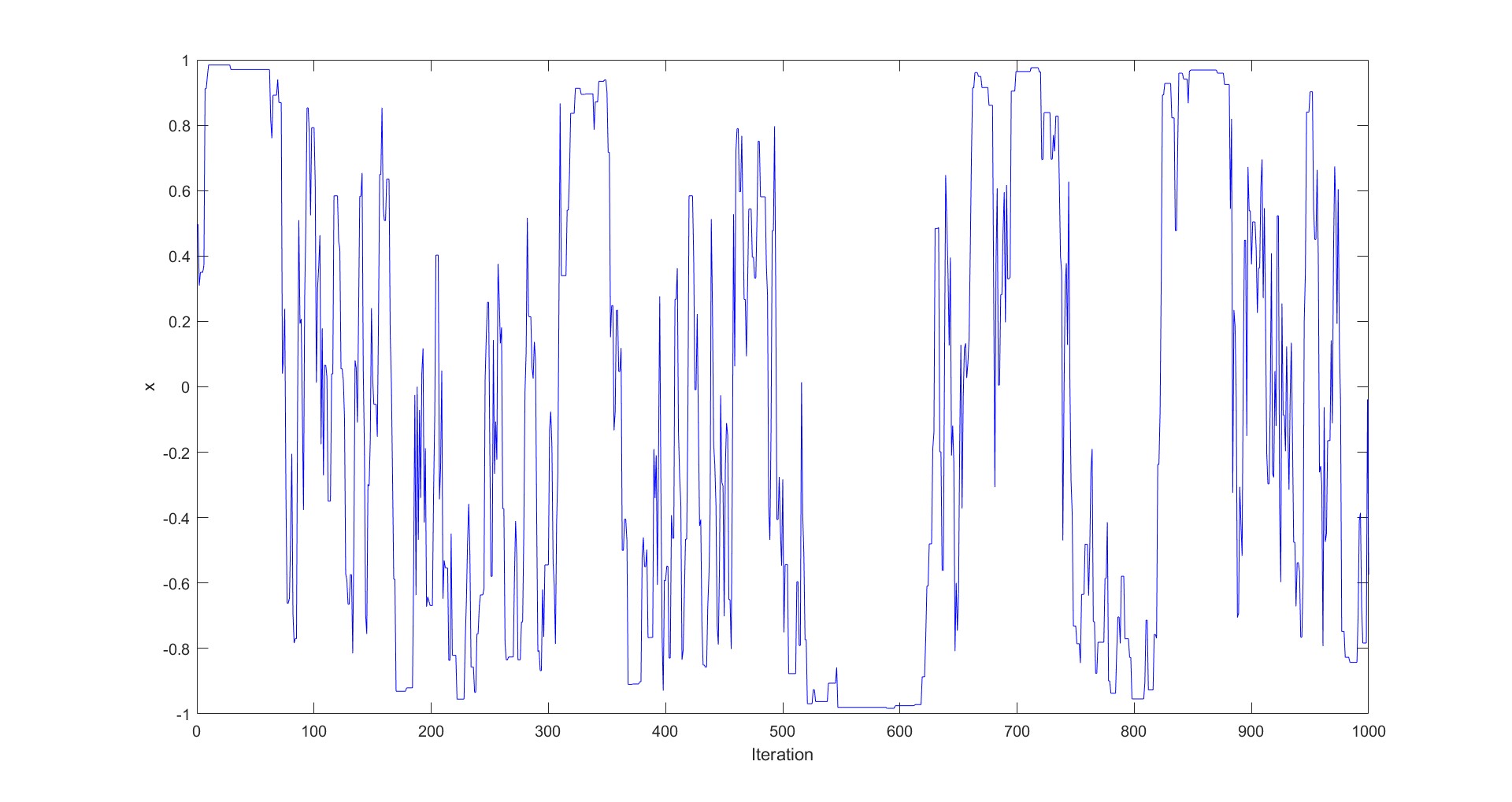}
    \hfill
    \includegraphics[width=0.48\textwidth]{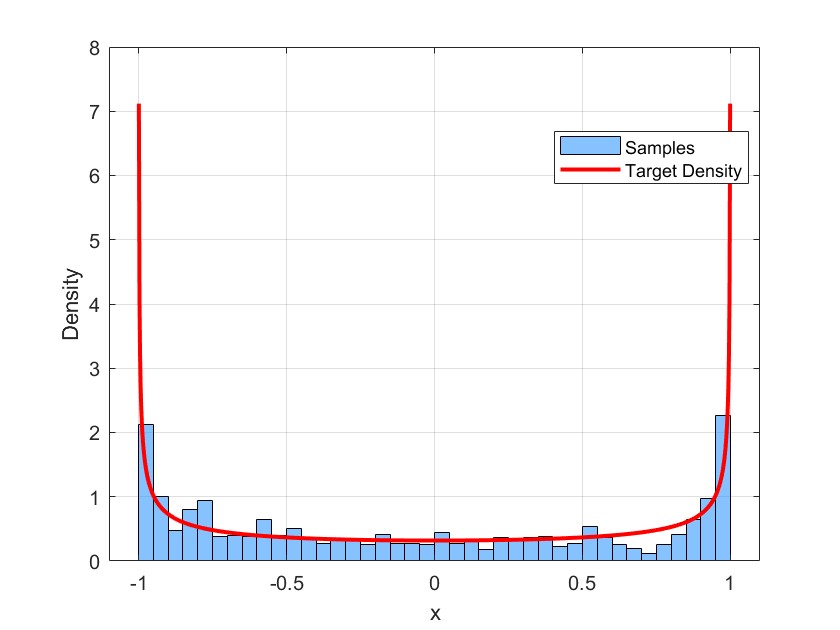}
    \caption{Barker traceplot and histogram on the one-dimensional target with exploding boundary at $-1$ and $1$ ($\pi(x) \propto \left( 1-x^2 \right)^{-\frac{1}{2}}$). Step-size $h=0.1$.}\label{fig:Barker.Exploding.Boundary}
\end{figure}

The second target is of the form (\ref{target.exploding.boundary:1}), supported on $(-1,1)$, and exploding at the boundary. The algorithm is run for $N = 1000$ iterations, with step-size $h = 0.1$, starting from $ 0.5$. We present the traceplot and histogram of the MALA algorithm in Figure \ref{fig:Barker.Exploding.Boundary}. Compared to Figure \ref{fig:MALA.Exploding.Boundary}, the Barker traceplot seems to get stuck less often at the boundary, while the histogram looks much more symmetric and tends to capture the shape of the target more more efficiently. We should emphasise here, that over independent realisations of the algorithm, the Barker chains showcased much more consistent behaviour when compared to MALA, which has frequent realisations that completely failed to explore the area away from the boundaries, and would presumably require drastically more iterations in order to converge reasonably.

\begin{figure}[ht]
    \centering
\includegraphics[width=11cm, height=5cm]{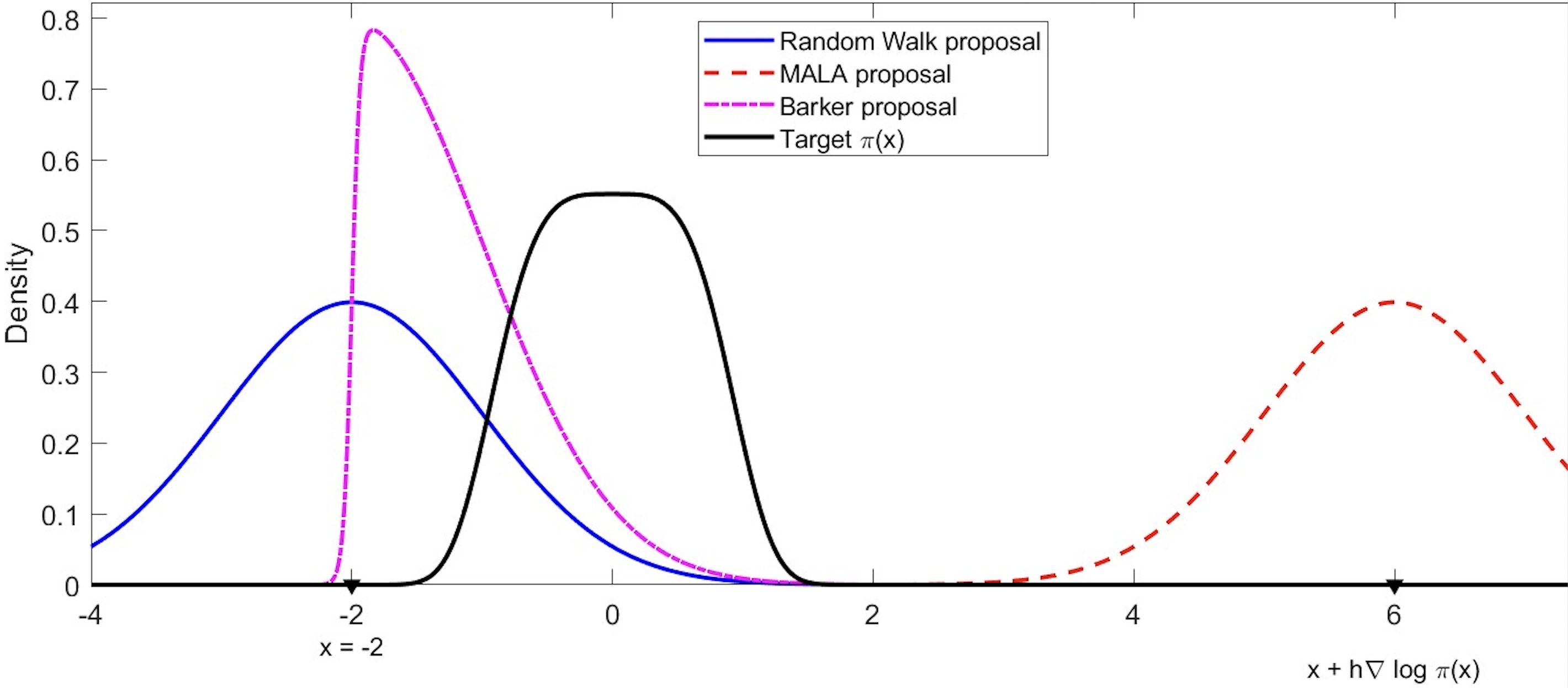}
    \caption{Density plots of Barker, MALA and Random Walk proposals. Target $\pi(x)$ is the log-quartic. Step-size $h=0.5$. Current position is $x=-2$.}\label{fig:Barker.and.other.densities}
\end{figure}

In order to give a visual representation of the Barker proposal, in Figure \ref{fig:Barker.and.other.densities} we plot the densities of the Barker, MALA and Random Walk proposal (Gaussian centered around the current position), when targeting the log-quartic target $\pi(x) \propto \exp \left( -x^4 \right)$. We use step-size $h=0.5$ and the current position for all three proposals is $x=-2$. It is evident that the Gaussian Random walk does not capture the shape of the target well, assigning most of its mass close to the current point $x=-2$. MALA on the other hand, failed to an ``overshooting'' phenomenon, jumping massively to the other side of the tails of the target (centered around $6$). The Barker seems to be the middle ground, centering its mass closer towards zero (the target mean).

%


\subsubsection{Hamiltonian Monte Carlo with Non-Quadratic Kinetic Energy}

Beyond Langevin-type MCMC methods, a particularly popular class of modern gradient-based MCMC algorithms are developed around \textit{Hamiltonian Monte Carlo} (HMC), a broad term which we take to encompass various implementable discretisations of the aforementioned Refreshed Hamiltonian Dynamics process of \cite{bou2017randomized}. Despite various conceptual differences to Langevin Monte Carlo (which include moving from stochastic dynamics to deterministic dynamics, from `one-step' proposal mechanisms to `trajectorial' proposals, and so on), in practice, Hamiltonian Monte Carlo suffers from the same qualitative shortcomings in the face of steep gradients; as soon as $\nabla \log \pi$ is of superlinear growth at infinity, the numerical stability of conventional numerical integrators can collapse. By analogy with Proximal Langevin algorithms, \cite{chaari2016hamiltonian} proposed to run HMC on a Moreau-Yosida-smoothed version of the target distribution; this appears not to have been taken up much in practice, though the recent appearance of the preprint \cite{shukla2025proximal} suggests that the door has not been closed entirely.

An interesting alternative approach was proposed by \cite{livingstone2019kinetic}, who leave the potential $U$ unchanged, and instead seek to modify the \emph{kinetic} energy associated to the underlying (fictitious) Hamiltonian system. More precisely, where the usual HMC samples from a joint target distribution $\mu$ (with states given by $z=(x,v)$) which satisfies
\begin{align*}
    -\log \mu \left( z \right) = U \left( x \right) + \frac{1}{2} \left\| v \right\|^2 +\text{const.},
\end{align*}
one can instead seek to modify the marginal distribution of the momentum\footnote{We hope that the expert reader will forgive us for sticking to `velocity-type' notation here, in the interests of cohesion with our presentation of PDMPs.} $v$ and sample according to
\begin{align*}
    -\log \mu \left( z \right) = U \left( x \right) + K \left( v \right) +\text{const.},
\end{align*}
for some other well-chosen function $K : \mathbf{R}^d \to \mathbf{R}$. In this setting, the Hamiltonian dynamics instead take the form
\begin{align*}
    \dot{x} = \nabla K \left( v \right), \qquad \dot{v} = - \nabla U \left( x \right),
\end{align*}
which exposes that steepness in $\nabla U$ might be alleviated by suitable combination with $\nabla K$. Indeed, the authors argue in favour of designing $K$ such that $\nabla K \circ \nabla U$ is asymptotically of linear growth as $\left\| x \right\| \to \infty$. This facilitates the numerical stability of `usual' discretisations of Hamiltonian dynamics, while also ensuring that the state $x$ ultimately has a linear drift back towards the `bulk' of the state space, i.e. that the dynamics are not flattened to the point of inefficiency. They then advocate for a `relativistic' kinetic energy choice (following \cite{lu2017relativistic}) of the form
\begin{align*}
    K \left( v \right) = \left( 1 + \left\| v \right\|^2 \right)^{a/2} - 1, \qquad a \geq 1
\end{align*}
(modulo various multiplicative scaling factors), which has the desired quadratic-like behaviour for small $\left\| v \right\|$, but grows like $\| v \|^a$ for large $\left\| v \right\|$, ensuring a gentle perturbation to conventional Hamiltonian dynamics in the bulk of the space, with a flexible regularising influence out in the tails. In Figure 18, we present the contour plots of the Hamiltonian dynamics for the one-dimensional log-quartic target (that is with potential $U(x)=x^4$) and two choices of kinetic energy. The horizontal axis represents the $x$-space, with the vertical axis as the velocity space. These plots show the curves along which the Hamiltonian energy $- \log \mu(x,v)$ remains constant, along which the dynamics move. In particular, if we start from $\left( x, v \right) = \left( 1, 0 \right)$, the path of the dynamics is indicated with bold.

\begin{figure}[ht]
    \centering

    \begin{subfigure}[t]{0.48\textwidth}
        \centering
        \includegraphics[width=\textwidth]{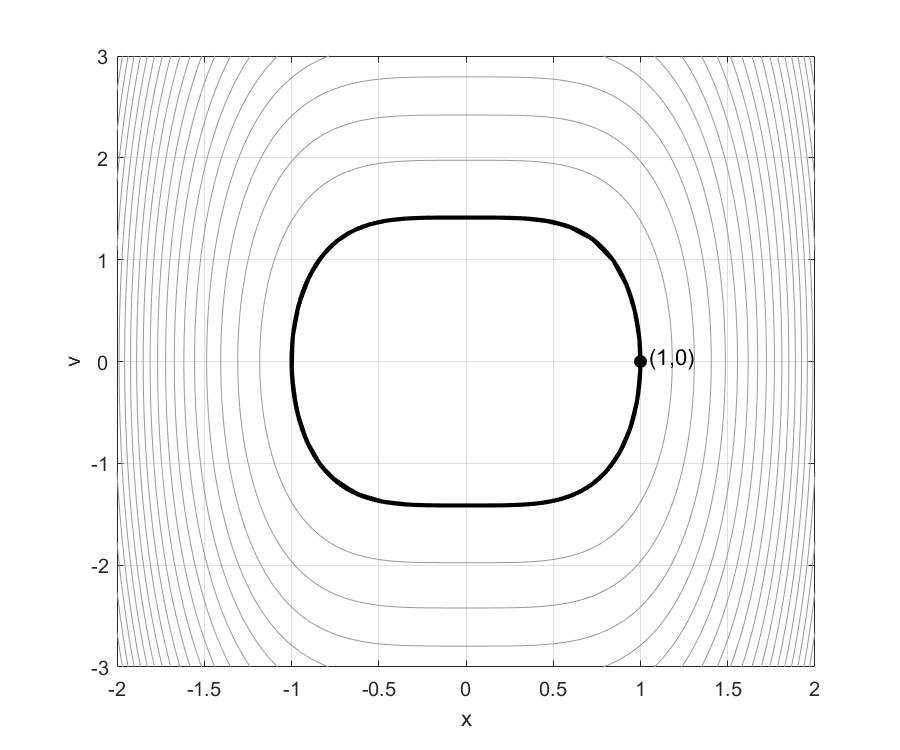}
        \caption{$K(v)=\frac{1}{2}v^2$}
    \end{subfigure}
    \hfill
    \begin{subfigure}[t]{0.48\textwidth}
        \centering
        \includegraphics[width=\textwidth]{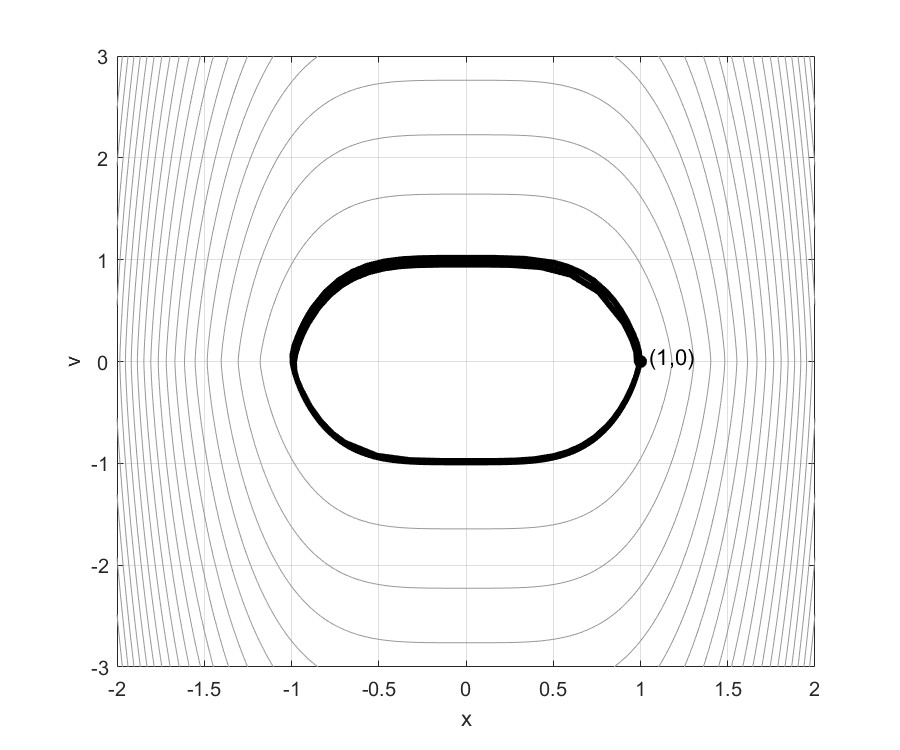}
        \caption{$K(v)= v^{4/3}$}
    \end{subfigure}
    \caption{Contour plots of the Hamiltonian dynamics with the quartic potential $U(x)=x^4$ and different kinetic energies. $x$- axis is the $x$-space, $y$-axis is the velocity space. With bold, the path of the process starting from $x=1$, $v=0$.}
    \label{fig:Contours.Hamiltonain}
\end{figure}

A specific construction with some fascinating properties involves taking $K \left( v \right) = \left\| v \right\|_1$ in the above construction, corresponding to momenta following the multivariate Laplace distribution. In this setting, it holds that $\nabla K \left( v \right) = \mathrm{sign}_. \left( v \right)$, so that (ignoring singularities at which the coordinates of $v$ change sign) one always has $\dot{x} \in \left\{ \pm 1 \right\}^d$, i.e. the velocity of the process lives on the discrete hypercube, and hence the $x$ component has a uniformly bounded (in fact, constant) speed, enabling stable discretisation in some generality.

This peculiarity of the velocities also hints to an intriguing connection with Piecewise-Deterministic Markov Processes, and in particular, to the Zig-Zag Process of \cite{bierkens2019zig}. Indeed, work of \cite{nishimura2025zigzag} explores this connection in greater depth, witnessing this `Hamiltonian Monte Carlo with Laplace-Distributed Momentum' as a `Non-Markovian Zig-Zag Process'. In this analogy, the times at which some momentum coordinate $v_i$ crosses $0$ correspond qualitatively to the `velocity coordinate flip' events of the Zig-Zag Process. While the system follows deterministic dynamics, randomness is induced via random refreshments of the momentum $v$, ensuring that appropriate stochasticity enters the system. 
Empirically, the authors observe that in the Hamiltonian formulation, the process is able to travel further in space for an equivalent number of `events', following the intuition that with these `more deterministic' dynamics, one can suppress unnecessary stochasticity and promote efficient spatial exploration.

\begin{figure}[ht]
    \centering
    \includegraphics[width=0.5\textwidth]{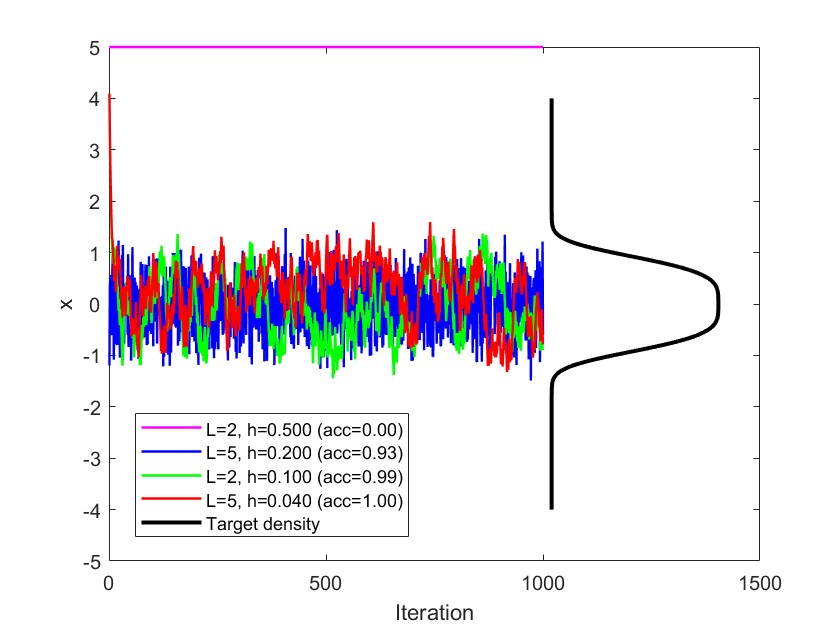}
    \caption{Hamiltonian Monte Carlo with kinetic energy $K(v) = | v |^\frac{4}{3}$. The target density ($\pi(x) \propto \exp \left( -x^4 \right)$) is overlaid on the right hand side of the graph. The legend indicates the various step-sizes ($h$) and leapfrog steps ($L$) along with the average acceptance probability. }\label{fig:Relativistic.HMC.various.stepsizes}
\end{figure}

We conclude with a numerical study of the light tailed one-dimensional log-quartic target of the form $\pi(x) \propto \exp \left( -x^4 \right)$, introduced in Example \ref{ex:steep.gradient}. We consider the Hamiltonian dynamics with kinetic energy of the form $K(v) = | v |^\frac{4}{3}$, in accordance with the suggestion of \cite{livingstone2019kinetic}. We use the Leapfrog algorithm to discretise the Hamiltonian dynamics, using step-size $h$ and $L$ leapfrog steps per move. This leads to discretisation errors which we correct via the usual Metropolis approach, ensuring that the chain targets the correct distribution. We ran the algorithm for $N=1000$ iterations and starting position $x_0=5$. Our results are summarised in Figure \ref{fig:Relativistic.HMC.various.stepsizes}, where we present the traceplots of four algorithms with various choices of step-size and leapfrog steps. On the legend, the various choices of $h$ and $L$ are described, along with the average acceptance probability of the proposed jumps. It is evident that, compared to MALA, or other algorithms previously mentioned, this HMC algorithm has the potential to reach the mode of the target and move around the space faster, an attribute of having momentum due to the Hamiltonian dynamics. On the other hand, there are now two tuning parameters, $h$ and $L$, instead of only $h$, and a wrong choice can lead to very inefficient algorithms (see the magenta plot).

\subsubsection{Piecewise-Deterministic Monte Carlo}\label{sec:Piecewise.Deterministic.Monte.Carlo}

For the preceding examples, the impact of working with `rough' potentials $U$ is felt most keenly through the difficulty of stably and accurately discretising the stiff dynamics of some associated stochastic process. Accordingly, the solutions focus on using more refined discretisation strategies, or seeking a modified stochastic process whose discretisation suffers less from these concerns of stiffness. A somewhat radical proposal is then to seek a new stochastic process for which discretisation is not only stable, but can even \emph{be avoided entirely}. Other things being equal, the appeal of such an approach is reasonably clear; on the other hand, the existence of such an approach is much less clear.

In this regard, Piecewise-Deterministic Markov Processes play a rather distinguished role. While the Langevin and Hamiltonian approaches to Monte Carlo simulation are built around the numerical approximation of a dynamical system which is rich with information about the target distribution, but challenging to construct exactly, Piecewise-Deterministic Monte Carlo methods like the Bouncy Particle Sampler (BPS) and Zig-Zag Process (ZZ) instead explore the target distribution using rather rudimentary dynamics which are largely uninformed by the geometry of the target distribution, and ensure their long-term stability by intermittently punctuating these continuous dynamics with discontinuous jumps in velocity space. The usual story is a simple and compelling one: pick a direction, run in that direction until it starts to look like a bad idea; once that happens, pick a new direction (`bounce') in a reasonable way, and carry on running in that new direction. The details of how to select directions well are more involved, but are by now reasonably well-understood.

Deferring discussion of the (admittedly involved) practical aspects of simulating PDMPs, we can begin to highlight some of their appealing properties, focusing on the interplay with non-smooth targets. Firstly, most PDMPs of practical interest move, by design, at a constant speed $s_\star$: no matter how strange an energy landscape one seeks to explore, there is no risk of spatial explosivity (though other pathologies concerning e.g. jumps require a more subtle discussion), and one even has the almost-sure bound $\left\| X_t - X_0 \right\| \leq s_\star \cdot t$. Secondly, while discretisation of steep drifts in ODEs and SDEs is often `direction-agnostic' in its effects -- a strongly-stabilising steep drift towards the center of the space can be as disastrous as a steep drift which causes the process to explode into the tails -- the feedback mechanism in PDMPs is of a different flavour. More specifically, when the process is moving in directions for which the potential $U$ is steeply decreasing, the process simply carries on in this favourable direction; by contrast, when the potential is steeply \emph{increasing}, one rapidly experiences a `bounce' event, and rectifies the situation. Non-smoothness and boundary effects are similarly muted in their impact on the movement of the process; if you start to move in a bad direction, then whatever the provenance of this badness, the response of the process is to propose a more fruitful direction, as quickly as possible.

\begin{figure}[ht]
    \centering

    \begin{subfigure}[t]{0.47\textwidth}
        \includegraphics[width=\linewidth]{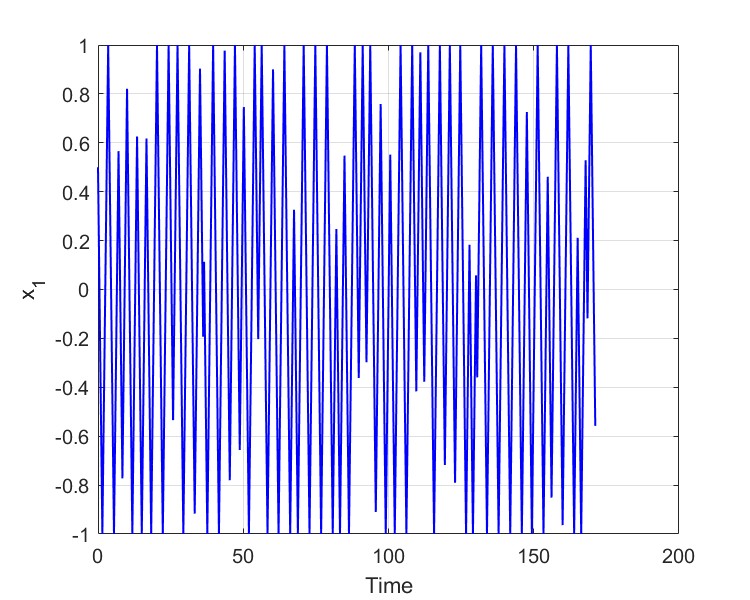}
        \caption{Traceplot of the first coordinate.}
        \label{fig:ZZ.Truncated.Gauss:part1}
    \end{subfigure}
    \hfill
    \begin{subfigure}[t]{0.47\textwidth}
    \includegraphics[width=\linewidth]{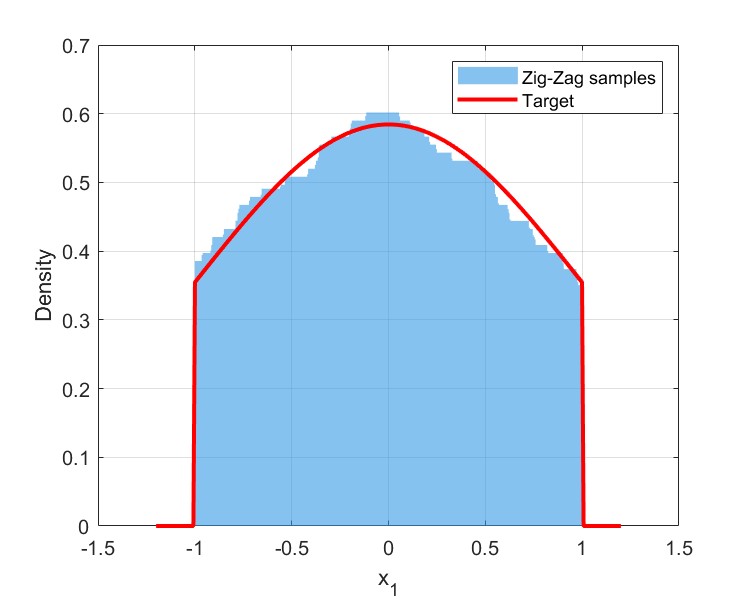}
        \caption{Histogram of the first coordinate}
        \label{fig:ZZ.Truncated.Gauss:part2}
    \end{subfigure}
    \vspace{1mm}
    \caption{Zig-Zag Sampler on the $10^4$-dimensional box-constrained Gaussian target (\ref{Truncated.Gaussian:1}), ran for $N=10^6$ direction switches.}
\label{fig:ZZ.Truncated.Gauss}
\end{figure}

In order to indicate the potential usefulness of these processes in sampling, we return to the setting of Example \ref{example:Truncated.Gauss} and the box-constrained Gaussian defined in (\ref{Truncated.Gaussian:1}). In Figure 
\ref{fig:ZZ.Truncated.Gauss}, we present the numerical results of simulating Zig-Zag process on this target, forcing the process to change direction whenever it hits the boundary of the box. For example, if the $i^\text{th}$ coordinate of the process moves in direction $v_i=+1$ and hits the boundary at $x_i=+1$, the velocity's $i^\text{th}$ coordinate will immediately switch to $v_i=-1$. We present the traceplot and the histogram of the first coordinate samples, having ran the process until $N=10^6$ changes of direction occurred, either by hitting the boundary, or due to a random direction switch. This makes the comparison fair to the MALA algorithm, as number of evaluations of the gradient of the log-density (which is one of the most computationally expensive parts of the algorithms) were roughly the same for both algorithms. Comparing Figure \ref{fig:ZZ.Truncated.Gauss} against Figure \ref{fig:MALA.Truncated.Gauss}, it is evident from the traceplot that the process explores the state space much faster while the histogram indicates that the process has created much more accurate samples from the target.

Due to their crucial historical role in the genesis of PDMC methods, we pause briefly to explicitly discuss the interaction between PDMPs and boundary effects, in the form of hard constraints. Some of the most impactful early developments in Piecewise-Deterministic Monte Carlo were made in the Statistical Physics literature under the name of `Event-Chain Monte Carlo' (ECMC), where they were used for (among other things) studying the dynamics of large ensembles of hard spheres; see \cite{bernard2009event, harland2017event, krauth2021event, maggs2022large, michel2014generalized} for some indicative references from this community. To be more precise, write $\mathbf{T}^d$ for the torus in dimension $d$ with associated metric $\mathsf{d}_\mathbf{T}$, pick some large `number of particles' $N \gg 1$, and consider some positive radius $R > 0$. One then studies the uniform distribution over collections of $N$ particles $\left( x_1, x_2, \cdots, x_N \right) \in \left( \mathbf{T}^d \right)^n$, subject to the constraint that for all $1 \leq i < j \leq N$, there holds the `exclusion principle' $\mathsf{d}_\mathbf{T} \left( x_i, x_j \right) \geq 2 \cdot R$. That is, viewing each $x_i \in \mathbf{T}^d$ as the centroid of a sphere of radius $R$, one imposes that any two distinct spheres cannot overlap. Such systems are of physical interest as a foundational model for phase transitions between the liquid and solid states of matter.

In this context, the `potential' for the system is essentially non-existent, and so the challenge of sampling is governed entirely by the non-convex hard-sphere constraint. The core ECMC strategy thus proceeds by selecting a single particle, assigning it a velocity, propagating it in the direction of that velocity until it collides with another stationary particle, and then `bounces' that particle into motion, transferring its velocity away. Interacting with the boundary no longer raises questions of how to tune an appropriate step-size to safely remain inside the feasible set, and becomes entirely a question of computing when the next collision occurs. Even without expanding upon the improved mixing behaviour conferred by the use of these piecewise-deterministic dynamics, this is already of substantial interest in view of the more pedestrian concern of how Markov process-based algorithms might interact more simply with hard constraints. In any case, the qualitative story here remains valid for other challenges which have been discussed here, including steep gradients, non-smooth targets, other boundary effects (including `stickiness'), and more. For some illustrative examples in statistical contexts, see e.g. \cite{goldman2022gradient, hardcastle2025diffusion, chevallier2024pdmp}.

Nevertheless, one must make some comments on numerical aspects of PDMPs. While discretisation \emph{per se} might be avoidable, there are various challenges associated with the simulation of PDMPs, chief among them being the simulation of the `bounce' events. For rather structured models, there are good strategies available for exact simulation by either harnessing the `shape' of the event rate (e.g. in the case of convexity assumptions; see Example 1 in \cite{bouchard2018bouncy}, or \cite{sutton2023concave}) or by effective use of bounds on the event rate within thinning procedures \cite{lewis1978simulation}. Various other numerical strategies have been proposed for making this task feasible for wider classes of model; these are arguably most successful at making `impossible' models `possible', rather than at making `possible' models faster; see e.g. \cite{fearnhead2024stochastic, pagani2024nuzz, corbella2022automatic, andral2024automated} for examples to this effect.

Moving beyond the difficulty of simulating these events, there is also the compounding concern of event frequency. While theoretical results reveal that PDMPs can converge to their invariant measure quite quickly when viewed as continuous-time processes (see e.g. \cite{durmus2020geometric, deligiannidis2021randomized, lu2020complexity, lu2022explicit}), the practical complexity of PDMPs is really measured by the `clock' of how many events have occurred (or in some cases, how many events have been proposed), which changes the picture somewhere. Some heuristics based on product-form model problems suggest that for sampling problems on $\mathbf{R}^d$, one might expect to witness $\Theta \left( d^\frac{1}{2} \right)$ events per unit time at stationarity, even in the best (reasonable) case. Along similar lines, for potentials which vary irregularly, one might expect to witness events at higher frequencies; see \cite{bierkens2025scaling} for a study of this phenomenon on ill-conditioned Gaussian target distributions. In any case, there is no free lunch, and the principle of `conservation of difficulty' is borne out to some extent: even for rough targets, PDMPs can admit stable dynamics, but one will eventually have to work hard to simulate those stable dynamics.

\subsection{Open Problems}

We here highlight some problems which are recurrent in this sub-literature, and whose resolution would be of some major practical and theoretical interest.

\subsubsection*{Robustness to Roughness and Hyperparameter Adaptation}

A number of works \cite{atchade2006adaptive, livingstone2022barker} make the empirical and intuitive comment that when using MCMC algorithms based on more `robust' proposals in this way, adaptation of algorithmic hyperparameters (e.g. step-size, covariance matrix, etc.) is more stable and efficient. It is very appealing to attach some theory to this observation, which in practice, turns out to provide a very strong argument in favour of such methods.

\subsubsection*{Robustness to Roughness without Overconservatism}

Some of the lightweight approaches to robustification effectively amount to replacing a superlinear drift with a bounded drift. We know that processes of bounded drift will typically not be any better than exponentially ergodic (e.g. it becomes difficult to satisfy a Logarithmic Sobolev Inequality). Can we constrain our drift more gently so that we retain numerical stability, but without hindering our ability to return to the bulk of the state space exponentially quickly? Proximal MCMC offers a partial solution here, but makes certain implicit assumptions about the feasibility of solving certain optimisation problems efficiently. A more generic and lightweight solution would be of substantial interest. We note some recent answers in this direction \citep{johnston2024strongly, lytras2025taming} which have been obtained by enriching the usual `taming' framework.

\subsubsection*{Robustness to Roughness by Local Adaptation}

Our solutions here have effectively focused on resolving instabilities with rather global changes to the dynamics. It would be of substantial interest to devise robustification strategies with a more localised character, i.e. adaptively detecting challenges in the local geometry, and modifying the dynamics only when needed. This characterisation is rather too vague to be useful at present, but it stands to reason that such a strategy could be made both precise and practical.

\subsubsection*{Robustness to Roughness by Discretisation}

Several proposed strategies have the flavour of discretising the spatial domain, and hence reduce the overall sampling problem to a sequence of discrete-space Markov chains; this is the case for the Barker-Langevin approach of \cite{livingstone2022barker} and some of its relatives (e.g. \cite{bou2018continuous, duffield2025lattice}). This simple strategy can be quite successful, even when done in an ad-hoc way. What are principles for performing such a discretisation `optimally'?

\subsubsection*{Model Problems for Roughness}

Finally, our classification of `rough' problems is rather binary in nature. It would be useful to develop a richer spectrum of model problems which reflect the variety of sampling challenges posed in modern applications, and to develop a more refined language for delineating between their distinct pathologies. This should then provide a clearer path towards resolving said challenges.

\section{Heavy-Tailedness}\label{sec:heavy:00}

\subsection{Formulation of the Pathology and Examples}\label{sec:heavy.pathology}

In this section we will focus on a class of targets that assign a significant amount of mass at the tails of the state space. We will be working with the following definition.

\begin{definition}
    We will call a density $\pi \in C^1 \left( \mathbf{R}^d \right)$ {\it heavy-tailed} if 
    \begin{equation*}
        \lim_{r \to \infty } \sup \left\{ \left\| \nabla \log \pi(x) \right\| : \left\| x \right\| \geq r \right\} =0.
    \end{equation*}
\end{definition}

Note that this is more restrictive than various other standard definitions of heavy-tailedness which focus only on the probability mass which is contained in the tails of the distribution, e.g. that $\int \pi \left( \mathrm{d} x \right) \exp \left( s \cdot \left\| x \right\| \right) = \infty$ for all $s > 0$. We choose to work with this formulation because of the central practical role of this gradient in many popular `local' MCMC algorithms. If one sought to further highlight this distinction, then one might rename our definition as `flat-tailed'.

Morally, one can think of this class as the densities with tails heavier than any exponential distribution. Observe, for example, that a symmetrised one-dimensional exponential distribution on $\mathbf{R}$ is of the form $\pi(x) = \frac{a}{2} \cdot \exp \left( -a \left| x \right| \right)$ for some $a > 0$, whereby it holds that $\left| \mathrm{D} \log \pi(x) \right| = a$, i.e. the gradient of the potential acts with a force of constant magnitude.


Distributions with heavy tails arise naturally in settings when one tries to model extreme or rare events (for example in finance \cite{embrechts.kluppelberg.mikosch:97}, see also \cite{limpert.eckhard.werner.markus:01} for more general applications). They also often arise in practical Bayesian statistics, for example in models that aim to account for the behaviour of outliers in a more robust manner (e.g. \cite{lange.little.taylor:89}). The resulting posteriors typically exhibit heavy tails and one needs to create samples from that posterior in order to perform inference. The same phenomenon can occur when one uses a shrinkage prior (e.g. the horseshoe prior \cite{yao.wei.yu:14}); this can be natural in settings wherein one is trying to estimate many parameters, of which only a small subset are expected to be significantly non-zero. 

From a sampling point of view, MCMC techniques tend to struggle with such targets, as the dynamics of the chain fail to systematically explore the tails of the distribution, wherein much of the probability mass resides. Furthermore, in contrast to the targets considered in the previous section, for targets with heavy tails, even ideal continuous processes (such as the Langevin dynamics) tend to be quite slow at exploring. For example, let us recall the example of Overdamped Langevin diffusion (\ref{overdamped.langevin:00}), given by
\begin{align*}
    \mathrm{d} X_t =\nabla \log \pi \left( X_t \right) \mathrm{d} t + \sqrt{2} \, \mathrm{d} W_t.
\end{align*}
Since $\pi$ is heavy-tailed, when $X_t$ is exploring the (appreciable) probability mass contained in the tails, it holds that $\nabla \log \pi(X_t) \approx 0$, and so the dynamics of $X_t$ are largely governed by the Brownian motion term, with little impact from the shape of $\pi$ itself. This results into backtracking dynamics and prohibits the process from returning to areas of higher density sufficiently fast, leading to inefficient space exploration. 


Similar random-walk behaviour on heavy-tailed targets is observed for other continuous-time processes discussed earlier in this work, such as the Bouncy Particle Sampler or the Zig-Zag process (see e.g. the discussion in \cite{vasdekis.roberts:23, vasdekis2022note}). The problem there is further emphasised by the fact that by construction, these PDMPs move around the space with constant speed, which precludes the possibility of transitioning between the tails and the centre of mass at a sufficiently fast rate to e.g. ensure exponential ergodicity. 

As a result of this inefficient behaviour, MCMC proposals that arise as numerical discretisations of the continuous-time processes explored earlier cannot hope to have reliable performance on heavy-tailed targets, as the underlying process which they seek to approximate fails to exhibit such a performance. We will showcase this with the following two examples.

\begin{example}[Cauchy distribution]\label{cauchy.example:00}
    Our first example in the heavy-tailed category is the Cauchy distribution, i.e. when the target distribution is of the form
    \begin{equation}\label{Cauchy.definition:00}
        \pi(x) \propto (1+x^2)^{-1}.
    \end{equation}
    Observe that under $\pi$, no moment of order $\geq 1$ exists, reflecting that $\pi$ places a great deal of mass around large values of $|x|$. Observe also that $\left| \mathrm{D} \log \pi \left(x\right) \right| \leq \min \left\{ 1, 2 \cdot \left| x \right|^{-1} \right\}$, i.e. for large $|x|$, the landscape of $\pi$ is quite flat.
\end{example}

\begin{figure}[ht]
    \centering
    \begin{subfigure}[t]{0.47\textwidth}
        \includegraphics[width=\linewidth]{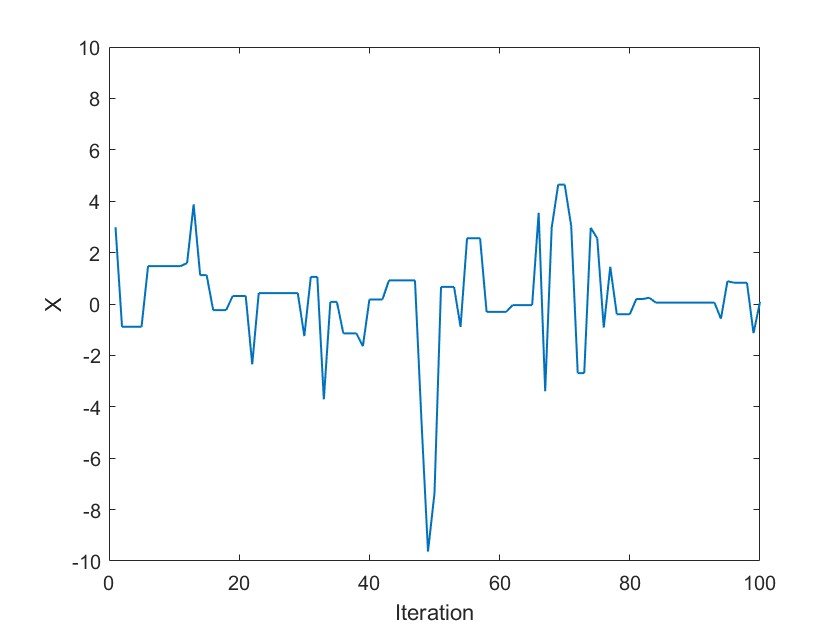}
        \caption{Traceplot (first 100 iterations)}
        \label{Fig:Cauchy.MALA:part1}
    \end{subfigure}
    \hfill
    \begin{subfigure}[t]{0.47\textwidth}
        \includegraphics[width=\linewidth]{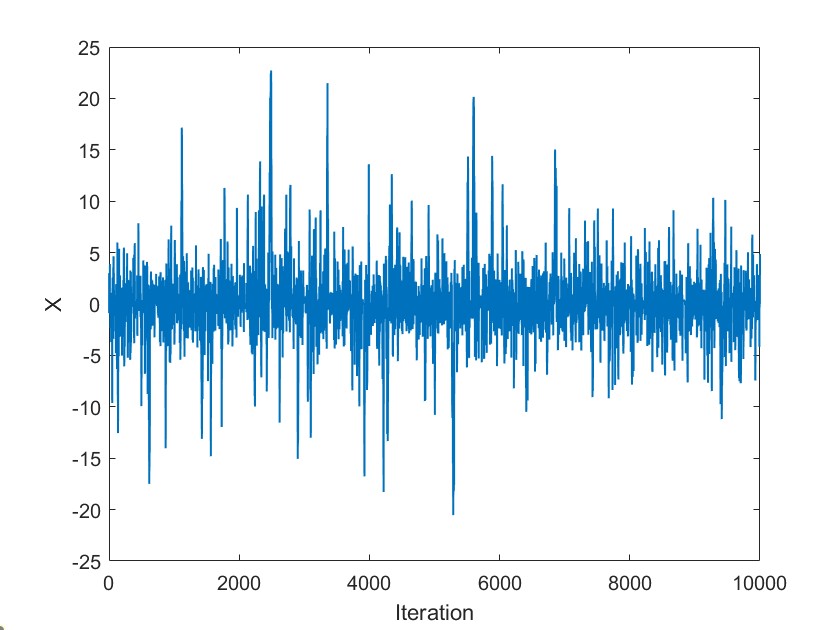}
        \caption{Traceplot}
        \label{Fig:Cauchy.MALA:part2}
    \end{subfigure}
    
    \vspace{0.5cm}  

  
    \begin{subfigure}[t]{0.47\textwidth}
    \includegraphics[width=\linewidth]{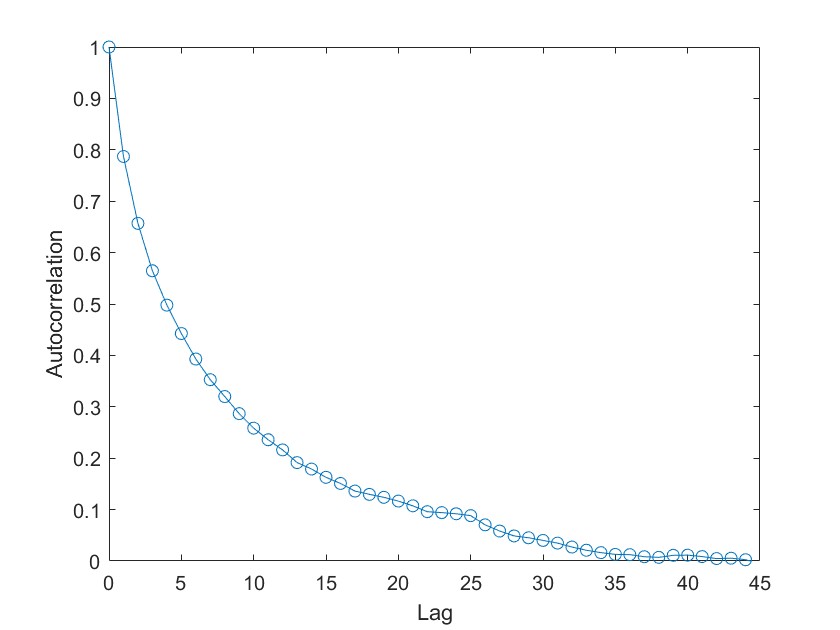}
        \caption{Autocorrelations}
        \label{Fig:Cauchy.MALA:part3}
    \end{subfigure}
     \hfill
    \begin{subfigure}[t]{0.47\textwidth}
    \includegraphics[width=\linewidth]{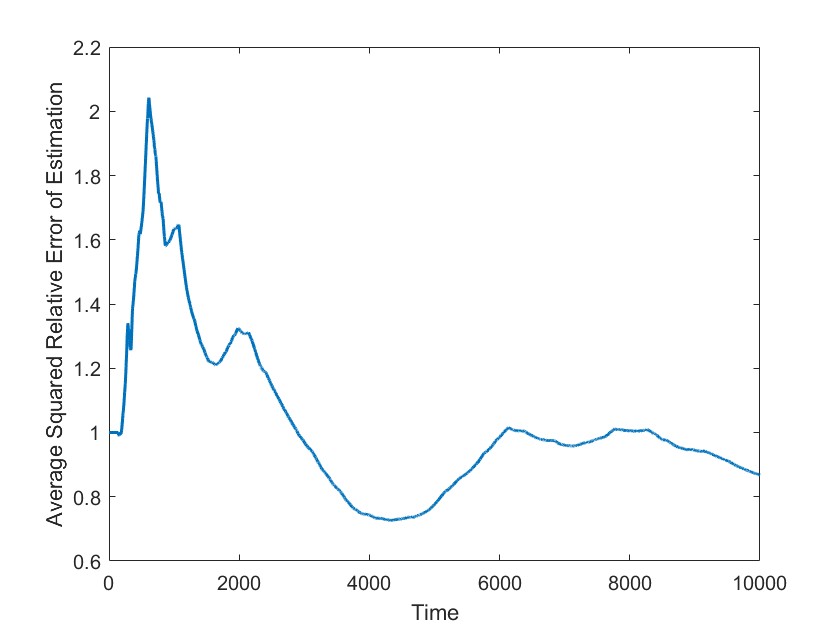}
        \caption{Squared relative error estimating the probability of a tail event (actual probability : 0.0627)}\label{Fig:Cauchy.MALA:part4}
    \end{subfigure}
  \vspace{1mm}
    \caption{MALA algorithm on a one-dimensional Cauchy target.}
    \label{Fig:Cauchy.MALA}
\end{figure}

To showcase the sampling problem in the setting of Example \ref{cauchy.example:00}, in Figure \ref{Fig:Cauchy.MALA} we present the results of a numerical simulation from the one-dimensional Cauchy distribution using the MALA algorithm. The first two plots show the traceplot. It can be seen that there are long and infrequent excursions at the tails, which typically indicate unstable algorithmic perfomance. We also present the autocorrelation plot, which shows how rapidly correlation between samples decays as a function of the time-distance between samples. Ideally, the correlation should decay fast, which would indicate a fast generation of independent samples, without the need to run the algorithm for many itarations. In this example, autocorrelations appear to decay quite fast, but we will later see variants of the algorithm, designed to target heavy tailed targets that will improve the autocorrelation decay for the same target. Finally, the last plot demonstrates the mean squared relative error in estimating the probability of the event $\{ X \geq 5 \}$, with $X$ following a Cauchy distribution. The plot reports the average squared relative errors over 100 independent realisations of the chain, as a function of algorithmic iterations. It is clear that the squared relative errors decay slowly as the algorithm evolves in time, barely reaching a value less than $0.8$ in the first $10000$ iterations. This is indicative of the fact that the algorithm is not well-suited to explore the tails of the distribution. We emphasise here that the step-size of the algorithm (in this case $h=10$) was carefully chosen to maximise the algorithmic performance in terms of Effective Sample Size (see e.g. \cite{brooks.gelman.joens.meng:11}), and the algorithmic behaviour can be considerably worse with under-optimised step-sizes.

We now present a somewhat more realistic statistical model which exhibits heavy tails both in the parameters and in the observations.

\begin{example}[Cauchy Regression with Horseshoe Prior]\label{ex:Cauchy.Regression}
    In the context of the linear model, one approach to robustly capture the behaviour of outliers (e.g. \cite{lange.little.taylor:89}) is to model the observational errors as following the Cauchy distribution. In this model, given parameters $\beta=(\beta_1, \beta_2, \dots, \beta_{p})^{\top}$, and design matrix $\mathbf{X} \in \mathbf{R}^{n \times p}$, the response vector $\mathbf{Y} \in \mathbf{R}^n$ is given by
\begin{equation}\label{regression.model:001}
        \mathbf{Y} = \mathbf{X} \beta + \epsilon,
    \end{equation}
    with the errors $\epsilon = \left( \epsilon_1, \dots, \epsilon_n \right)^\top$ following i.i.d. Cauchy distributions as in (\ref{Cauchy.definition:00}). In order to allow additional flexibility to the model, a widely used prior for the parameters $\beta$ is the horseshoe prior, which is defined hierarchically by setting
\begin{equation}\label{horseshoe}
        \left\{ \beta_j \right\} \mid \left\{ \lambda_j \right\} \overset{\mathrm{ind}}{\sim} \mathcal{N} \left( 0, \lambda_j \right), \qquad \lambda_j \overset{\mathrm{iid}}{\sim} \mathsf{Cauchy}^+.
    \end{equation}
    where $\mathsf{Cauchy}^+$ is the standard Cauchy random variable, conditioned on being positive. This has the effect of allowing some of the regression coefficients to take values close to zero, thus potentially reducing the effective dimensionality of the parameter space, with other coefficients free to take much larger values (due to the relatively heavy marginal tails). In this case, the joint posterior over $\left( \beta, \lambda \right)$ takes the form
    \begin{equation}\label{regression.posterior:001}
        \pi \left( \beta, \lambda | \mathbf{X, Y} \right) \propto  \left( \prod_{j=1}^{p}   \frac{1}{ 1+ \lambda_j^2} \right) \cdot \left( \prod_{j=1}^{p} \frac{1}{\sqrt{2\lambda_j}} \exp \left( - \frac{1}{2\lambda_j} \beta_j^2 \right) \right) \cdot \left( \prod_{i=1}^n \frac{1}{1 + \left(Y_i-\sum_{j=1}^{p}\beta_jX_{i,j}\right)^2} \right),
    \end{equation}
    which presents heavy tails for the parameters $\lambda$ and for the marginal of $\beta$ (see e.g. \cite{carvalho2009handling}).
    \end{example}
    
\begin{figure}[ht]
    \centering
    \begin{subfigure}[t]{0.47\textwidth}
        \includegraphics[width=\linewidth]{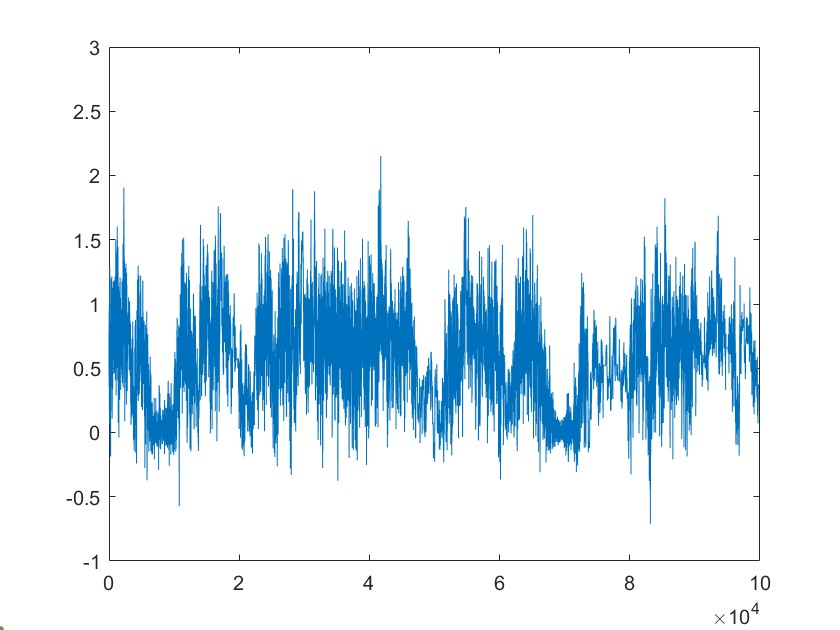}
        \caption{Traceplot of the first coordinate}
        \label{Fig:Cauchy.Regression.MALA.part1}
    \end{subfigure}
    \hfill
    \begin{subfigure}[t]{0.47\textwidth}
        \includegraphics[width=\linewidth]{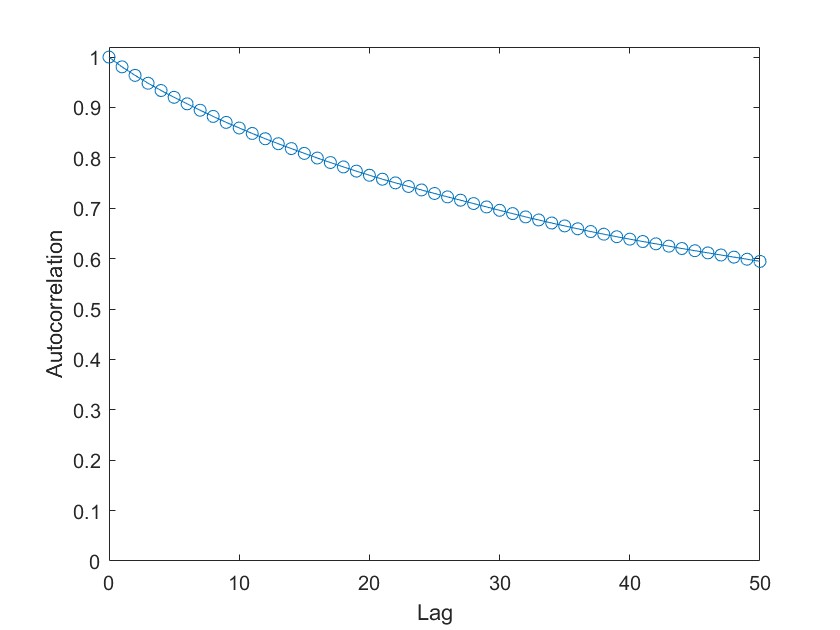}
        \caption{Autocorrelations of the first coordinate}
        \label{Fig:Cauchy.Regression.MALA.part2}
    \end{subfigure}
    \caption{MALA algorithm on a 5-dimensional Cauchy-Regression target.}\label{Fig:Cauchy.Regression.MALA}
\end{figure}

    The slow behaviour of MALA in Example \ref{ex:Cauchy.Regression} is showcased in Figure \ref{Fig:Cauchy.Regression.MALA}, where we summarise the findings of a numerical simulation with $p = 5$ parameters and $n = 20$ observations. The true values of $\beta_j$ were drawn from a horseshoe distribution (\ref{horseshoe}), the covariates $X_{i,j}$'s were drawn from the uniform distribution in $\{ 0, 1, 2, 3\}$, and conditionally upon these values, response data $Y_i$ was drawn from the model (\ref{regression.model:001}). We used the MALA algorithm to draw samples from the marginal posterior (\ref{regression.posterior:001}). Figure \ref{Fig:Cauchy.Regression.MALA.part1} shows the traceplot of the first coordinate $\beta_1$, while Figure \ref{Fig:Cauchy.Regression.MALA.part2} shows the autocorrelation plot of $\beta_1$. Evidently, the autocorrelations decay as a very slow rate, indicating large dependence between the generated samples, and poor mixing.

In the next section, we will review contemporary approaches that aim to produce more robust and efficient algorithms for this setting. We note that by contrast with the case of rough targets, the solutions are necessarily somewhat less generic. This can be understood as a symptom of the general phenomenon that in order to alleviate the effects of heavy tails, one often requires an understanding of the degree of heavy-tailedness in the problem. With this being said, the overall strategies are reasonably general in spirit, but do require a greater degree of instance-specific tuning than is typical of the light-tailed world.


\subsection{Proposed Solutions}

We will discuss two approaches that have been proposed in order to improve the performance of MCMC algorithms on heavy tails. These involve transforming either the state space or the time scale in which the process runs. In the first approach, one transforms the state space via a function $f$ which, if carefully chosen, induces a push-forward density $\tilde{\pi}(x)$ with significantly lighter tails than $\pi$. The idea is to then run a more efficient MCMC algorithm to target the light-tailed $\tilde{\pi}$, and then push the generated samples back through the function $f^{-1}$ in order to recover approximate samples from $\pi$. The second approach also involves changing the target to a density $\tilde{\pi}$ with lighter tails, but instead of filtering the samples through a space transformation, one changes the time-scale in which the process moves, effectively discarding some of the samples. This biases the $\tilde{\pi}$-generated samples in an appropriate fashion so that the remaining samples are distributed according to $\pi$.

\subsubsection{Space Transformations}\label{sec:space.transforms}

In the context of heavy-tailed sampling, the core idea behind space-transformation strategies is to identify an appropriate invertible function $f: \mathbf{R}^d \rightarrow X$, $X \subset \mathbf{R}^d$ so that the push-forward measure $\tilde{\pi}$ on $X$, defined by $\tilde{\pi}(A) = \pi(f^{-1}(A))$ has lighter tails than $\pi$. One then takes advantage of MCMC algorithms that are more efficient when targeting $\tilde{\pi}$, in order to construct samples $X_1,\dots, X_n$ approximately distributed according to $\tilde{\pi}$. Pulling these samples back to the original space by application of $f^{-1}$ then yields samples  $f^{-1} (X_1), \dots , f^{-1} (X_n)$ that are approximately distributed according to $\pi$.

\begin{example}[Cauchy target]\label{ex:Cauchy}
To showcase how this approach can be useful, let us consider the one-dimensional Cauchy distribution from Example \ref{cauchy.example:00}, along with the space transformation 
\begin{equation}\label{sign.log.space.transform:1}
    f(x) = \mathrm{sign} (x) \cdot \log\left( 1 + \left| x \right| \right).
\end{equation} 
Consider the push-forward measure $\tilde{\pi}(A)=\pi(f^{-1}(A))$, where $\pi$ is the Cauchy distribution. For the density of $\tilde{\pi}$, the change-of-variables formula yields that
\begin{align*}
    \tilde{\pi} \left( y \right) = \pi \left( f^{-1} \left( y \right) \right) \cdot \left| \mathrm{D} f^{-1} \left( y \right) \right| = \frac{\pi \left( f^{-1} \left( y \right) \right)}{\left| \mathrm{D} f \circ f^{-1} \left( y \right) \right|}.
\end{align*}
We calculate that $f^{-1} \left( y \right) = \mathrm{sign} \left( x \right) \cdot \left( \exp \left| y \right| - 1 \right)$, and $\mathrm{D}f (x)= \left( 1 + \left| x  \right| \right)^{-1}$, so that one obtains
\begin{align*}
    \pi \left( f^{-1} \left( y \right) \right) \propto \frac{1}{1+\left( f^{-1} \left( y \right) \right)^2}, \qquad  \mathrm{D} f \circ f^{-1} \left( y \right) = \exp \left| y \right|,
\end{align*}
and therefore
\begin{align*}
    \tilde{\pi} \left( y \right) &\propto \frac{\exp \left| y \right|}{1+\left( \exp \left| y \right| -1 \right)^2} = \left( 2 \cosh y - 1 \right)^{-1} = \exp \left( -\left| y \right| + O\left( 1 \right) \right)
\end{align*}
with lighter tails, similar to those of Laplace distribution. One can then use an MCMC algorithm to target the distribution $\tilde{\pi}$, in order to efficiently generate samples $Y_1, \dots, Y_n$ that are approximately distributed according to $\tilde{\pi}$, and then interpret the samples $X_i := f^{-1}(Y_i) = \mathrm{sign} \left( Y_i \right) \cdot \left( \exp \left| Y_i \right| - 1 \right)$, $i = 1,\dots , n$ as approximate samples from $\pi$.
%
%
\end{example}

There have been various suggestions on which space transformation to use, and this may depend on the MCMC algorithm one uses to generate samples from the transformed target $\tilde{\pi}$. For target distributions whose tails decay at a polynomial rate, one can see that transformations of logarithmic growth (such as the one considered in (\ref{sign.log.space.transform:1})) can be effective choices, as they push the target forward onto a distribution which is asymptotically approximately log-concave, without impacting the smoothness properties of the implied potential too aggressively. 

Multivariate examples of this form of transformations were explored in \cite{johnson2012variable}, were the authors suggested the use of isotropic transforms, i.e. functions of the form
\begin{equation}\label{isotropic.tranform:1}
    f \left( x \right) = h \left( \left\| x \right\| \right) \cdot \frac{x}{\left\| x \right\|} , \ \ x \in \mathbf{R}^d
\end{equation}
for some appropriate increasing $h \in C^1 \left( \mathbf{R}_+ \right)$.

\begin{figure}[ht]
    \centering
    \begin{subfigure}[t]{0.47\textwidth}
        \includegraphics[width=\linewidth]{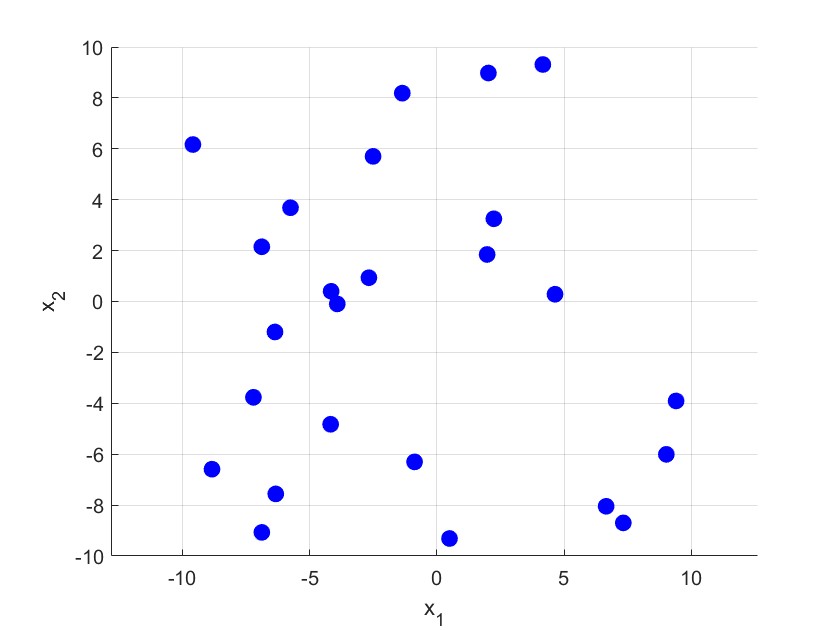}
        \caption{Uniformly distributed points}
        \label{Fig:Uniform.Transformed.Square:part1}
    \end{subfigure}
    \hfill
    \begin{subfigure}[t]{0.47\textwidth}
        \includegraphics[width=\linewidth]{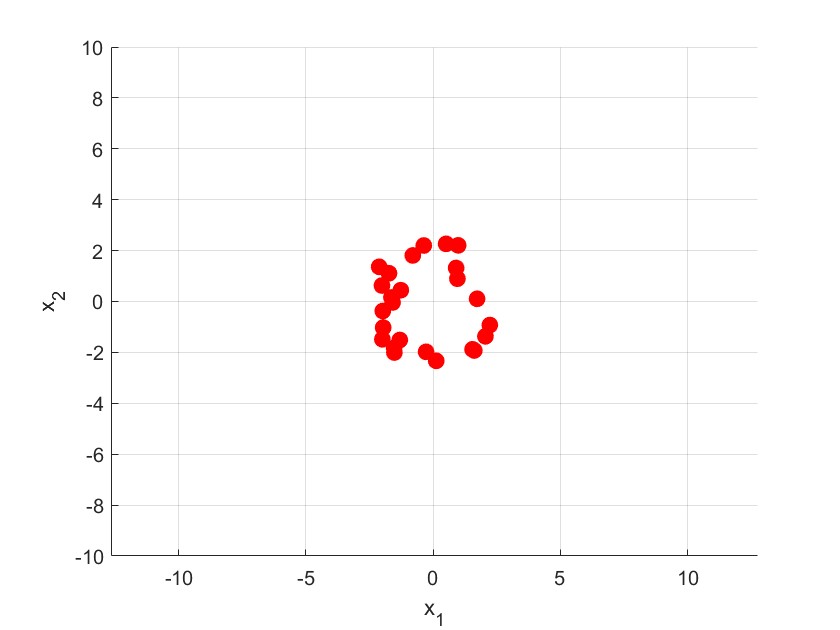}
        \caption{Transformed points}
        \label{Fig:Uniform.Transformed.Square:part2}
    \end{subfigure}
    \caption{Uniformly distributed points on the square $[-10 , 10]^2$ and the transformed points after applying the transformation $f(x)= \log \left( 1+ \| x \| \right) \cdot x / \| x \|$.}
\label{Fig:Uniform.Transformed.Square}
\end{figure}

In order to visualise the impact of such a space transformation, on the left plot of Figure \ref{Fig:Uniform.Transformed.Square} we simulated 25 points uniformly at random in the box $[-10,10]^2$. We then considered the function $f$ as in (\ref{isotropic.tranform:1}), with $h(r) = \log \left( 1+ r \right)$, essentially generalising (\ref{sign.log.space.transform:1}) to two dimensions. We apply the function $f$ to the uniformly simulated points, and the resulting points are shown in the right plot of Figure \ref{Fig:Uniform.Transformed.Square}. It is evident that the transformed points are much more concentrated towards the origin, indicating that their distribution has lighter tails.

Parameterising $f$ following Equation \ref{isotropic.tranform:1} offers some simplicity, in the sense that it reduces the task of searching for an effective high-dimensional transport map into a problem which is essentially univariate in nature. In concordance with the discussion above, the authors of \cite{johnson2012variable} suggest that for target distributions with polynomially-decaying tails, one ought to take $h$ which grows logarithmically for large values of $\left\| x \right\|$. They then show that when using the Random Walk Metropolis algorithm (see e.g. \cite{sherlock2010random}) on the transformed target, one can obtain much-improved rates of convergence, compared to applying the same method to the original target. In particular, the authors prove that the transformed algorithm is exponentially ergodic in scenarios where the base method is not.

Similarly, \cite{he2023analysis} introduced the Transformed Unadjusted Langevin Algorithm. The space transformation is also isotropic, i.e. of the form of (\ref{isotropic.tranform:1}), and the function $h$ is taken to have polylogarithmic growth for large arguments. The authors then make use of ULA (\ref{ULA:00}) to explore the transformed target, and demonstrate theoretically that by a careful design of the transform map, one can gain improved rates of convergence. 

\begin{figure}[ht]
    \centering
    \begin{subfigure}[t]{0.47\textwidth}
        \includegraphics[width=\linewidth]{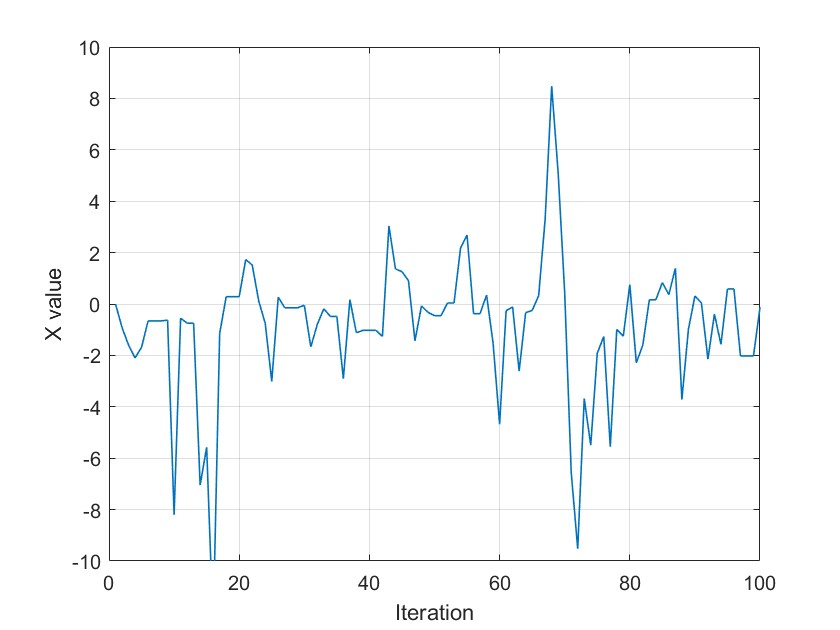}
        \caption{Traceplot (first 100 iterations)}
        \label{Fig:Transformed.Cauchy.MALA:part1}
    \end{subfigure}
    \hfill
    \begin{subfigure}[t]{0.47\textwidth}
        \includegraphics[width=\linewidth]{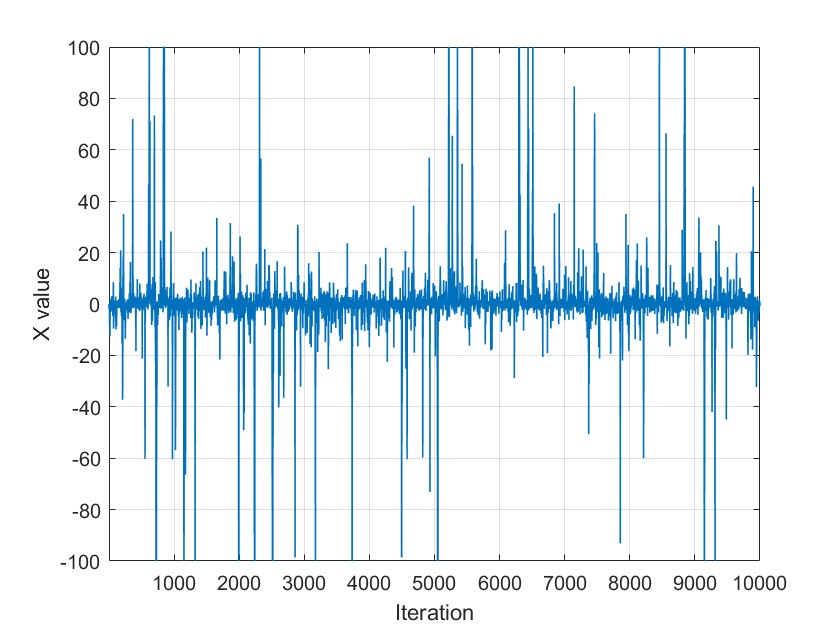}
        \caption{Traceplot}
        \label{Fig:Transformed.Cauchy.MALA:part2}
    \end{subfigure}
    
    \vspace{0.5cm}  

  
    \begin{subfigure}[t]{0.47\textwidth}
    \includegraphics[width=\linewidth]{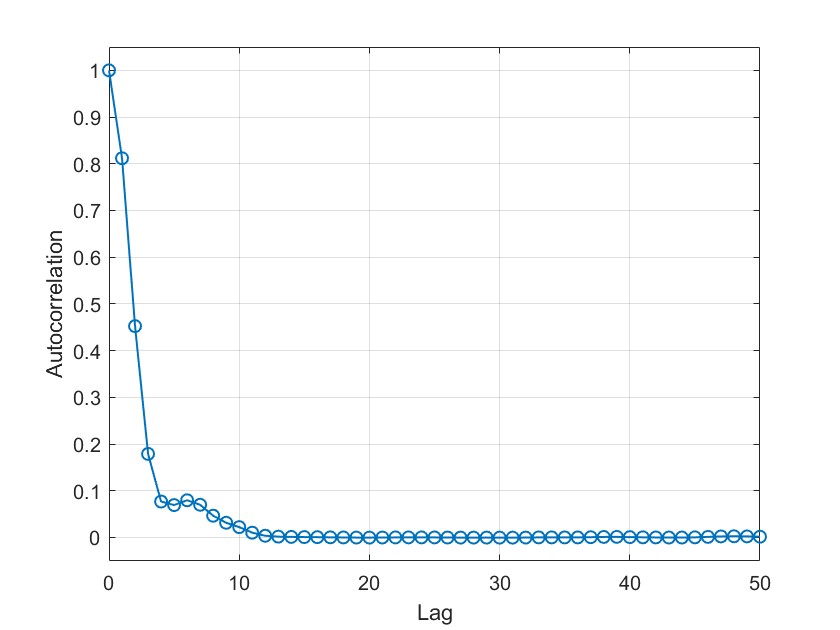}
        \caption{Autocorrelations}
        \label{Fig:Transformed.Cauchy.MALA:part3}
    \end{subfigure}
     \hfill
    \begin{subfigure}[t]{0.47\textwidth}
    \includegraphics[width=\linewidth]{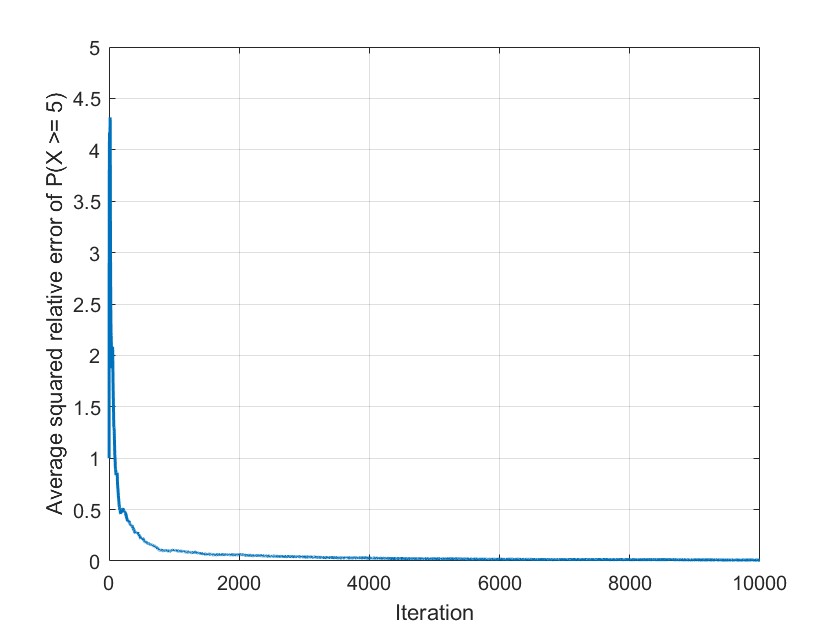}
        \caption{Squared relative error estimating the probability of a tail event (actual probability : 0.0627)}
\label{Fig:Transformed.Cauchy.MALA:part4}
    \end{subfigure}
  \vspace{1mm}
    \caption{Transformed MALA algorithm on a one-dimensional Cauchy target.}
    \label{Fig:Transformed.Cauchy.MALA}
\end{figure}

Similarly, by introducing a Metropolis-adjustment, one can run an instance of MALA on the transformed target, ensuring that the samples are asymptotically consistent, before transferring the samples back to the original space via the inverse transformation. We use this approach in our next simulation, which uses a space transformation with $f$ as in (\ref{sign.log.space.transform:1}) to target the Cauchy distribution from Example \ref{ex:Cauchy}. As explained in that Example, the resulting push-forward measure $\tilde{\pi}$ will have lighter tails, therefore, it will be easier for MALA to explore the area where most of the target mass concentrates. We present our results in Figure \ref{Fig:Transformed.Cauchy.MALA}. This can be compared against Figure \ref{Fig:Cauchy.MALA}, where we ran MALA directly on the heavy tailed target. The first two plots show the traceplot, and in contrast to Figure \ref{Fig:Cauchy.MALA}, the algorithm seems to have much more frequent and short tail excursions, indicating that the process explores the state space sufficiently fast. This is further emphasised by the autocorrelation plot, which shows a much faster decay of autocorrelations, compared to the decay observed in Figure \ref{Fig:Cauchy.MALA}. Finally, the last plot demonstrates the mean squared relative error in estimating the probability of the tail event $\{ X \geq 5 \}$, with $X$ following a Cauchy distribution. We use the same approach as the one discussed when running a MALA algorithm directly on the Cauchy target, reporting the average squared relative errors over 100 independent realisations of the chain. Comparing with Figure \ref{Fig:Cauchy.MALA}, it is evident the errors decay much faster, having reached a value less than $0.1$ before $1000$ iterations, and staying consistently small thereafter. As a note, we comment that the step-size of the algorithm did not influence the algorithmic performance as extensively as in the previous case, where MALA was targeting the Cauchy target directly. Here we have picked the step-size to be $h=0.5$.

\begin{figure}[ht]
    \centering
    \begin{subfigure}[t]{0.47\textwidth}
        \includegraphics[width=\linewidth]{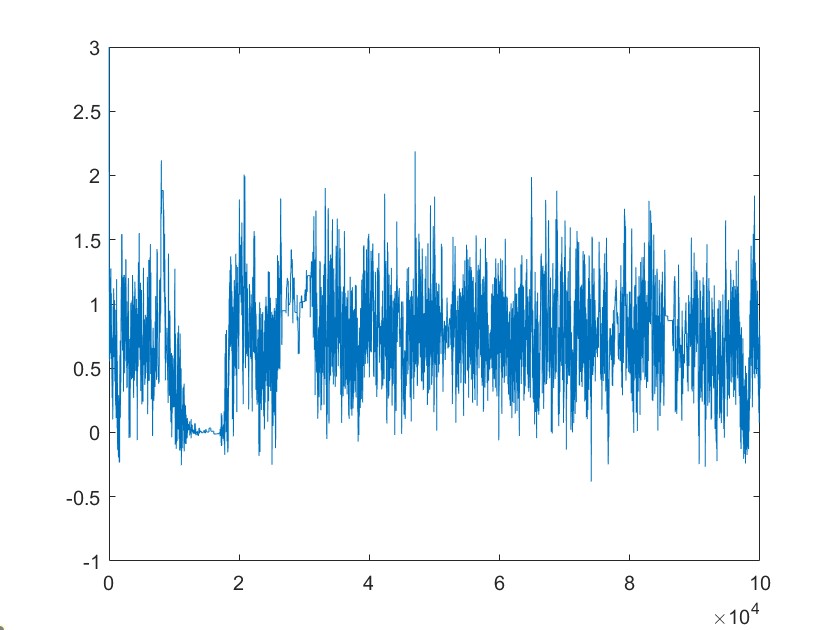}
        \caption{Traceplot of the first coordinate}
        \label{Fig:Cauchy.Regression.Transformed.MALA.part1}
    \end{subfigure}
    \hfill
    \begin{subfigure}[t]{0.47\textwidth}
        \includegraphics[width=\linewidth]{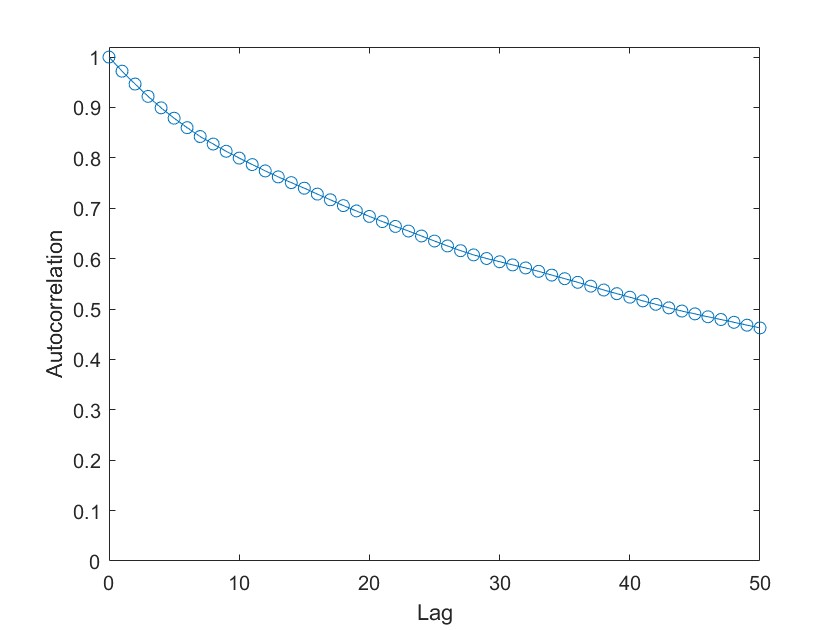}
        \caption{Autocorrelations of the first coordinate}
\label{Fig:Cauchy.Regression.Transformed.MALA.part2}
    \end{subfigure}
    \caption{Transformed MALA algorithm on a 5-dimensional Cauchy-Regression target.}\label{Fig:Cauchy.Regression.Transformed.MALA}
\end{figure}

We also consider the Transformed MALA on the Cauchy regression model, with Horseshoe prior, as in Example \ref{ex:Cauchy.Regression}. As in Section \ref{sec:heavy.pathology}, we take $p=5$ and $n=20$ observations. We again use a spherically symmetric transformation $f$ as in (\ref{isotropic.tranform:1}), with $h(r) = \log(1+r)$, as in Figure \ref{Fig:Uniform.Transformed.Square}. We ran MALA on the transformed target and obtain our primal samples by applying the inverse transformation. The step-size was chosen so that the average acceptance rate was between $50-60 \%$. As in Example \ref{ex:Cauchy.Regression}, we focus on the first coordinate of the process $\beta_1$ and present the traceplot and the autocorrelation plot of the algorithm in Figure \ref{Fig:Cauchy.Regression.Transformed.MALA}. In this case, the traceplot seems a bit more stable compared to Figure \ref{Fig:Cauchy.Regression.MALA} where MALA was directly targeting the posterior. On the other hand, while there is a small improvement on the rate of decay of the autocorrelations (which reach the value $0.45$ after 50-lag, compared to the value of $0.60$ the MALA had reached), the autocorrelation decay is still extremely slow. This indicates that while space transformation can improve the performance of an algorithm, the multi-dimensionality of the problem can interact with the transformation to create additional problems. One problem that is of particular interest in this setting is that different directions in the target density can exhibit different rates of decay, which one expects to be somewhat generic behaviour in complex, high-dimensional contexts. The goal of the transformation $f$ is then to create a transformed target with rates of decay which are as uniform as possible, and a spherically-symmetric transform is limited in its ability to achieve this. In this regard, some extra care is needed when designing transformations for high-dimensional problems.

A strategy with a slightly different flavour is to instead construct space transformations which push $\mathbf{R}^d$ into a compact set, and then run the MCMC there. One benefit of this strategy is that if the compactifying map leads to a reasonably well-behaved pushed-forward target, then there is reason to expect that the convergence profile of the chain will depend much more gently on the initialisation, on the basis that all points within the support of the target are within a certain \emph{bounded} distance of one another.

\begin{figure}[ht]
    \centering
    \includegraphics[width=\textwidth, height=9cm]{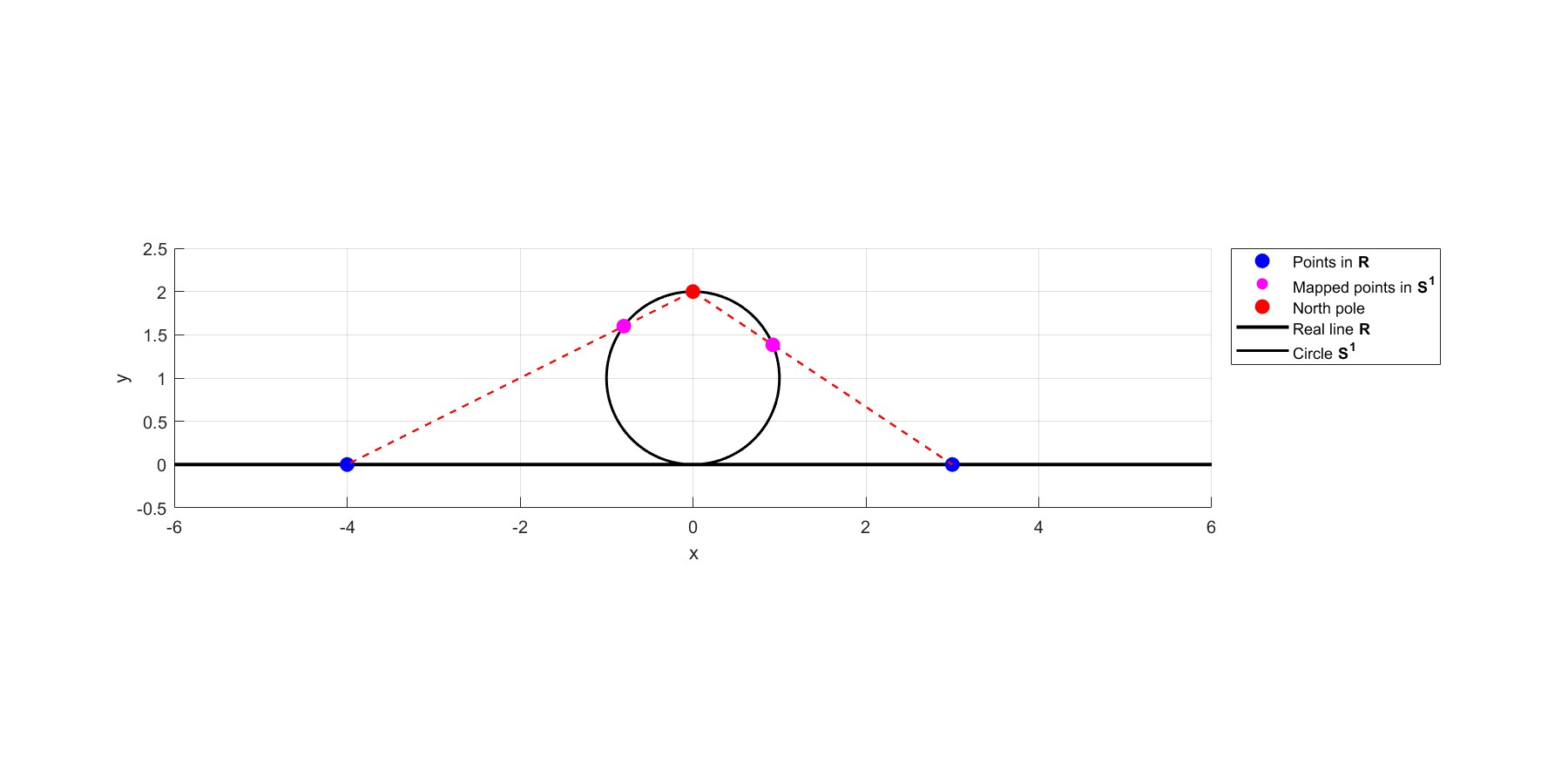}
    \vspace{-2.5cm}
    \caption{Stereographic projection from $\mathbf{R}$ to the circle $\mathbf{S}^1$. The x-axis is the real line. The figure shows how each point in $\mathbf{R}$ (blue points) is mapped to the circle (magenta points). The magenta points are at the intersection between the circle and the line connecting each real (blue) point to the north pole (red point).}\label{fig:Stereograpic.Projection}
\end{figure}

In this direction, \cite{yang.latuszynski.roberts:24} recently suggested a space transformation that maps $\mathbf{R}^d$ to the unit sphere in one dimension higher, i.e. $X = \mathbf{S}^d \subset\mathbf{R}^{d+1}$. This is done using the stereographic projection (see e.g. \cite{lee:13}) depicted in Figure \ref{fig:Stereograpic.Projection}. A relatively standard MCMC algorithm is then used to sample from the transformed target on the sphere, and the authors show that under reasonable conditions (which accommodate various degrees of heavy-tailedness), one recovers exponentially-fast convergence to the target in total variation distance. Perhaps even more impressively, this rate of convergence in total variation distance can be bounded \emph{independently of the starting position} of the algorithm, i.e. the process is {\it uniformly ergodic}. On the other hand, due to the dramatic fashion in which the stereographic projection warps the state space, some care is again required in centering and scaling the target distribution. For further work which studies the challenges associated with this adaptation, see also \cite{bell.latuszynski.roberts:25}.

Finally, we note that other more elaborate space transformations have been suggested in the MCMC literature and applied with some success (see e.g. \cite{hoffman2019neutralizingbadgeometryhamiltonian, osmundsen.kleppe.leisenfeld:21, brooks.gelman.joens.meng:11,  hird.livingstone:24}), but often without focusing on heavy-tailed target applications. In view of our focus here, we therefore omit further discussion of these approaches.

\subsubsection{Time Transformations}\label{sec:time.transform}

Rather than transforming the state space, a second approach to deal with the heavy-tailedness of the target is to apply a time-transformation (also known as a \emph{time change}) to the process, whereby the evolution of time itself within the algorithm is altered. Equivalently, one introduces a state-dependent `speed function' which dictates how quickly the process moves given its current position. Choosing an appropriate speed function can thus allow the process to move around the space faster and explore the tails more efficiently. For example, the long but infrequent excursions of the process at the tails (showcased in Section \ref{sec:heavy.pathology}), can become much shorter but persistent when the movement is regulated through a speed function, leading to more consistent estimation.

\begin{figure}[ht]
    \centering
\includegraphics[width=0.6\textwidth]{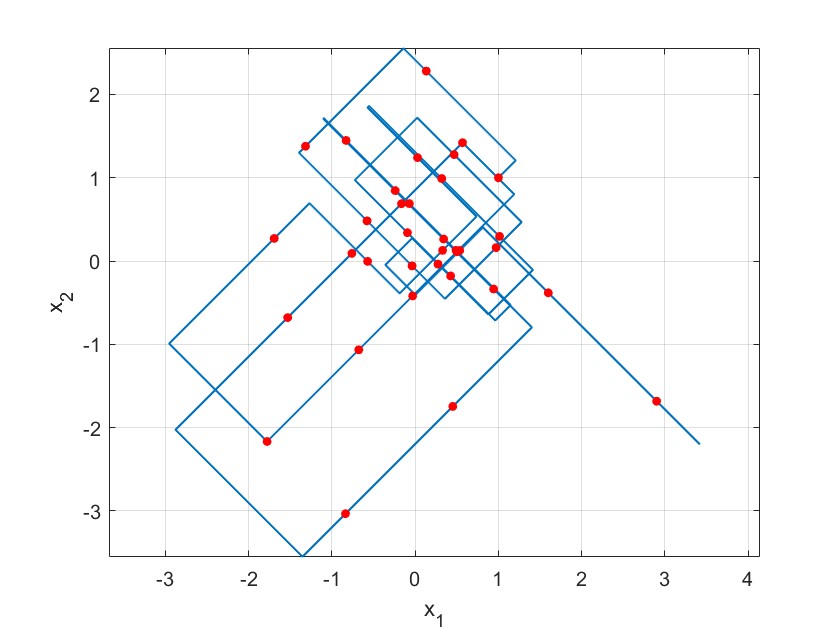}
\caption{Scatter plot of a time-changed Zig-Zag process. The red points denote the position of the process every $h = 0.5$ time units. Target: 2-dimensional Cauchy, $\pi(x) \propto \left(1+x_1^2+x_2^2\right)^{-\frac{3}{2}}$. Speed function: $s(x)=\left(  1+ x_1^2+x_2^2\right)^\frac{1}{2}$.}\label{fig:time.change.2d.Cauchy.scatter}
\end{figure}

A general framework for how to define a time-changed process can be found in Chapter 6 of \cite{ethier.kurtz:09}, while a unifying framework for employing these processes for MCMC sampling was developed in \cite{bertazzi.vasdekis:25}. A process $(X_t)_{t \geq 0}$ on $\mathbf{R}^d$ is defined to be a {\it time-change} of a process $(Y_t)_{t \geq 0}$ if for all $t \geq 0$
\begin{equation}\label{eq:time_change}
	X_t =Y_{r(t)}, \text{ where } \,\,	r(t) := \int_0^t s(X_u) \mathrm{d} u
\end{equation}
for some function $s: \mathbf{R}^d \rightarrow (0,+\infty)$. The process $Y$ is called the {\it base process} and the function $s$ is the {\it speed function}. Note that on a heuristic level, it holds that
\begin{equation*}
    \frac{\mathrm{d} X_t}{\mathrm{d} t} = \frac{\mathrm{d} Y_{r(t)}}{\mathrm{d} r(t)}\cdot \frac{\mathrm{d} r(t)}{\mathrm{d} t}= \dot{Y}_{r(t)} \cdot s(X_t).
\end{equation*}
Thus, the process $X$ traverses the path of the process $Y$ but does so `$s$ times faster'. Figure \ref{fig:time.change.2d.Cauchy.scatter} illustrates this phenomenon. The background path is the one of a `canonical' Zig-Zag process (which here plays the role of the base process $Y$), moving with constant speed. The red points denote the position of the time-changed process $X$ at times $0.5,1,1.5,2,2.5,3,\dots$, after applying a time-change with speed $s(x)=\left(1 + |x|^2\right)^\frac{1}{2}$. While these points are equi-distant in time, it is evident that they are not equi-distant in space, since the process $X$ moves with varying speed, depending on the current position. The time-changed process is targeting a bivariate, isotropic Cauchy distribution, i.e.
\begin{equation*}
    \pi(x) \propto \left(  1+x_1^2+x_2^2 \right)^{-\frac{3}{2}}.
\end{equation*}

Conceptually, using a time-changed process $X$ has a lot in common with the space-transformations discussed in Section \ref{sec:space.transforms}. 
A useful perspective to consider when devising time-changed processes for the purpose of sampling is the following: assuming that $s$ is $\pi$-integrable, continuous and bounded below by a positive constant, it holds that
\begin{equation}
    X \text{ is } \pi\text{-invariant} \iff Y
\text{ is } \tilde{\pi}\text{-invariant},
\end{equation}
where the measure $\tilde{\pi}$ is defined as
\begin{equation}\label{time.changed.measure}
    \tilde{\pi} \left( \mathrm{d}x \right) = \frac{1}{\pi(s)} s(x) \, \pi \left( \mathrm{d}x \right)
\end{equation}
with $\pi(s) = \int_{\mathbf{R}^d} s(y) \, \pi \left( \mathrm{d}y \right)$ the induced normalising constant.

Bearing this in mind, if one wants to use the time-changed process $X$ to create samples from $\pi$, one ought to proceed as follows: construct a base process $Y$ that targets the modified measure $\tilde{\pi}$, and then apply an appropriate time-change to adjust the speed of the process, ensuring that one spends more time in relevant areas of the space, ultimately biasing the samples towards $\pi$. In this regard, at a high level, one can say that both space- and time-transformation address the problem of the target having heavy tails by targeting a different measure $\tilde{\pi}$ and then `correcting' the samples. 

An important distinction between the two approaches, however, lies in the type of tails induced on the transformed distribution $\tilde{\pi}$. Space transformations typically lead to a lighter-tailed $\tilde{\pi}$, under the expectation that the process targeting $\tilde{\pi}$ will explore the target more efficiently, thereby yielding higher-quality samples after applying the inverse transform. By contrast, when applying time transformations, the tails of $\tilde{\pi}$ as in (\ref{time.changed.measure}) become even heavier, provided that $s(x) \xrightarrow{|x| \to \infty} \infty$. This allows the base process $Y$ (targeting $\tilde{\pi}$) to exhibit long excursions in the tails. These regions are subsequently visited more rapidly by $X$, leading to improved tail exploration. Indeed, \cite{bertazzi.vasdekis:25} shows that under assumptions on the speed function $s$, applying a time-change can lead to exponential and even uniform ergodicity, even for heavy tailed targets, where the base process $Y$ would converge much slower.

On the other hand, using a time-transformed process can be challenging from a practical point of view and can significantly increase the computational cost of the algorithm. This increased difficulty arises from the fact that one must be able to analytically integrate the function $s$ along the path of the process (see, for example, the definition of the quantity $r$ in (\ref{eq:time_change})). From this perspective, base processes that move in simple ways can be more promising candidates for use in a time-transformed setting, at least from a computational standpoint. Two prominent examples are the Zig-Zag process and the Bouncy Particle Sampler, introduced in Section \ref{sec:Piecewise.Deterministic.Monte.Carlo}, since between jumping events these processes move in straight lines. As shown in \cite{vasdekis.roberts:23}, where a time-transformed Zig-Zag process was introduced under the name “Speed Up Zig-Zag,” between jumping events, determining the position of the time-changed process at a given time amounts to solving a one-dimensional ordinary differential equation (ODE), which the authors argue can be done exactly for speed functions of the form
\begin{equation*}
s(x) = \left( 1 + |x|^2 \right)^\frac{1+k}{2}, \quad k \in \mathbf{N}.
\end{equation*}
Of course in a practical setting, when the target is high dimensional, one might want to use other speed functions. For example, when the tails of the target have different rates of decay across different directions, a well-designed speed function ought to reflect this.

\begin{figure}[ht]
    \centering
    \begin{subfigure}[t]{0.47\textwidth}
        \includegraphics[width=\linewidth]{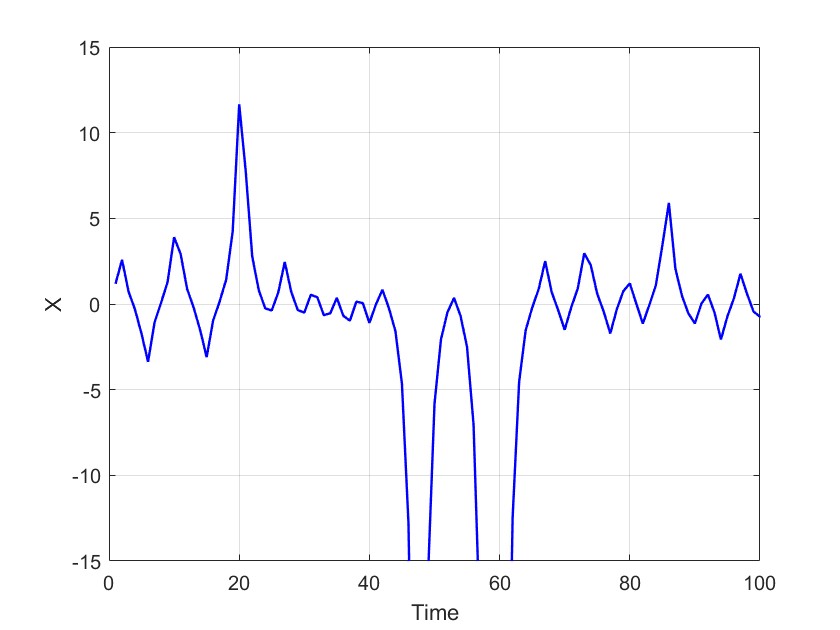}
        \caption{Traceplot (first 100 iterations)}\label{Fig:Time.ZZ.Cauchy:part1}
    \end{subfigure}
    \hfill
    \begin{subfigure}[t]{0.47\textwidth}
        \includegraphics[width=\linewidth]{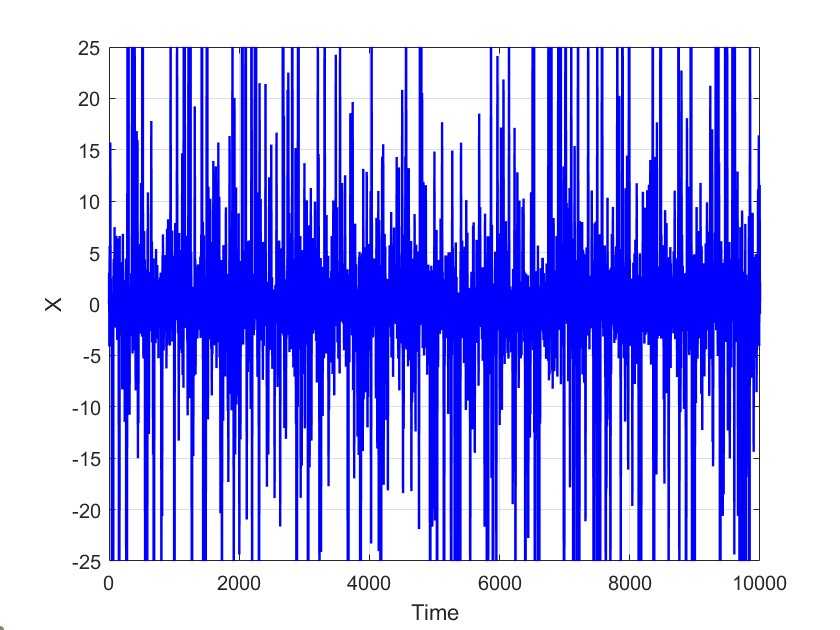}
        \caption{Traceplot}\label{Fig:Time.ZZ.Cauchy:part2}
    \end{subfigure}
    
    \vspace{0.5cm}  

  
    \begin{subfigure}[t]{0.47\textwidth}
    \includegraphics[width=\linewidth]{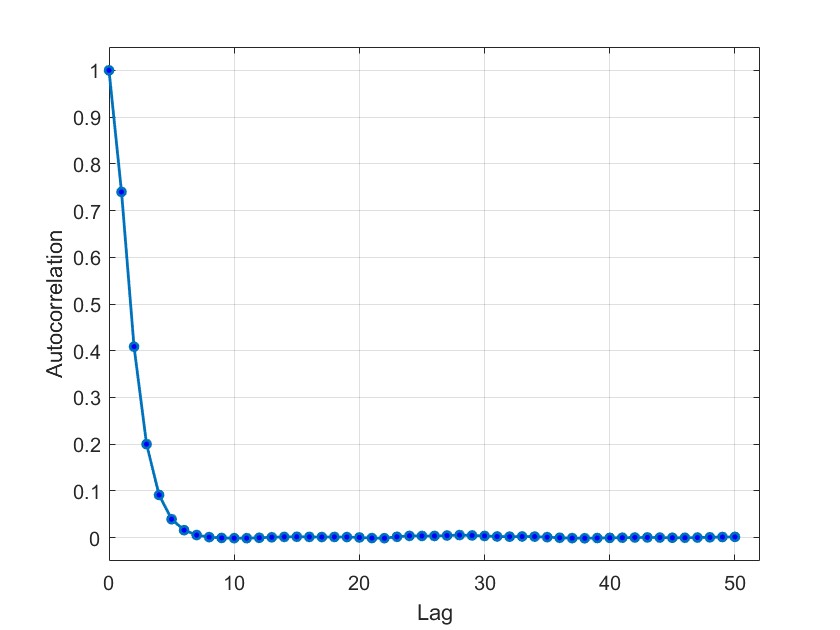}
        \caption{Autocorrelations}\label{Fig:Time.ZZ.Cauchy:part3}
    \end{subfigure}
     \hfill
    \begin{subfigure}[t]{0.47\textwidth}
    \includegraphics[width=\linewidth]{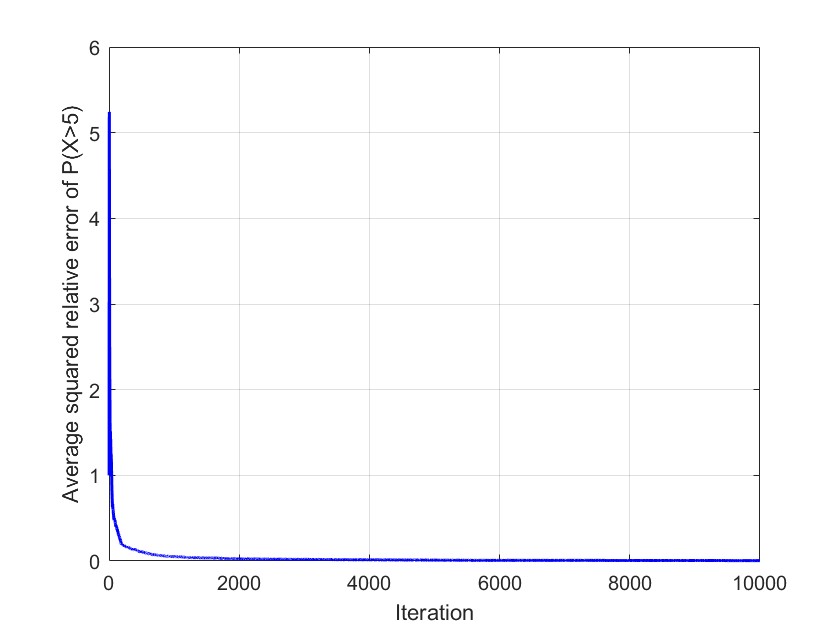}
        \caption{Squared relative error estimating the probability of a tail event (actual probability : 0.0627)}\label{Fig:Time.ZZ.Cauchy:part4}
    \end{subfigure}
  \vspace{1mm}
    \caption{Time-transformed Zig-Zag algorithm on a one-dimensional Cauchy target. Speed function: $s(x)=\left(  1+x^2 \right)^\frac{1}{2}$.}\label{Fig:Time.ZZ.Cauchy}
\end{figure}

To showcase the performance of time-transformed processes, we revisit our examples from Sections \ref{sec:heavy.pathology} and \ref{sec:space.transforms}. We first consider the Cauchy target from Example \ref{cauchy.example:00}. We run a time-transformed Zig-Zag process, using the speed function $s(x)=\left( 1+x^2 \right)^\frac{1}{2}$. For a somewhat equal comparison with Figure \ref{Fig:Cauchy.MALA} we ran the algorithm for the same number of target evaluations as in Section \ref{sec:heavy.pathology}. We present our results in Figure \ref{Fig:Time.ZZ.Cauchy}, presenting four plots of the same type as when we ran MALA directly on the target (Figure \ref{Fig:Cauchy.MALA}). It is evident from the traceplot that the process explores the space well, having frequent and short tail excursions, while the autocorrelation plot decays sufficiently fast. Estimation of the tail event $\left\{ X \geq 5 \right\}$ also becomes very accurate after a few algorithmic iterations.

We also consider the time-transformed Zig-Zag on Example \ref{ex:Cauchy.Regression}, the Cauchy regression model with a Horseshoe prior, with 
$p=5$ and $n=20$ observations. As in the one-dimensional Cauchy target, we ran the algorithm for the same number of target evaluations as in Section \ref{sec:heavy.pathology}. We note, however, that the time-changed process required substantially more wall-clock computation time to run (approximately two orders of magnitude more), as the routine must repeatedly optimise complicated functions.

\begin{figure}[ht]
    \centering
    \begin{subfigure}[t]{0.47\textwidth}
        \includegraphics[width=\linewidth]{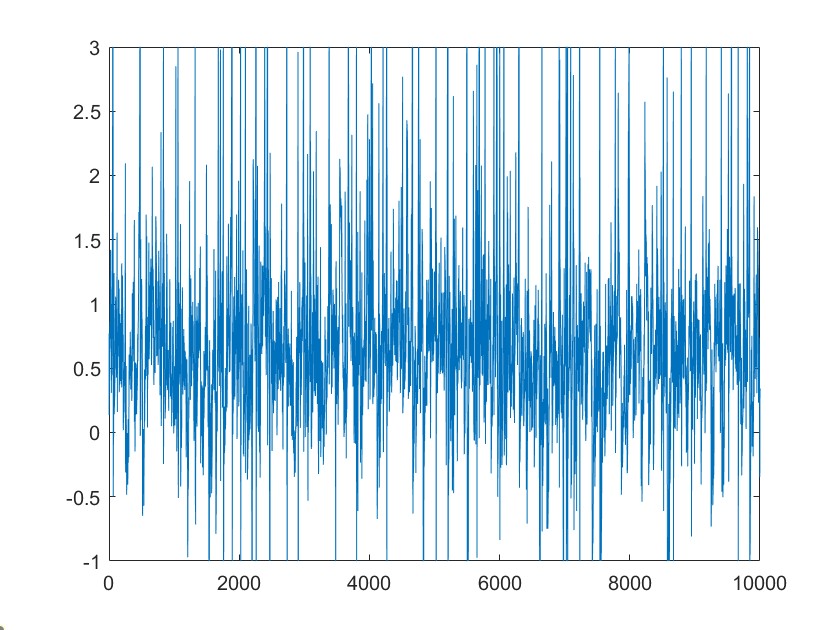}
        \caption{Traceplot of the first coordinate}\label{Fig:Cauchy.Regression.Time.ZZ.part1}
    \end{subfigure}
    \hfill
    \begin{subfigure}[t]{0.47\textwidth}
        \includegraphics[width=\linewidth]{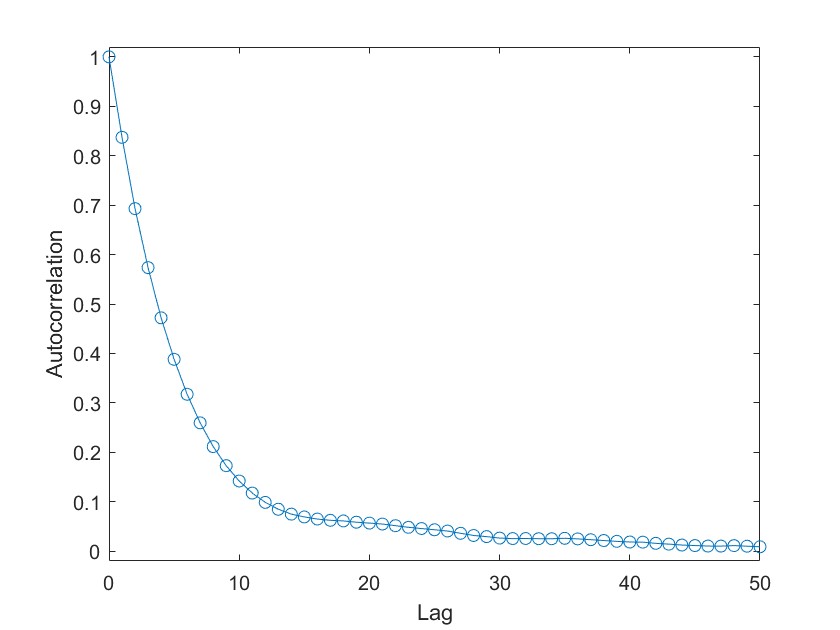}
        \caption{Autocorrelations of the first coordinate}
\label{Fig:Cauchy.Regression.Time.ZZ.part2}
    \end{subfigure}
    \caption{Time-transformed Zig-Zag algorithm on a 5-dimensional Cauchy-Regression target. Speed function: $s(x)=\left(  1+x^2 \right)^\frac{1}{2}$.}\label{Fig:Cauchy.Regression.Time.ZZ}
\end{figure}

Nevertheless, focusing on the first coordinate $\beta_1$, we present the trace plot and the autocorrelation plot of the algorithm in Figure \ref{Fig:Cauchy.Regression.Time.ZZ}. Compared with Figures \ref{Fig:Cauchy.Regression.MALA} and \ref{Fig:Cauchy.Regression.Transformed.MALA}, the trace plot appears substantially more stable, and the autocorrelation exhibits a markedly faster decay.

Time-transforming a process for sampling was considered in \cite{roberts.stramer:02} were the authors studied time-changes of diffusion processes, using a speed function of the form $s(x)=\pi(x)^{-a}$, $a \in (0,1)$. For more recent uses of time-changing diffusion for sampling, see also \cite{leimkuhler.lohmann.whalley:25, leroy2024adaptive}. There are also connections with the idea of self-normalised importance sampling (see e.g. \cite{robert.casella:05}), in the sense that rescaling time according to a speed function can have a similar effect to adjusting the weight of particular samples from the base process. In that direction, \cite{andral.douc.marival.robert:24} use as base process a discrete time Markov chain with stationary distribution different than the target and adjust it by letting it spend a random number of iterations at each state, with the expected occupation time playing a similar role to the (reciprocal) speed function; see also \cite{malefaki.iliopoulos:08}.

We note also that there exist numerical integrators for SDEs which involve `adaptive time-stepping' routines, where the time-discretisation parameter $h$ is effectively taken to depend on the location of the process; see e.g. \cite{foster2023convergence} for a nice recent exposition. While these are not based on time-changing per se -- indeed, the ultimate goal of such methods is to make a strong approximation of the original SDE -- there is at the very least some conceptual common ground in their construction. Conversely, we observe that such methods appear to have garnered most interest in the context of rough targets, with the application to flat targets an apparently more recent development.






\subsection{Open Problems}

We again pause to discuss some problems which are not entirely addressed in the present literature, but would be of some interest to resolve.

\subsubsection*{Heavy Tails in High Dimension}

High dimensions are understood to make generic sampling tasks more challenging, and this is particularly true in the case of heavy-tailed problems. One heuristic in this direction is that in high dimension, there are many more ways in which to be heavy-tailed (or not). In particular, heterogeneity of tail-heaviness across directions is a real challenge; see the literature on variational inference for some relevant discussion \citep{jaini20, liang22, hickling25}. There are also qualitative differences between spherically-symmetric tails and product-form tails which are well-understood in some contexts, but are rather under-discussed in the sampling literature at present. The language of functional inequalities and isoperimetry gives some guidance in this direction (see e.g. \cite{bobkov2007large, gozlan2010isoperimetry}), but does not really tackle the practical question of detecting `what the problem is' in a specific instance. Along similar lines, it appears important to define a richer class of model problems which are reflective of practical challenges, but also tractable enough that their analysis might yield some actionable insight.

\subsubsection*{Heavy Tails with Unknown Heaviness}

Many of these problems are reasonably solvable when we know precisely how heavy the tails are. In a practical setting, this is hard to do without solving the sampling problem well. It would be useful to find adaptive strategies for introducing the tails more gradually, and resolving problems as they appear. Techniques for tail index estimation from the extremes literature (e.g. the classical Hill estimator \cite{hill1975simple} and its extensions) could potentially be useful tools in either case, having similarly been applied successfully in the context of density estimation \cite{laszkiewicz22}.

\subsubsection*{Reliable Transformations for Negating Heavy Tails}

Effective transformations of space or time are difficult to come up with in the absence of strong a priori information. Existing works have proposed the use of radial transformations, and it would also be rather natural to propose coordinate-wise transformations. These transformations are plausibly learnable in practice, but also flexible enough to accommodate an interesting range of target behaviours. It would be of great value to identify other families of structured target distributions for which `good' transformations naturally suggest themselves.

\subsubsection*{Negating Heavy Tails without Inducing Roughness}

While confining the target more can resolve the issues with tail heaviness, a careless approach to confinement can easily exacerbate issues with roughness. It seems to be widely-known at an informal level that `one must be careful' in such scenarios, but it appears difficult to automate this intuition. Even for the relatively safe and non-intrusive strategy of applying a transformation which coincides with the identity mapping on a large ball of radius $R \gg 0$, and then compresses mass much more tightly outside of this ball, there do not appear to be default solutions in the literature for setting this critical radius $R$. Practical resolutions to these challenges could be of great utility for navigating heavy-tailed problems.

\section{Conclusion}


While classical MCMC still `works', there is an increasing realisation that the sampling problems arising in contemporary applications often fall outside of the scope of the `usual' assumptions which we have in mind when developing methods and theory. There has been substantial progress on deriving algorithms which achieve a degree of robustness with respect to some of these pathologies, while remaining performant in `benign' settings. We have reviewed a number of existing solutions and identified some areas for potential contributions and growth.

Our focus has been on sampling problems for which `size' is somehow the primary issue, whether through the richness of information about the geometry of the target distribution for rough targets, or the sparsity of information provided by flat targets. This has allowed for a relatively clean treatment, whereby careful inflation or attenuation of this information can lead to more stable algorithms. Nevertheless, despite this relative simplicity, various challenges remain, largely relating to the difficulty of estimating these features `on-the-fly' as part of the sampling process, rather than by a priori mathematical computations. We see this as an interesting avenue for future contributions.

In any case, it bears mentioning that there are many other sampling problems of practical interest for which `size' is a relatively minor concern, and `shape' is entirely crucial. One instance of this is the task of multi-modal sampling, for which the literature on multi-canonical or `tempering / annealing' strategies is vast. Another slightly more vague instance would be the phenomenon of target distributions with `heterogeneous local geometry', for which different parts of the state space require drastically different `shapes' of dynamics to explore efficiently; rigorous progress on this topic is still rather nascent. We certainly do not suggest that the developments detailed in this survey are designed with such (arguably more difficult) challenges in mind, though one hopes that they will at least be compatible with them.

The past decade or so has been immensely fruitful from the perspective of establishing theoretical guarantees for MCMC in the `well-conditioned, log-concave' setting, with the quantitative analysis of various algorithms coming along in leaps and bounds during this period. In search of similar foundational guarantees for methods which apply to a broader class of sampling problems, we are optimistic that this burgeoning literature on `robust' MCMC will provide ample inspiration, leading to improved practical solutions with a firm theoretical basis. We are excited to see how things progress from here.

\bibliography{main}

@book{anderson2015stochastic,
  title={Stochastic analysis of biochemical systems},
  author={Anderson, David F and Kurtz, Thomas G},
  volume={674},
  year={2015},
  publisher={Springer}
}

@article{andral.douc.marival.robert:24,
title = {The importance {M}arkov chain},
journal = {Stochastic Processes and their Applications},
volume = {171},
pages = {104316},
year = {2024},
issn = {0304-4149},
doi = {https://doi.org/10.1016/j.spa.2024.104316},
url = {https://www.sciencedirect.com/science/article/pii/S030441492400022X},
author = {Charly Andral and Randal Douc and Hugo Marival and Christian P. Robert}
}

@article{andral2024automated,
  title={Automated Techniques for Efficient Sampling of {P}iecewise-{D}eterministic {M}arkov {P}rocesses},
  author={Andral, Charly and Kamatani, Kengo},
  journal={arXiv preprint arXiv:2408.03682},
  year={2024}
}

@article{andrieu2020general,
  title={A general perspective on the {M}etropolis-{H}astings kernel},
  author={Andrieu, Christophe and Lee, Anthony and Livingstone, Sam},
  journal={arXiv preprint arXiv:2012.14881},
  year={2020}
}

@article{andrieu2023weak,
    title={Weak {P}oincar\'{e} Inequalities for {M}arkov chains: theory and applications},
    author={Andrieu, Christophe and Lee, Anthony and Power, Sam and Wang, Andi Q},
    journal={The Annals of Applied Probability},
    year = {to appear}
}

@article{andrieu2024explicit,
  title={Explicit convergence bounds for {M}etropolis {M}arkov chains: Isoperimetry, spectral gaps and profiles},
  author={Andrieu, Christophe and Lee, Anthony and Power, Sam and Wang, Andi Q},
  journal={The Annals of Applied Probability},
  volume={34},
  number={4},
  pages={4022--4071},
  year={2024},
  publisher={Institute of Mathematical Statistics}
}

@article{ascolani2024entropy,
  title={Entropy contraction of the {G}ibbs sampler under log-concavity},
  author={Ascolani, Filippo and Lavenant, Hugo and Zanella, Giacomo},
  journal={arXiv preprint arXiv:2410.00858},
  year={2024}
}

@article{atchade2006adaptive,
  title={An adaptive version for the {M}etropolis-adjusted {L}angevin algorithm with a truncated drift},
  author={Atchad{\'e}, Yves F},
  journal={Methodology and Computing in Applied Probability},
  volume={8},
  number={2},
  pages={235--254},
  year={2006},
  publisher={Springer}
}

@article{barker1965monte,
  title={Monte {C}arlo calculations of the radial distribution functions for a proton electron plasma},
  author={Barker, Anthony Alfred},
  journal={Australian Journal of Physics},
  volume={18},
  number={2},
  pages={119--134},
  year={1965},
  publisher={CSIRO Publishing}
}

@article{belisle1993hit,
	title        = {Hit-and-{R}un algorithms for generating multivariate distributions},
	author       = {B{\'e}lisle, Claude JP and Romeijn, H Edwin and Smith, Robert L},
	year         = {1993},
	journal      = {Mathematics of Operations Research},
	publisher    = {INFORMS},
	volume       = {18},
	number       = {2},
	pages        = {255--266}
}

@misc{bell.latuszynski.roberts:25,
      title={Adaptive {S}tereographic {MCMC}}, 
      author={Cameron Bell and Krzystof Łatuszyński and Gareth O. Roberts},
      year={2025},
      eprint={2408.11780},
      archivePrefix={arXiv},
      primaryClass={stat.CO},
      url={https://arxiv.org/abs/2408.11780}, 
}

@inproceedings{benko2025langevin,
  title        = {Langevin {M}onte {C}arlo Beyond {L}ipschitz {G}radient {C}ontinuity},
  author       = {Benko, Matej and Chlebicka, Iwona and Endal, Jorgen and Miasojedow, B{\l}a{\.z}ej},
  year         = {2025},
  booktitle    = {Proceedings of the AAAI Conference on Artificial Intelligence},
  volume       = {39},
  pages        = {15541--15549}
}

@article{bernard2009event,
  title={Event-chain {M}onte {C}arlo algorithms for hard-sphere systems},
  author={Bernard, Etienne P and Krauth, Werner and Wilson, David B},
  journal={Physical Review E—Statistical, Nonlinear, and Soft Matter Physics},
  volume={80},
  number={5},
  pages={056704},
  year={2009},
  publisher={APS}
}

@article{bertazzi.bierkens.dobson:22,
	title        = {Approximations of {P}iecewise-{D}eterministic {M}arkov {P}rocesses and their convergence properties},
	author       = {Andrea Bertazzi and Joris Bierkens and Paul Dobson},
	year         = {2022},
	journal      = {Stochastic Processes and their Applications},
	volume       = {154},
	pages        = {91--153},
	doi          = {https://doi.org/10.1016/j.spa.2022.09.004},
	issn         = {0304-4149},
	url          = {https://www.sciencedirect.com/science/article/pii/S0304414922001958}
}

@misc{bertazzi.vasdekis:25,
	title        = {Sampling with time-changed {M}arkov processes},
	author       = {Andrea Bertazzi and Giorgos Vasdekis},
	year         = {2025},
	url          = {https://arxiv.org/abs/2501.15155},
	eprint       = {2501.15155},
	archiveprefix = {arXiv},
	primaryclass = {stat.CO}
}

@article{besag1994comments,
  title={Comments on “{R}epresentations of knowledge in complex systems” by {U} {G}renander and {MI} {M}iller},
  author={Besag, Julian},
  journal={J. Roy. Statist. Soc. Ser. B},
  volume={56},
  number={591-592},
  pages={4},
  year={1994}
}

@article{beskos2005exact,
author = {Alexandros Beskos and Gareth O. Roberts},
title = {{Exact simulation of diffusions}},
volume = {15},
journal = {The Annals of Applied Probability},
number = {4},
publisher = {Institute of Mathematical Statistics},
pages = {2422 -- 2444},
year = {2005},
doi = {10.1214/105051605000000485},
URL = {https://doi.org/10.1214/105051605000000485}
}

@article{beskos2006exact,
  title={Exact and computationally efficient likelihood-based estimation for discretely observed diffusion processes (with discussion)},
  author={Beskos, Alexandros and Papaspiliopoulos, Omiros and Roberts, Gareth O and Fearnhead, Paul},
  journal={Journal of the Royal Statistical Society Series B: Statistical Methodology},
  volume={68},
  number={3},
  pages={333--382},
  year={2006},
  publisher={Oxford University Press}
}

@article{beskos2006retrospective,
  title={Retrospective exact simulation of diffusion sample paths with applications},
  author={Beskos, Alexandros and Papaspiliopoulos, Omiros and Roberts, Gareth O},
  journal={Bernoulli},
  volume={12},
  number={6},
  pages={1077--1098},
  year={2006},
  publisher={Bernoulli Society for Mathematical Statistics and Probability}
}

@misc{betancourt:18,
      title={A conceptual introduction to {H}amiltonian {M}onte {C}arlo}, 
      author={Michael Betancourt},
      year={2018},
      eprint={1701.02434},
      archivePrefix={arXiv},
      primaryClass={stat.ME},
      url={https://arxiv.org/abs/1701.02434}, 
}

@article{bierkens.duncan:17,
title={Limit theorems for the {Z}ig-{Z}ag {P}rocess}, 
volume={49},
DOI={10.1017/apr.2017.22}, 
number={3},
journal={Advances in Applied Probability},
author={Bierkens, Joris and Duncan, Andrew},
year={2017},
pages={791–825}
}

@article{bierkens2019zig,
author = {Joris Bierkens and Paul Fearnhead and Gareth Roberts},
title = {{The {Z}ig-{Z}ag {P}rocess and super-efficient sampling for {B}ayesian analysis of big data}},
volume = {47},
journal = {The Annals of Statistics},
number = {3},
publisher = {Institute of Mathematical Statistics},
pages = {1288 -- 1320},
year = {2019},
doi = {10.1214/18-AOS1715},
URL = {https://doi.org/10.1214/18-AOS1715}
}

@article{bierkens2025scaling,
  title={Scaling of piecewise deterministic {M}onte {C}arlo for anisotropic targets},
  author={Bierkens, Joris and Kamatani, Kengo and Roberts, Gareth O},
  journal={Bernoulli},
  volume={31},
  number={3},
  pages={2323--2350},
  year={2025},
  publisher={Bernoulli Society for Mathematical Statistics and Probability}
}

@article{blanchet2020exact,
  title={Exact simulation for multivariate {I}t{\^o} diffusions},
  author={Blanchet, Jose and Zhang, Fan},
  journal={Advances in Applied Probability},
  volume={52},
  number={4},
  pages={1003--1034},
  year={2020},
  publisher={Cambridge University Press}
}

@article{bobkov2007large,
author = {Sergey Bobkov},
title = {{Large deviations and isoperimetry over convex probability measures with heavy tails}},
volume = {12},
journal = {Electronic Journal of Probability},
publisher = {Institute of Mathematical Statistics and Bernoulli Society},
pages = {1072--1100},
year = {2007},
doi = {10.1214/EJP.v12-440},
URL = {https://doi.org/10.1214/EJP.v12-440}
}

@article{bou2010pathwise,
title={Pathwise accuracy and ergodicity of {M}etropolized integrators for {SDE}s},
  author={Bou-Rabee, Nawaf and Vanden-Eijnden, Eric},
  journal={Communications on Pure and Applied Mathematics: A Journal Issued by the Courant Institute of Mathematical Sciences},
  volume={63},
  number={5},
  pages={655--696},
  year={2010},
  publisher={Wiley Online Library}
}

@article{bou2013nonasymptotic,
  title={Nonasymptotic mixing of the {MALA} algorithm},
  author={Bou-Rabee, Nawaf and Hairer, Martin},
  journal={IMA Journal of Numerical Analysis},
  volume={33},
  number={1},
  pages={80--110},
  year={2013},
  publisher={OUP}
}

@article{bou2017randomized,
    ISSN = {10505164},
    URL = {http://www.jstor.org/stable/26361544},
    author = {Nawaf Bou-Rabee and Jesús María Sanz-Serna},
    journal = {The Annals of Applied Probability},
    number = {4},
    pages = {2159--2194},
    publisher = {Institute of Mathematical Statistics},
    title = {RANDOMIZED {H}AMILTONIAN {M}ONTE {C}ARLO},
    urldate = {2025-10-07},
    volume = {27},
    year = {2017}
}

@book{bou2018continuous,
  author    = {Bou-Rabee, Nawaf and Vanden{-}Eijnden, Eric},
  title     = {Continuous--Time Random Walks for the Numerical Solution of Stochastic Differential Equations},
  series    = {Memoirs of the American Mathematical Society},
  volume    = {256, no. 1228},
  publisher = {American Mathematical Society},
  address   = {Providence, RI},
  year      = {2018},
  pages     = {iv + 124},
  isbn      = {978-1-4704-3181-5},
  doi       = {10.1090/memo/1228},
}

@article{bouchard2018bouncy,
  title={The {B}ouncy {P}article {S}ampler: A nonreversible, rejection-free {M}arkov chain {M}onte {C}arlo method},
  author={Bouchard-C{\^o}t{\'e}, Alexandre and Vollmer, Sebastian J and Doucet, Arnaud},
  journal={Journal of the American Statistical Association},
  volume={113},
  number={522},
  pages={855--867},
  year={2018},
  publisher={Taylor \& Francis}
}

@book{boyd2004convex,
  title={Convex {O}ptimization},
  author={Boyd, Stephen P and Vandenberghe, Lieven},
  year={2004},
  publisher={Cambridge university press}
}

@book{brooks.gelman.joens.meng:11,
  title     = {Handbook of {M}arkov {C}hain {M}onte {C}arlo},
  editor    = {Brooks, Steve and Gelman, Andrew and Jones, Galin and Meng, Xiao-Li},
  year      = {2011},
  publisher = {Chapman \& Hall/CRC},
  address   = {New York},
  doi       = {10.1201/b10905},
  isbn      = {9780429138508},
  pages     = {619}
}

@article{brosse2019tamed,
	title        = {The tamed unadjusted {L}angevin algorithm},
	author       = {Brosse, Nicolas and Durmus, Alain and Moulines, {\'E}ric and Sabanis, Sotirios},
	year         = {2019},
	journal      = {Stochastic Processes and their Applications},
	publisher    = {Elsevier},
	volume       = {129},
	number       = {10},
	pages        = {3638--3663}
}

@article{bubeck2015convex,
  title={Convex optimization: {A}lgorithms and complexity},
  author={Bubeck, S{\'e}bastien},
  journal={Foundations and Trends{\textregistered} in Machine Learning},
  volume={8},
  number={3-4},
  pages={231--357},
  year={2015},
  publisher={Now Publishers, Inc.}
}

@inproceedings{carvalho2009handling,
	title        = {Handling sparsity via the horseshoe},
	author       = {Carvalho, Carlos M and Polson, Nicholas G and Scott, James G},
	year         = {2009},
	booktitle    = {Artificial intelligence and statistics},
	pages        = {73--80},
	organization = {PMLR}
}

@article{chaari2016hamiltonian,
  title={A {H}amiltonian {M}onte {C}arlo method for non-smooth energy sampling},
  author={Chaari, Lotfi and Tourneret, Jean-Yves and Chaux, Caroline and Batatia, Hadj},
  journal={IEEE Transactions on Signal Processing},
  volume={64},
  number={21},
  pages={5585--5594},
  year={2016},
  publisher={IEEE}
}

@inproceedings{cheng2018underdamped,
  title={Underdamped {L}angevin {MCMC}: {A} non-asymptotic analysis},
  author={Cheng, Xiang and Chatterji, Niladri S and Bartlett, Peter L and Jordan, Michael I},
  booktitle={Conference on learning theory},
  pages={300--323},
  year={2018},
  organization={PMLR}
}

@article{chevallier2024pdmp,
  title={{PDMP} {M}onte {C}arlo methods for piecewise-smooth densities},
  author={Chevallier, Augustin and Power, Sam and Wang, Andi Q and Fearnhead, Paul},
  journal={Advances in Applied Probability},
  volume={56},
  number={4},
  pages={1153--1194},
  year={2024},
  publisher={Cambridge University Press}
}

@inproceedings{chewi2021optimal,
  title={Optimal dimension dependence of the {M}etropolis-adjusted {L}angevin algorithm},
  author={Chewi, Sinho and Lu, Chen and Ahn, Kwangjun and Cheng, Xiang and Le Gouic, Thibaut and Rigollet, Philippe},
  booktitle={Conference on Learning Theory},
  pages={1260--1300},
  year={2021},
  organization={PMLR}
}

@article{chewi2024analysis,
  title={Analysis of {L}angevin {M}onte {C}arlo: From {P}oincar\'{e} to {L}og-{S}obolev},
  author={Chewi, Sinho and Erdogdu, Murat A and Li, Mufan and Shen, Ruoqi and Zhang, Matthew S},
  journal={Foundations of Computational Mathematics},
  pages={1--51},
  year={2024},
  publisher={Springer}
}

@article{corbella2022automatic,
  title={Automatic {Z}ig-{Z}ag sampling in practice},
  author={Corbella, Alice and Spencer, Simon EF and Roberts, Gareth O},
  journal={Statistics and Computing},
  volume={32},
  number={6},
  pages={107},
  year={2022},
  publisher={Springer}
}

@article{crucinio2025optimal,
    title       = {Optimal scaling results for {M}oreau-{Y}osida {M}etropolis-adjusted {L}angevin algorithms},
    author      = {Crucinio, Francesca R and Durmus, Alain and Jim{\'e}nez, Pablo and Roberts, Gareth O},
    journal     = {Bernoulli},
    volume      = {31},
    number      = {3},
    pages       = {1889--1907},
    year        = {2025},
    publisher   = {Bernoulli Society for Mathematical Statistics and Probability}
}

@article{dalalyan2017theoretical,
  title={Theoretical guarantees for approximate sampling from smooth and log-concave densities},
  author={Dalalyan, Arnak S},
  journal={Journal of the Royal Statistical Society Series B: Statistical Methodology},
  volume={79},
  number={3},
  pages={651--676},
  year={2017},
  publisher={Oxford University Press}
}

@article{davis1984piecewise,
  title={Piecewise-{D}eterministic {M}arkov processes: A general class of non-diffusion stochastic models},
  author={Davis, Mark HA},
  journal={Journal of the Royal Statistical Society: Series B (Methodological)},
  volume={46},
  number={3},
  pages={353--376},
  year={1984},
  publisher={Wiley Online Library}
}

@article{deligiannidis2021randomized,
  title={Randomized {H}amiltonian {M}onte {C}arlo as scaling limit of the {B}ouncy {P}article {S}ampler and dimension-free convergence rates},
  author={Deligiannidis, George and Paulin, Daniel and Bouchard-C{\^o}t{\'e}, Alexandre and Doucet, Arnaud},
  journal={The Annals of Applied Probability},
  volume={31},
  number={6},
  pages={2612--2662},
  year={2021},
  publisher={JSTOR}
}

@article{duffield2025lattice,
  title={Lattice {R}andom {W}alk {D}iscretisations of {S}tochastic {D}ifferential {E}quations},
  author={Duffield, Samuel and Aifer, Maxwell and Melanson, Denis and Belateche, Zach and Coles, Patrick J},
  journal={arXiv preprint arXiv:2508.20883},
  year={2025}
}

@article{durmus2017nonasymptotic,
    author = {Alain Durmus and {\'E}ric Moulines},
    title = {{Nonasymptotic convergence analysis for the unadjusted {L}angevin algorithm}},
    volume = {27},
    journal = {The Annals of Applied Probability},
    number = {3},
    publisher = {Institute of Mathematical Statistics},
    pages = {1551 -- 1587},
    year = {2017},
    doi = {10.1214/16-AAP1238},
    URL = {https://doi.org/10.1214/16-AAP1238}
}

@article{durmus2019high,
    author = {Alain Durmus and {\'E}ric Moulines},
    title = {{High-dimensional {B}ayesian inference via the unadjusted {L}angevin algorithm}},
    volume = {25},
    journal = {Bernoulli},
    number = {4A},
    publisher = {Bernoulli Society for Mathematical Statistics and Probability},
    pages = {2854 -- 2882},
    year = {2019},
    doi = {10.3150/18-BEJ1073},
    URL = {https://doi.org/10.3150/18-BEJ1073}
}

@article{durmus2020geometric,
  title={Geometric ergodicity of the {B}ouncy {P}article {S}ampler},
  author={Durmus, Alain and Guillin, Arnaud and Monmarch{\'e}, Pierre},
  journal={The Annals of Applied Probability},
  volume={30},
  number={5},
  pages={2069--2098},
  year={2020},
  publisher={JSTOR}
}

@article{durmus2024asymptotic,
  title={Asymptotic bias of inexact {M}arkov chain {M}onte {C}arlo methods in high dimension},
  author={Durmus, Alain and Eberle, Andreas},
  journal={The Annals of Applied Probability},
  volume={34},
  number={4},
  pages={3435--3468},
  year={2024},
  publisher={Institute of Mathematical Statistics}
}

@book{e2021applied,
  title={Applied {S}tochastic {A}nalysis},
  author={E, Weinan and Li, Tiejun and Vanden-Eijnden, Eric},
  volume={199},
  year={2021},
  publisher={American Mathematical Soc.}
}

@book{embrechts.kluppelberg.mikosch:97,
  author    = {Paul Embrechts and Claudia Kl{\"u}ppelberg and Thomas Mikosch},
  title     = {Modelling {E}xtremal {E}vents: for {I}nsurance and {F}inance},
  series    = {Stochastic Modelling and Applied Probability},
  volume    = {33},
  publisher = {Springer‐Verlag},
  address   = {Berlin},
  year      = {1997},
  edition   = {1st},
  pages     = {xvi+645},
  isbn      = {3-540-60931-8},
  doi       = {10.1007/978-3-642-33483-2}
}

@book{ethier.kurtz:09,
	title={Markov processes: characterization and convergence},
	author={Ethier, Stewart N and Kurtz, Thomas G},
	volume={282},
	year={2009},
	publisher={John Wiley \& Sons}
}

@article{fearnhead2008particle,
  title={Particle filters for partially-observed diffusions},
  author={Fearnhead, Paul and Papaspiliopoulos, Omiros and Roberts, Gareth O},
  journal={Journal of the Royal Statistical Society Series B: Statistical Methodology},
  volume={70},
  number={4},
  pages={755--777},
  year={2008},
  publisher={Oxford University Press}
}

@article{fearnhead2024stochastic,
  title={Stochastic {G}radient {P}iecewise {D}eterministic {M}onte {C}arlo {S}amplers},
  author={Fearnhead, Paul and Grazzi, Sebastiano and Nemeth, Chris and Roberts, Gareth O},
  journal={arXiv preprint arXiv:2406.19051},
  year={2024}
}

@article{foster2023convergence,
  title={On the convergence of adaptive approximations for stochastic differential equations},
  author={Foster, James and Jelin{\v{c}}i{\v{c}}, Andra{\v{z}}},
  journal={arXiv preprint arXiv:2311.14201},
  year={2023}
}

@article{gelfand1991weak,
  title={Weak convergence of {M}arkov chain sampling methods and annealing algorithms to diffusions},
  author={Gelfand, Saul Brian and Mitter, Sanjoy K},
  journal={Journal of Optimization Theory and Applications},
  volume={68},
  number={3},
  pages={483--498},
  year={1991},
  publisher={Springer}
}

@article{gelfand2000gibbs,
  title={Gibbs sampling},
  author={Gelfand, Alan E},
  journal={Journal of the American statistical Association},
  volume={95},
  number={452},
  pages={1300--1304},
  year={2000},
  publisher={Taylor \& Francis}
}

@article{gelman1997weak,
  title={Weak convergence and optimal scaling of random walk {M}etropolis algorithms},
  author={Gelman, Andrew and Gilks, Walter R and Roberts, Gareth O},
  journal={The annals of applied probability},
  volume={7},
  number={1},
  pages={110--120},
  year={1997},
  publisher={Institute of Mathematical Statistics}
}

@article{giles2015multilevel,
  title={Multilevel {M}onte {C}arlo methods},
  author={Giles, Michael B},
  journal={Acta numerica},
  volume={24},
  pages={259--328},
  year={2015},
  publisher={Cambridge University Press}
}

@article{gillespie1977exact,
  title={Exact stochastic simulation of coupled chemical reactions},
  author={Gillespie, Daniel T},
  journal={The journal of physical chemistry},
  volume={81},
  number={25},
  pages={2340--2361},
  year={1977},
  publisher={ACS Publications}
}

@article{gillespie2001approximate,
  title={Approximate accelerated stochastic simulation of chemically reacting systems},
  author={Gillespie, Daniel T},
  journal={The Journal of chemical physics},
  volume={115},
  number={4},
  pages={1716--1733},
  year={2001},
  publisher={AIP Publishing}
}

@article{goldman2022gradient,
  title={Gradient-based {M}arkov chain {M}onte {C}arlo for {B}ayesian inference with non-differentiable priors},
  author={Goldman, Jacob Vorstrup and Sell, Torben and Singh, Sumeetpal Sidhu},
  journal={Journal of the American Statistical Association},
  volume={117},
  number={540},
  pages={2182--2193},
  year={2022},
  publisher={Taylor \& Francis}
}

@article{gozlan2010isoperimetry,
  title={Isoperimetry for product of heavy tails distributions},
  author={Gozlan, Nathael and Roberto, Cyril and Samson, Paul-Marie},
  journal={Progress in analysis and its applications},
  pages={470--478},
  year={2010},
  publisher={World Sci. Publ., Hackensack, NJ}
}

@article{hairer2003geometric,
  title={Geometric numerical integration illustrated by the {S}t{\"o}rmer--{V}erlet method},
  author={Hairer, Ernst and Lubich, Christian and Wanner, Gerhard},
  journal={Acta numerica},
  volume={12},
  pages={399--450},
  year={2003},
  publisher={Cambridge University Press}
}

@article{hardcastle2025diffusion,
	title        = {Diffusion {P}iecewise-{E}xponential models for survival extrapolation using {P}iecewise-{D}eterministic {M}onte {C}arlo},
	author       = {Hardcastle, Luke and Livingstone, Samuel and Baio, Gianluca},
	year         = {2025},
	journal      = {arXiv preprint arXiv:2505.05932}
}

@article{harland2017event,
  title={Event-chain {M}onte {C}arlo algorithms for three-and many-particle interactions},
  author={Harland, Julian and Michel, Manon and Kampmann, Tobias A and Kierfeld, Jan},
  journal={Europhysics Letters},
  volume={117},
  number={3},
  pages={30001},
  year={2017},
  publisher={IOP Publishing}
}

@article{hastings1970monte,
    ISSN = {00063444, 14643510},
    URL = {http://www.jstor.org/stable/2334940},
    author = {W. K. Hastings},
    journal = {Biometrika},
    number = {1},
    pages = {97--109},
    publisher = {[Oxford University Press, Biometrika Trust]},
    title = {Monte {C}arlo {S}ampling {M}ethods {U}sing {M}arkov {C}hains and {T}heir {A}pplications},
    urldate = {2025-10-07},
    volume = {57},
    year = {1970}
}

@article{he2023analysis,
  title={An analysis of transformed unadjusted {L}angevin algorithm for heavy-tailed sampling},
  author={He, Ye and Balasubramanian, Krishnakumar and Erdogdu, Murat A},
  journal={IEEE Transactions on Information Theory},
  volume={70},
  number={1},
  pages={571--593},
  year={2023},
  publisher={IEEE}
}

@InProceedings{hickling25,
  title = 	 {Flexible Tails for Normalizing Flows},
  author =       {Hickling, Tennessee and Prangle, Dennis},
  booktitle = 	 {Proceedings of the 42nd International Conference on Machine Learning},
  pages = 	 {23155--23178},
  year = 	 {2025},
  editor = 	 {Singh, Aarti and Fazel, Maryam and Hsu, Daniel and Lacoste-Julien, Simon and Berkenkamp, Felix and Maharaj, Tegan and Wagstaff, Kiri and Zhu, Jerry},
  volume = 	 {267},
  series = 	 {Proceedings of Machine Learning Research},
  month = 	 {13--19 Jul},
  publisher =    {PMLR},
  url = 	 {https://proceedings.mlr.press/v267/hickling25a.html}
}

@article{higham2001algorithmic,
  title={An algorithmic introduction to numerical simulation of stochastic differential equations},
  author={Higham, Desmond J},
  journal={SIAM review},
  volume={43},
  number={3},
  pages={525--546},
  year={2001},
  publisher={SIAM}
}

@book{higham2021introduction,
  title={An introduction to the numerical simulation of stochastic differential equations},
  author={Higham, Desmond and Kloeden, Peter},
  year={2021},
  publisher={SIAM}
}

@article{hill1975simple,
  title={A simple general approach to inference about the tail of a distribution},
  author={Hill, Bruce M},
  journal={The annals of statistics},
  pages={1163--1174},
  year={1975},
  publisher={JSTOR}
}

@article{hird.livingstone:24,
  author  = {Max Hird and Samuel Livingstone},
  title   = {Quantifying the {E}ffectiveness of {L}inear {P}reconditioning in {M}arkov {C}hain {M}onte {C}arlo},
  journal = {Journal of Machine Learning Research},
  year    = {2025},
  volume  = {26},
  number  = {119},
  pages   = {1--51},
  url     = {http://jmlr.org/papers/v26/23-1633.html}
}

@InProceedings{hird.livingstone.zanella:22,
author="Hird, Max
and Livingstone, Samuel
and Zanella, Giacomo",
editor="Keller, Alexander",
title="A fresh {T}ake on `{B}arker {D}ynamics' for {MCMC}",
booktitle="Monte Carlo and Quasi-Monte Carlo Methods",
year="2022",
publisher="Springer International Publishing",
address="Cham",
pages="169--184",
isbn="978-3-030-98319-2"
}

@book{hiriart.lemarechal:93,
  author    = {Jean‑Baptiste Hiriart‑Urruty and Claude Lemaréchal},
  title     = {Convex {A}nalysis and {M}inimization {A}lgorithms II: {A}dvanced {T}heory and {B}undle {M}ethods},
  series    = {Grundlehren der mathematischen Wissenschaften},
  volume    = {306},
  publisher = {Springer-Verlag},
  year      = {1993},
  address   = {Berlin, Heidelberg},
  isbn      = {978-3-662-06409-2, 978-3-540-56852-0},
  doi       = {10.1007/978-3-662-06409-2},
  url       = {https://link.springer.com/book/10.1007/978-3-662-06409-2}
}

@article{hodgkinson2021implicit,
  title={Implicit {L}angevin algorithms for sampling from log-concave densities},
  author={Hodgkinson, Liam and Salomone, Robert and Roosta, Fred},
  journal={Journal of Machine Learning Research},
  volume={22},
  number={136},
  pages={1--30},
  year={2021}
}

@misc{hoffman2019neutralizingbadgeometryhamiltonian,
      title={Neu{T}ra-lizing {B}ad {G}eometry in {H}amiltonian {M}onte {C}arlo Using {N}eural {T}ransport}, 
      author={Matthew Hoffman and Pavel Sountsov and Joshua V. Dillon and Ian Langmore and Dustin Tran and Srinivas Vasudevan},
      year={2019},
      eprint={1903.03704},
      archivePrefix={arXiv},
      primaryClass={stat.CO},
      url={https://arxiv.org/abs/1903.03704}, 
}

@article{hutzenthaler2011strong,
  title={Strong and weak divergence in finite time of {E}uler's method for stochastic differential equations with non-globally {L}ipschitz continuous coefficients},
  author={Hutzenthaler, Martin and Jentzen, Arnulf and Kloeden, Peter E},
  journal={Proceedings of the Royal Society A: Mathematical, Physical and Engineering Sciences},
  volume={467},
  number={2130},
  pages={1563--1576},
  year={2011},
  publisher={The Royal Society Publishing}
}

@article{hutzenthaler2012strong,
    title       ={Strong convergence of an explicit numerical method for {SDE}s with nonglobally {L}ipschitz continuous coefficients},
    author      ={Hutzenthaler, Martin and Jentzen, Arnulf and Kloeden, Peter E},
    journal     = {Annals of Applied Probability},
    year        ={2012}
}

@InProceedings{jaini20,
  title = 	 {Tails of {L}ipschitz Triangular Flows},
  author =       {Jaini, Priyank and Kobyzev, Ivan and Yu, Yaoliang and Brubaker, Marcus},
  booktitle = 	 {Proceedings of the 37th International Conference on Machine Learning},
  pages = 	 {4673--4681},
  year = 	 {2020},
  editor = 	 {III, Hal Daumé and Singh, Aarti},
  volume = 	 {119},
  series = 	 {Proceedings of Machine Learning Research},
  month = 	 {13--18 Jul},
  publisher =    {PMLR},
  url = 	 {https://proceedings.mlr.press/v119/jaini20a.html}
}

@article{johnson2012variable,
    title={Variable transformation to obtain geometric ergodicity in the random-walk {M}etropolis algorithm},
    author={Johnson, Leif T and Geyer, Charles J},
    journal={The Annals of Statistics},
    pages={3050--3076},
    year={2012},
    publisher={JSTOR}
}

@article{johnston2024strongly,
  title={A strongly monotonic polygonal Euler scheme},
  author={Johnston, Tim and Sabanis, Sotirios},
  journal={Journal of Complexity},
  volume={80},
  pages={101801},
  year={2024},
  publisher={Elsevier}
}

@article{krauth2021event,
  title={Event-chain {M}onte {C}arlo: Foundations, applications, and prospects},
  author={Krauth, Werner},
  journal={Frontiers in Physics},
  volume={9},
  pages={663457},
  year={2021},
  publisher={Frontiers Media SA}
}

@article{lange.little.taylor:89,
    ISSN = {01621459, 1537274X},
    URL = {http://www.jstor.org/stable/2290063},
    author = {Kenneth L. Lange and Roderick J. A. Little and Jeremy M. G. Taylor},
    journal = {Journal of the American Statistical Association},
    number = {408},
    pages = {881--896},
    publisher = {[American Statistical Association, Taylor & Francis, Ltd.]},
    title = {Robust {S}tatistical {M}odeling {U}sing the t {D}istribution},
    urldate = {2025-07-30},
    volume = {84},
    year = {1989}
}

@InProceedings{laszkiewicz22,
  title = 	 {Marginal {T}ail-{A}daptive {N}ormalizing {F}lows},
  author =       {Laszkiewicz, Mike and Lederer, Johannes and Fischer, Asja},
  booktitle = 	 {Proceedings of the 39th International Conference on Machine Learning},
  pages = 	 {12020--12048},
  year = 	 {2022},
  editor = 	 {Chaudhuri, Kamalika and Jegelka, Stefanie and Song, Le and Szepesvari, Csaba and Niu, Gang and Sabato, Sivan},
  volume = 	 {162},
  series = 	 {Proceedings of Machine Learning Research},
  month = 	 {17--23 Jul},
  publisher =    {PMLR},
  url = 	 {https://proceedings.mlr.press/v162/laszkiewicz22a.html}
}

@book{lee:13,
  title     = {Introduction to {S}mooth {M}anifolds},
  author    = {Lee, John M.},
  edition   = {2nd},
  series    = {Graduate Texts in Mathematics},
  volume    = {218},
  publisher = {Springer},
  address   = {New York, NY},
  year      = {2013},
  doi       = {10.1007/978-1-4419-9982-5},
  isbn      = {978-1-4419-9981-8}
}

@misc{leimkuhler.lohmann.whalley:25,
      title={A {L}angevin sampling algorithm inspired by the {A}dam optimizer}, 
      author={Benedict Leimkuhler and René Lohmann and Peter Whalley},
      year={2025},
      eprint={2504.18911},
      archivePrefix={arXiv},
      primaryClass={stat.CO},
      url={https://arxiv.org/abs/2504.18911}, 
}

@article{leroy2024adaptive,
	title        = {Adaptive stepsize algorithms for {L}angevin dynamics},
	author       = {Leroy, Alix and Leimkuhler, Benedict and Latz, Jonas and Higham, Desmond J},
	year         = {2024},
	journal      = {SIAM Journal on Scientific Computing},
	publisher    = {SIAM},
	volume       = {46},
	number       = {6},
	pages        = {A3574--A3598}
}

@inproceedings{lewis1978simulation,
author = {Lewis, Peter A.W. and Shedler, Gerald S.},
title = {Simulation methods for Poisson processes in nonstationary systems},
year = {1978},
publisher = {IEEE Press},
booktitle = {Proceedings of the 10th Conference on Winter Simulation - Volume 1},
pages = {155–163},
numpages = {9},
location = {Miami Beach, FL},
series = {WSC '78}
}

@InProceedings{liang22,
  title = 	 {{F}at{–}{T}ailed {V}ariational {I}nference with {A}nisotropic {T}ail {A}daptive {F}lows},
  author =       {Liang, Feynman and Mahoney, Michael and Hodgkinson, Liam},
  booktitle = 	 {Proceedings of the 39th International Conference on Machine Learning},
  pages = 	 {13257--13270},
  year = 	 {2022},
  editor = 	 {Chaudhuri, Kamalika and Jegelka, Stefanie and Song, Le and Szepesvari, Csaba and Niu, Gang and Sabato, Sivan},
  volume = 	 {162},
  series = 	 {Proceedings of Machine Learning Research},
  month = 	 {17--23 Jul},
  publisher =    {PMLR},
  url = 	 {https://proceedings.mlr.press/v162/liang22a.html}
}

@article{limpert.eckhard.werner.markus:01,
    author = {Limpert, Eckhard and Stahel, Werner A. and Abbt, Markus},
    title = {Log-normal {D}istributions across the Sciences: {K}eys and {C}lues},
    journal = {BioScience},
    volume = {51},
    number = {5},
    pages = {341-352},
    year = {2001},
    month = {05},
    issn = {0006-3568},
    doi = {10.1641/0006-3568(2001)051[0341:LNDATS]2.0.CO;2},
    url = {https://doi.org/10.1641/0006-3568(2001)051[0341:LNDATS]2.0.CO;2},
    eprint = {https://academic.oup.com/bioscience/article-pdf/51/5/341/26891292/51-5-341.pdf},
}

@misc{livingstone.nusken.vasdekis.zhang:24,
	title        = {Skew-symmetric schemes for stochastic differential equations with non-{L}ipschitz drift: an unadjusted {B}arker algorithm},
	author       = {Samuel Livingstone and Nikolas Nüsken and Giorgos Vasdekis and Rui-Yang Zhang},
	year         = {2024},
	url          = {https://arxiv.org/abs/2405.14373},
	eprint       = {2405.14373},
	archiveprefix = {arXiv},
	primaryclass = {math.PR}
}

@article{livingstone2019kinetic,
	title        = {Kinetic energy choice in {H}amiltonian/{H}ybrid {M}onte {C}arlo},
	author       = {Livingstone, Samuel and Faulkner, Michael F and Roberts, Gareth O},
	year         = {2019},
	journal      = {Biometrika},
	publisher    = {Oxford University Press},
	volume       = {106},
	number       = {2},
	pages        = {303--319}
}

@article{livingstone2022barker,
	title        = {The {B}arker proposal: Combining robustness and efficiency in gradient-based {MCMC}},
	author       = {Livingstone, Samuel and Zanella, Giacomo},
	year         = {2022},
	journal      = {Journal of the Royal Statistical Society Series B: Statistical Methodology},
	publisher    = {Oxford University Press},
	volume       = {84},
	number       = {2},
	pages        = {496--523}
}

@inproceedings{lu2017relativistic,
  title={Relativistic {M}onte {C}arlo},
  author={Lu, Xiaoyu and Perrone, Valerio and Hasenclever, Leonard and Teh, Yee Whye and Vollmer, Sebastian},
  booktitle={Artificial Intelligence and Statistics},
  pages={1236--1245},
  year={2017},
  organization={PMLR}
}

@article{lu2020complexity,
  title={Complexity of zigzag sampling algorithm for strongly log-concave distributions},
  author={Lu, Jianfeng and Wang, Lihan},
  journal={Statistics and Computing},
  volume={32},
  number={3},
  pages={48},
  year={2022},
  publisher={Springer}
}

@article{lu2022explicit,
  title={On explicit $\mathrm{L}^2$-convergence rate estimate for piecewise deterministic {M}arkov processes in {MCMC} algorithms},
  author={Lu, Jianfeng and Wang, Lihan},
  journal={The Annals of Applied Probability},
  volume={32},
  number={2},
  pages={1333--1361},
  year={2022},
  publisher={Institute of Mathematical Statistics}
}

@article{lytras2024tamed,
  title={Tamed {L}angevin sampling under weaker conditions},
  author={Lytras, Iosif and Mertikopoulos, Panayotis},
  journal={arXiv preprint arXiv:2405.17693},
  year={2024}
}

@article{lytras2025taming,
  title={Taming under isoperimetry},
  author={Lytras, Iosif and Sabanis, Sotirios},
  journal={Stochastic Processes and their Applications},
  pages={104684},
  year={2025},
  publisher={Elsevier}
}

@article{lytras2025ktula,
  title={k{TULA}: A {L}angevin sampling algorithm with improved {KL} bounds under super-linear log-gradients},
  author={Lytras, Iosif and Sabanis, Sotirios and Zhang, Ying},
  journal={arXiv preprint arXiv:2506.04878},
  year={2025}
}

@article{maggs2022large,
  title={Large-scale dynamics of event-chain {M}onte {C}arlo},
  author={Maggs, AC and Krauth, Werner},
  journal={Physical Review E},
  volume={105},
  number={1},
  pages={015309},
  year={2022},
  publisher={APS}
}

@article{malefaki.iliopoulos:08,
title = {On convergence of properly weighted samples to the target distribution},
journal = {Journal of Statistical Planning and Inference},
volume = {138},
number = {4},
pages = {1210-1225},
year = {2008},
issn = {0378-3758},
doi = {https://doi.org/10.1016/j.jspi.2007.05.030},
url = {https://www.sciencedirect.com/science/article/pii/S0378375807002261},
author = {Sonia Malefaki and George Iliopoulos}
}

@inproceedings{mauri2024robust,
  title={Robust approximate sampling via stochastic gradient {B}arker dynamics},
  author={Mauri, Lorenzo and Zanella, Giacomo},
  booktitle={International Conference on Artificial Intelligence and Statistics},
  pages={2107--2115},
  year={2024},
  organization={PMLR}
}

@article{metropolis1953equation,
  title={Equation of state calculations by fast computing machines},
  author={Metropolis, Nicholas and Rosenbluth, Arianna W and Rosenbluth, Marshall N and Teller, Augusta H and Teller, Edward},
  journal={The journal of chemical physics},
  volume={21},
  number={6},
  pages={1087--1092},
  year={1953},
  publisher={American Institute of Physics}
}

@article{michel2014generalized,
  title={Generalized event-chain {M}onte {C}arlo: Constructing rejection-free global-balance algorithms from infinitesimal steps},
  author={Michel, Manon and Kapfer, Sebastian C and Krauth, Werner},
  journal={The Journal of chemical physics},
  volume={140},
  number={5},
  year={2014},
  publisher={AIP Publishing}
}

@book{milstein2004stochastic,
  title={Stochastic numerics for mathematical physics},
  author={Milstein, Grigori N and Tretyakov, Michael V},
  volume={39},
  year={2004},
  publisher={Springer}
}

@article{milstein2005numerical,
  title={Numerical integration of stochastic differential equations with nonglobally {L}ipschitz coefficients},
  author={Milstein, Grigori N and Tretyakov, Michael V},
  journal={SIAM journal on numerical analysis},
  volume={43},
  number={3},
  pages={1139--1154},
  year={2005},
  publisher={SIAM}
}

@article{natarovskii2021quantitative,
    ISSN = {10505164, 21688737},
    URL = {https://www.jstor.org/stable/27174878},
    author = {Viacheslav Natarovskii and Daniel Rudolf and Björn Sprungk},
    journal = {The Annals of Applied Probability},
    number = {2},
    pages = {pp. 806--825},
    publisher = {Institute of Mathematical Statistics},
    title = {QUANTITATIVE SPECTRAL GAP ESTIMATE AND {W}ASSERSTEIN CONTRACTION OF SIMPLE SLICE SAMPLING},
    urldate = {2025-10-07},
    volume = {31},
    year = {2021}
}

@article{neal2003slice,
	title        = {Slice sampling},
	author       = {Neal, Radford M},
	year         = {2003},
	journal      = {The annals of statistics},
	publisher    = {Institute of Mathematical Statistics},
	volume       = {31},
	number       = {3},
	pages        = {705--767}
}

@article{neal2011mcmc,
    title={{MCMC} using {H}amiltonian dynamics},
    author={Neal, Radford M},
    journal={Handbook of {M}arkov {C}hain {M}onte {C}arlo},
    volume={2},
    number={11},
    pages={2},
    year={2011},
    publisher={Chapman and Hall/CRC}
}

@article{nishimura2025zigzag,
    title={Zigzag path connects two {M}onte {C}arlo samplers: {H}amiltonian counterpart to a {P}iecewise-{D}eterministic {M}arkov process},
    author={Nishimura, Akihiko and Zhang, Zhenyu and Suchard, Marc A},
    journal={Journal of the American Statistical Association},
    volume={120},
    number={550},
    pages={1077--1089},
    year={2025},
    publisher={Taylor \& Francis}
}

@article{osmundsen.kleppe.leisenfeld:21,
    author = {Kjartan Kloster Osmundsen and Tore Selland Kleppe and Roman Liesenfeld},
    title = {Importance {S}ampling-{B}ased {T}ransport {M}ap {H}amiltonian {M}onte {C}arlo for {B}ayesian {H}ierarchical {M}odels},
    journal = {Journal of Computational and Graphical Statistics},
    volume = {30},
    number = {4},
    pages = {906--919},
    year = {2021},
    publisher = {ASA Website},
    doi = {10.1080/10618600.2021.1923519},
    URL = {https://doi.org/10.1080/10618600.2021.1923519},
    eprint = {https://doi.org/10.1080/10618600.2021.1923519}
}

@article{pagani2024nuzz,
  title={Nu{ZZ}: {N}umerical {Z}ig-{Z}ag for general models},
  author={Pagani, Filippo and Chevallier, Augustin and Power, Sam and House, Thomas and Cotter, Simon},
  journal={Statistics and Computing},
  volume={34},
  number={1},
  pages={61},
  year={2024},
  publisher={Springer}
}

@incollection{papaspiliopoulos2009methodological,
  title       = {Monte {C}arlo Probabilistic Inference for Diffusion Processes: A Methodological Framework},
  author      = {Papaspiliopoulos, Omiros},
  booktitle   = {Bayesian Time Series Models},
  editor      = {Barber, David and Cemgil, A. Taylan and Chiappa, Silvia},
  publisher   = {Cambridge University Press},
  address     = {Cambridge},
  year        = {2011},
  pages       = {82--103}
}

@article{pereyra2016proximal,
	title        = {Proximal {M}arkov chain {M}onte {C}arlo algorithms},
	author       = {Pereyra, Marcelo},
	year         = {2016},
	journal      = {Statistics and Computing},
	publisher    = {Springer},
	volume       = {26},
	pages        = {745--760}
}

@article{pereyra2020accelerating,
  title={Accelerating proximal {M}arkov chain {M}onte {C}arlo by using an explicit stabilized method},
  author={Pereyra, Marcelo and Mieles, Luis Vargas and Zygalakis, Konstantinos C},
  journal={SIAM Journal on Imaging Sciences},
  volume={13},
  number={2},
  pages={905--935},
  year={2020},
  publisher={SIAM}
}

@article{pillai.stuart.thiery:12,
  title        = {Optimal scaling and diffusion limits for the {L}angevin algorithm in high dimensions},
  author       = {Pillai, Natesh S. and Stuart, Andrew M. and Thiéry, Alexandre H.},
  journal      = {The Annals of Applied Probability},
  volume       = {22},
  number       = {6},
  pages        = {2320--2356},
  year         = {2012},
  doi          = {10.1214/11-AAP828},
}

@article{platen1999introduction,
  title={An introduction to numerical methods for stochastic differential equations},
  author={Platen, Eckhard},
  journal={Acta numerica},
  volume={8},
  pages={197--246},
  year={1999},
  publisher={Cambridge University Press}
}

@article{power2024weak,
  title={Weak {P}oincar\'{e} inequality comparisons for ideal and hybrid slice sampling},
  author={Power, Sam and Rudolf, Daniel and Sprungk, Bj{\"o}rn and Wang, Andi Q},
  journal={arXiv preprint arXiv:2402.13678},
  year={2024}
}

@book{robert.casella:05, author = {Robert, Christian P. and Casella, George}, title = {{{M}onte {C}arlo {S}tatistical {M}ethods}}, year = {2005}, isbn = {0387212396}, publisher = {Springer-Verlag}, address = {Berlin, Heidelberg} }

@article{roberts.rosenthal:02,
    author = {Roberts, Gareth O. and Rosenthal, Jeffrey S.},
    title = {Optimal {S}caling of {D}iscrete {A}pproximations to {L}angevin {D}iffusions},
    journal = {Journal of the Royal Statistical Society Series B: Statistical Methodology},
    volume = {60},
    number = {1},
    pages = {255-268},
    year = {2002},
    month = {01},
    issn = {1369-7412},
    doi = {10.1111/1467-9868.00123},
    url = {https://doi.org/10.1111/1467-9868.00123},
    eprint = {https://academic.oup.com/jrsssb/article-pdf/60/1/255/49589077/jrsssb_60_1_255.pdf},
}

@article{roberts.stramer:02,
	title ={{Langevin {D}iffusions and {M}etropolis-{H}astings {A}lgorithms}},
	author={Gareth O. Roberts and Osnat Stramer},
	journal={Methodology and Computing in Applied Probability},
	year={2002}
}

@article{roberts.tweedie:96,
    author = {Gareth O. Roberts and Richard L. Tweedie},
    title = {{Exponential convergence of {L}angevin distributions and their discrete approximations}},
    volume = {2},
    journal = {Bernoulli},
    number = {4},
    publisher = {Bernoulli Society for Mathematical Statistics and Probability},
    pages = {341 -- 363},
    year = {1996},
}

@article{rossky1978brownian,
    title={Brownian dynamics as smart {M}onte {C}arlo simulation},
    author={Rossky, Peter J and Doll, Jimmie D and Friedman, Harold L},
    journal={The Journal of Chemical Physics},
    volume={69},
    number={10},
    pages={4628--4633},
    year={1978},
    publisher={American Institute of Physics}
}

@incollection{rudolf2013hit,
  title={Hit-and-{R}un for numerical integration},
  author={Rudolf, Daniel},
  booktitle={Monte Carlo and Quasi-Monte Carlo Methods 2012},
  pages={597--612},
  year={2013},
  publisher={Springer}
}

@article{sherlock2010random,
author = {Chris Sherlock and Paul Fearnhead and Gareth O. Roberts},
title = {{The {R}andom {W}alk {M}etropolis: {L}inking {T}heory and {P}ractice {T}hrough a {C}ase {S}tudy}},
volume = {25},
journal = {Statistical Science},
number = {2},
publisher = {Institute of Mathematical Statistics},
pages = {172 -- 190},
year = {2010},
doi = {10.1214/10-STS327},
URL = {https://doi.org/10.1214/10-STS327}
}

@article{shukla2025mcmc,
	title        = {{MCMC} {I}mportance {S}ampling via {M}oreau-{Y}osida {E}nvelopes},
	author       = {Shukla, Apratim and Vats, Dootika and Chi, Eric C},
	year         = {2025},
	journal      = {arXiv preprint arXiv:2501.02228}
}

@article{shukla2025proximal,
  title={Proximal {H}amiltonian {M}onte {C}arlo},
  author={Shukla, Apratim and Vats, Dootika and Chi, Eric C},
  journal={arXiv preprint arXiv:2510.22252},
  year={2025}
}

@article{sutton2023concave,
  title={Concave-convex {PDMP}-based sampling},
  author={Sutton, Matthew and Fearnhead, Paul},
  journal={Journal of Computational and Graphical Statistics},
  volume={32},
  number={4},
  pages={1425--1435},
  year={2023},
  publisher={Taylor \& Francis}
}

@article{szpruch2013v,
  title={V-stable tamed {E}uler schemes},
  author={Szpruch, Lukasz},
  journal={arXiv preprint arXiv:1310.0785},
  year={2013}
}

@article{tierney1998note,
  title={A note on {M}etropolis-{H}astings kernels for general state spaces},
  author={Tierney, Luke},
  journal={Annals of applied probability},
  pages={1--9},
  year={1998},
  publisher={JSTOR}
}

@article{vasdekis2022note,
	title        = {A note on the polynomial ergodicity of the one-dimensional {Z}ig-{Z}ag process},
	author       = {Vasdekis, Giorgos and Roberts, Gareth O},
	year         = {2022},
	journal      = {Journal of Applied Probability},
	publisher    = {Cambridge University Press},
	volume       = {59},
	number       = {3},
	pages        = {895--903}
}

@article{vasdekis.roberts:23,
	title        = {{Speed up {Z}ig-{Z}ag}},
	author       = {Vasdekis, Giorgos and Roberts, Gareth O},
	year         = {2023},
	journal      = {The Annals of Applied Probability},
	publisher    = {Institute of Mathematical Statistics},
	volume       = {33},
	number       = {6A},
	pages        = {4693 -- 4746},
	doi          = {10.1214/23-AAP1930},
	url          = {https://doi.org/10.1214/23-AAP1930}
}

@article{vogrinc2023optimal,
  title={Optimal design of the {B}arker proposal and other locally balanced {M}etropolis--{H}astings algorithms},
  author={Vogrinc, Jure and Livingstone, Samuel and Zanella, Giacomo},
  journal={Biometrika},
  volume={110},
  number={3},
  pages={579--595},
  year={2023},
  publisher={Oxford University Press}
}

@article{vogrinc.kendall:19,
    author = {Jure Vogrinc and Wilfrid S. Kendall},
    title = {{Counterexamples for optimal scaling of {M}etropolis–{H}astings chains with rough target densities}},
    volume = {31},
    journal = {The Annals of Applied Probability},
    number = {2},
    publisher = {Institute of Mathematical Statistics},
    pages = {972 -- 1019},
    year = {2021},
    doi = {10.1214/20-AAP1612},
    URL = {https://doi.org/10.1214/20-AAP1612}
}

@article{wibisono2019proximal,
  title={Proximal {L}angevin algorithm: {R}apid convergence under isoperimetry},
  author={Wibisono, Andre},
  journal={arXiv preprint arXiv:1911.01469},
  year={2019}
}

@article{wu2022minimax,
  title={Minimax mixing time of the {M}etropolis-adjusted {L}angevin algorithm for log-concave sampling},
  author={Wu, Keru and Schmidler, Scott and Chen, Yuansi},
  journal={Journal of Machine Learning Research},
  volume={23},
  number={270},
  pages={1--63},
  year={2022}
}

@article{yang.latuszynski.roberts:24,
	title        = {{Stereographic {M}arkov chain {M}onte {C}arlo}},
	author       = {Jun Yang and Krzysztof Łatuszyński and Gareth O. Roberts},
	year         = {2024},
	journal      = {The Annals of Statistics},
	publisher    = {Institute of Mathematical Statistics},
	volume       = {52},
	number       = {6},
	pages        = {2692 -- 2713},
	doi          = {10.1214/24-AOS2426},
	url          = {https://doi.org/10.1214/24-AOS2426}
}

@article{yao.wei.yu:14,
    title = {Robust mixture regression using the t-distribution},
    journal = {Computational Statistics and Data Analysis},
    volume = {71},
    pages = {116-127},
    year = {2014},
    issn = {0167-9473},
    doi = {https://doi.org/10.1016/j.csda.2013.07.019},
    url = {https://www.sciencedirect.com/science/article/pii/S0167947313002648},
    author = {Weixin Yao and Yan Wei and Chun Yu}
}

@article{zanella2020informed,
    title={Informed proposals for local {MCMC} in discrete spaces},
    author={Zanella, Giacomo},
    journal={Journal of the American Statistical Association},
    volume={115},
    number={530},
    pages={852--865},
    year={2020},
    publisher={Taylor \& Francis}
}
\bibliographystyle{plain}

\end{document}